\newif\ifdraft
\newif\iffull
\newif\ifcomment
\newif\iflatexdiff
\newif\ifbibtex
\newif\ifpreprint
\latexdifftrue   
\preprinttrue    
\fulltrue        
\newif\ifplbpaper
\newif\ifapspaper
\apspapertrue

\ifpreprint

\fi
\def\dvers{v2.0}       
\def\dtitle{Centrality Dependence of $\pi$, K, p Production in \PbPb\ Collisions at \snn~=~2.76~TeV}  
\def\stitle{Centrality Dependence of $\pi$, K, p in \PbPb\ at \snn~=~2.76~TeV} 
\def\phnum{2013-019}     
\def\phdat{16 Feb 2013}  
\def\bibstname{apsrev4-1}   
\ifpreprint
\documentclass[ALICE,manyauthors,12pt]{cernphprep}
\usepackage[comma,square,numbers,sort&compress]{natbib}
\newcommand{\mywidth}{0.5\linewidth}
\else 
\newcommand{\mywidth}{0.9\linewidth}
\ifplbpaper
\documentclass[10pt,aps,prc,twocolumn,superscriptaddress,altaffilletter,nobibnotes,nofootinbib,floatfix]{revtex4-1}
\biboptions{comma,square,numbers,sort&compress}
\fi
\ifapspaper
\documentclass[10pt,aps,prc,twocolumn,superscriptaddress,altaffilletter,nobibnotes,nofootinbib,floatfix]{revtex4-1}
\newenvironment{frontmatter}{}{\maketitle}
\fi
\fi
\usepackage{graphicx}  
\usepackage{dcolumn}   
\usepackage{bm}        
\usepackage{amssymb}   
\usepackage{amsfonts}
\usepackage{graphics}
\usepackage{grffile}   
\usepackage{epsfig}
\usepackage{units}
\usepackage{hyperref}
\usepackage[usenames]{color}
\usepackage[normalem]{ulem} 
\usepackage{color}
\usepackage[utf8]{inputenc}
\usepackage[T1]{fontenc}
\usepackage{multirow}
\usepackage{svn-multi}
\svnidlong 
{$HeadURL: svn+ssh://svn.cern.ch/reps/mfPapers/2010SpectraID/trunk/PRCPbPb/prcSpectra.tex $} 
{$LastChangedDate: 2013-07-25 08:50:29 +0200 (Thu, 25 Jul 2013) $} 
{$LastChangedRevision: 1002 $} 
{$LastChangedBy: mfloris $}
\svnid
{$Id: prcSpectraUniversal.tex 815 2013-01-30 09:59:14Z mfloris $}
\usepackage{rotating}

\iflatexdiff
\RequirePackage{color}\definecolor{RED}{rgb}{1,0,0}\definecolor{BLUE}{rgb}{0,0,1}

\fi

\newcommand{\VZERO}        {\rm{VZERO}}

\newcommand{\pbar}         {$\rm\overline{p}$}

\newcommand{\dedx}         {\ensuremath{\mathrm{d}E/\mathrm{d}x}}

\newcommand{\PbPb}         {\mbox{Pb--Pb}}
\newcommand{\AuAu}         {\mbox{Au--Au}}

\newcommand{\dNdeta}       {\ensuremath{\mathrm{d}N_\mathrm{ch}/\mathrm{d}\eta}}

\newcommand{\dcaxy}        {\ensuremath{{\rm DCA}_{xy}}}
\newcommand{\dcaz}         {\ensuremath{{\rm DCA}_{z}}}
\newcommand{\EcrossB}      {E$\times$B}

\newcommand{\pt}           {\ensuremath{p_{\rm T}}}
\newcommand{\ppi}          {\ensuremath{{\rm p}/\pi}}
\newcommand{\kpi}          {\ensuremath{{\rm K}/\pi}}
\newcommand{\mt}           {\ensuremath{m_{\rm T}}}
\newcommand{\snn}          {\ensuremath{\sqrt{s_{\rm NN}}}}

\newcommand{\Tfo}          {\ensuremath{{T}_{\rm kin}}}
\newcommand{\Tch}          {\ensuremath{{T}_{\rm ch}}}

\newcommand{\avbT}         {\ensuremath{\left< \beta_{\rm T}\right>}}
\newcommand{\avpT}         {\ensuremath{\left< \pt \right>}}
\newcommand{\muB}          {\ensuremath{\mu_{B}}}

\newcommand{\gevc}[1]      {\ensuremath{{\rm GeV}/c}}

\newcommand{\warn}[1]      {}

\graphicspath{{./img/}}
\ifdraft
\usepackage{lineno}
\linenumbers
\modulolinenumbers[5]
\usepackage{fancyhdr}
\fi
\begin{document}
\ifpreprint
\begin{titlepage}
\PHnumber{\phnum}    
\PHdate{\phdat}
\title{\dtitle}
\ShortTitle{\stitle}
\iffull
\Collaboration{ALICE Collaboration~\thanks{See Appendix~\ref{app:collab} for the list of collaboration members}}
\else
\Collaboration{ALICE Collaboration}
\fi
\ShortAuthor{ALICE Collaboration} 
\ifdraft
\begin{center}
\today\\ \color{red}DRAFT \dvers\ \hspace{0.3cm} \$Revision: 1002 $\color{white}:$\$\color{black}\vspace{0.3cm}
\end{center}
\fi
\else
\begin{frontmatter}
\title{\dtitle}
\iffull
\ifpreprint

\begingroup
\small
\begin{flushleft}
B.~Abelev\Irefn{org1234}\And
J.~Adam\Irefn{org1274}\And
D.~Adamov\'{a}\Irefn{org1283}\And
A.M.~Adare\Irefn{org1260}\And
M.M.~Aggarwal\Irefn{org1157}\And
G.~Aglieri~Rinella\Irefn{org1192}\And
M.~Agnello\Irefn{org1313}\textsuperscript{,}\Irefn{org1017688}\And
A.G.~Agocs\Irefn{org1143}\And
A.~Agostinelli\Irefn{org1132}\And
Z.~Ahammed\Irefn{org1225}
% ...
\renewcommand\labelenumi{\textsuperscript{\theenumi}~}
\section*{Affiliation notes}
\renewcommand\theenumi{\roman{enumi}}
\begin{Authlist}
\item \Adef{0}Deceased
\item \Adef{M.V.Lomonosov Moscow State University, D.V.Skobeltsyn Institute of Nuclear Physics, Moscow, Russia}Also at: M.V.Lomonosov Moscow State University, D.V.Skobeltsyn Institute of Nuclear Physics, Moscow, Russia
\item \Adef{University of Belgrade, Faculty of Physics and Vinvca Institute of Nuclear Sciences, Belgrade, Serbia}Also at: University of Belgrade, Faculty of Physics and Vinvca Institute of Nuclear Sciences, Belgrade, Serbia
\item \Adef{Institute of Theoretical Physics, University of Wroclaw, Wroclaw, Poland}Also at: Institute of Theoretical Physics, University of Wroclaw, Wroclaw, Poland
\end{Authlist}
\section*{Collaboration Institutes}
\renewcommand\theenumi{\arabic{enumi}~}
\begin{Authlist}
\item \Idef{org1332}A. I. Alikhanyan National Science Laboratory (Yerevan Physics Institute) Foundation, Yerevan, Armenia
\item \Idef{org1279}Benem\'{e}rita Universidad Aut\'{o}noma de Puebla, Puebla, Mexico
\item \Idef{org1220}Bogolyubov Institute for Theoretical Physics, Kiev, Ukraine
\item \Idef{org20959}Bose Institute, Department of Physics and Centre for Astroparticle Physics and Space Science (CAPSS), Kolkata, India
\item \Idef{org1262}Budker Institute for Nuclear Physics, Novosibirsk, Russia
\item \Idef{org1292}California Polytechnic State University, San Luis Obispo, California, United States
\item \Idef{org1329}Central China Normal University, Wuhan, China
\item \Idef{org14939}Centre de Calcul de l'IN2P3, Villeurbanne, France
\item \Idef{org1197}Centro de Aplicaciones Tecnol\'{o}gicas y Desarrollo Nuclear (CEADEN), Havana, Cuba
\item \Idef{org1242}Centro de Investigaciones Energ\'{e}ticas Medioambientales y Tecnol\'{o}gicas (CIEMAT), Madrid, Spain
\item \Idef{org1244}Centro de Investigaci\'{o}n y de Estudios Avanzados (CINVESTAV), Mexico City and M\'{e}rida, Mexico
\item \Idef{org1335}Centro Fermi - Museo Storico della Fisica e Centro Studi e Ricerche ``Enrico Fermi'', Rome, Italy
\item \Idef{org17347}Chicago State University, Chicago, United States
\item \Idef{org1288}Commissariat \`{a} l'Energie Atomique, IRFU, Saclay, France
\item \Idef{org15782}COMSATS Institute of Information Technology (CIIT), Islamabad, Pakistan
\item \Idef{org1294}Departamento de F\'{\i}sica de Part\'{\i}culas and IGFAE, Universidad de Santiago de Compostela, Santiago de Compostela, Spain
\item \Idef{org1106}Department of Physics Aligarh Muslim University, Aligarh, India
\item \Idef{org1121}Department of Physics and Technology, University of Bergen, Bergen, Norway
\item \Idef{org1162}Department of Physics, Ohio State University, Columbus, Ohio, United States
\item \Idef{org1300}Department of Physics, Sejong University, Seoul, South Korea
\item \Idef{org1268}Department of Physics, University of Oslo, Oslo, Norway
\item \Idef{org1145}Dipartimento di Fisica dell'Universit\`{a} and Sezione INFN, Cagliari, Italy
\item \Idef{org1312}Dipartimento di Fisica dell'Universit\`{a} and Sezione INFN, Turin, Italy
\item \Idef{org1315}Dipartimento di Fisica dell'Universit\`{a} and Sezione INFN, Trieste, Italy
\item \Idef{org1285}Dipartimento di Fisica dell'Universit\`{a} `La Sapienza' and Sezione INFN, Rome, Italy
\item \Idef{org1154}Dipartimento di Fisica e Astronomia dell'Universit\`{a} and Sezione INFN, Catania, Italy
\item \Idef{org1132}Dipartimento di Fisica e Astronomia dell'Universit\`{a} and Sezione INFN, Bologna, Italy
\item \Idef{org1270}Dipartimento di Fisica e Astronomia dell'Universit\`{a} and Sezione INFN, Padova, Italy
\item \Idef{org1290}Dipartimento di Fisica `E.R.~Caianiello' dell'Universit\`{a} and Gruppo Collegato INFN, Salerno, Italy
\item \Idef{org1103}Dipartimento di Scienze e Innovazione Tecnologica dell'Universit\`{a} del Piemonte Orientale and Gruppo Collegato INFN, Alessandria, Italy
\item \Idef{org1114}Dipartimento Interateneo di Fisica `M.~Merlin' and Sezione INFN, Bari, Italy
\item \Idef{org1237}Division of Experimental High Energy Physics, University of Lund, Lund, Sweden
\item \Idef{org1192}European Organization for Nuclear Research (CERN), Geneva, Switzerland
\item \Idef{org1227}Fachhochschule K\"{o}ln, K\"{o}ln, Germany
\item \Idef{org1122}Faculty of Engineering, Bergen University College, Bergen, Norway
\item \Idef{org1136}Faculty of Mathematics, Physics and Informatics, Comenius University, Bratislava, Slovakia
\item \Idef{org1274}Faculty of Nuclear Sciences and Physical Engineering, Czech Technical University in Prague, Prague, Czech Republic
\item \Idef{org1229}Faculty of Science, P.J.~\v{S}af\'{a}rik University, Ko\v{s}ice, Slovakia
\item \Idef{org1184}Frankfurt Institute for Advanced Studies, Johann Wolfgang Goethe-Universit\"{a}t Frankfurt, Frankfurt, Germany
\item \Idef{org1215}Gangneung-Wonju National University, Gangneung, South Korea
\item \Idef{org20958}Gauhati University, Department of Physics, Guwahati, India
\item \Idef{org1212}Helsinki Institute of Physics (HIP) and University of Jyv\"{a}skyl\"{a}, Jyv\"{a}skyl\"{a}, Finland
\item \Idef{org1203}Hiroshima University, Hiroshima, Japan
\item \Idef{org1254}Indian Institute of Technology Bombay (IIT), Mumbai, India
\item \Idef{org36378}Indian Institute of Technology Indore, Indore, India (IITI)
\item \Idef{org1266}Institut de Physique Nucl\'{e}aire d'Orsay (IPNO), Universit\'{e} Paris-Sud, CNRS-IN2P3, Orsay, France
\item \Idef{org1277}Institute for High Energy Physics, Protvino, Russia
\item \Idef{org1249}Institute for Nuclear Research, Academy of Sciences, Moscow, Russia
\item \Idef{org1320}Nikhef, National Institute for Subatomic Physics and Institute for Subatomic Physics of Utrecht University, Utrecht, Netherlands
\item \Idef{org1250}Institute for Theoretical and Experimental Physics, Moscow, Russia
\item \Idef{org1230}Institute of Experimental Physics, Slovak Academy of Sciences, Ko\v{s}ice, Slovakia
\item \Idef{org1127}Institute of Physics, Bhubaneswar, India
\item \Idef{org1275}Institute of Physics, Academy of Sciences of the Czech Republic, Prague, Czech Republic
\item \Idef{org1139}Institute of Space Sciences (ISS), Bucharest, Romania
\item \Idef{org27399}Institut f\"{u}r Informatik, Johann Wolfgang Goethe-Universit\"{a}t Frankfurt, Frankfurt, Germany
\item \Idef{org1185}Institut f\"{u}r Kernphysik, Johann Wolfgang Goethe-Universit\"{a}t Frankfurt, Frankfurt, Germany
\item \Idef{org1177}Institut f\"{u}r Kernphysik, Technische Universit\"{a}t Darmstadt, Darmstadt, Germany
\item \Idef{org1256}Institut f\"{u}r Kernphysik, Westf\"{a}lische Wilhelms-Universit\"{a}t M\"{u}nster, M\"{u}nster, Germany
\item \Idef{org1246}Instituto de Ciencias Nucleares, Universidad Nacional Aut\'{o}noma de M\'{e}xico, Mexico City, Mexico
\item \Idef{org1247}Instituto de F\'{\i}sica, Universidad Nacional Aut\'{o}noma de M\'{e}xico, Mexico City, Mexico
\item \Idef{org1308}Institut Pluridisciplinaire Hubert Curien (IPHC), Universit\'{e} de Strasbourg, CNRS-IN2P3, Strasbourg, France
\item \Idef{org1182}Joint Institute for Nuclear Research (JINR), Dubna, Russia
\item \Idef{org1199}Kirchhoff-Institut f\"{u}r Physik, Ruprecht-Karls-Universit\"{a}t Heidelberg, Heidelberg, Germany
\item \Idef{org20954}Korea Institute of Science and Technology Information, Daejeon, South Korea
\item \Idef{org1017642}KTO Karatay University, Konya, Turkey
\item \Idef{org1160}Laboratoire de Physique Corpusculaire (LPC), Clermont Universit\'{e}, Universit\'{e} Blaise Pascal, CNRS--IN2P3, Clermont-Ferrand, France
\item \Idef{org1194}Laboratoire de Physique Subatomique et de Cosmologie (LPSC), Universit\'{e} Joseph Fourier, CNRS-IN2P3, Institut Polytechnique de Grenoble, Grenoble, France
\item \Idef{org1187}Laboratori Nazionali di Frascati, INFN, Frascati, Italy
\item \Idef{org1232}Laboratori Nazionali di Legnaro, INFN, Legnaro, Italy
\item \Idef{org1125}Lawrence Berkeley National Laboratory, Berkeley, California, United States
\item \Idef{org1234}Lawrence Livermore National Laboratory, Livermore, California, United States
\item \Idef{org1251}Moscow Engineering Physics Institute, Moscow, Russia
\item \Idef{org1322}National Centre for Nuclear Studies, Warsaw, Poland
\item \Idef{org1140}National Institute for Physics and Nuclear Engineering, Bucharest, Romania
\item \Idef{org1017626}National Institute of Science Education and Research, Bhubaneswar, India
\item \Idef{org1165}Niels Bohr Institute, University of Copenhagen, Copenhagen, Denmark
\item \Idef{org1109}Nikhef, National Institute for Subatomic Physics, Amsterdam, Netherlands
\item \Idef{org1283}Nuclear Physics Institute, Academy of Sciences of the Czech Republic, \v{R}e\v{z} u Prahy, Czech Republic
\item \Idef{org1264}Oak Ridge National Laboratory, Oak Ridge, Tennessee, United States
\item \Idef{org1189}Petersburg Nuclear Physics Institute, Gatchina, Russia
\item \Idef{org1170}Physics Department, Creighton University, Omaha, Nebraska, United States
\item \Idef{org1157}Physics Department, Panjab University, Chandigarh, India
\item \Idef{org1112}Physics Department, University of Athens, Athens, Greece
\item \Idef{org1152}Physics Department, University of Cape Town and  iThemba LABS, National Research Foundation, Somerset West, South Africa
\item \Idef{org1209}Physics Department, University of Jammu, Jammu, India
\item \Idef{org1207}Physics Department, University of Rajasthan, Jaipur, India
\item \Idef{org1200}Physikalisches Institut, Ruprecht-Karls-Universit\"{a}t Heidelberg, Heidelberg, Germany
\item \Idef{org1017688}Politecnico di Torino, Turin, Italy
\item \Idef{org1325}Purdue University, West Lafayette, Indiana, United States
\item \Idef{org1281}Pusan National University, Pusan, South Korea
\item \Idef{org1176}Research Division and ExtreMe Matter Institute EMMI, GSI Helmholtzzentrum f\"ur Schwerionenforschung, Darmstadt, Germany
\item \Idef{org1334}Rudjer Bo\v{s}kovi\'{c} Institute, Zagreb, Croatia
\item \Idef{org1298}Russian Federal Nuclear Center (VNIIEF), Sarov, Russia
\item \Idef{org1252}Russian Research Centre Kurchatov Institute, Moscow, Russia
\item \Idef{org1224}Saha Institute of Nuclear Physics, Kolkata, India
\item \Idef{org1130}School of Physics and Astronomy, University of Birmingham, Birmingham, United Kingdom
\item \Idef{org1338}Secci\'{o}n F\'{\i}sica, Departamento de Ciencias, Pontificia Universidad Cat\'{o}lica del Per\'{u}, Lima, Peru
\item \Idef{org1313}Sezione INFN, Turin, Italy
\item \Idef{org1115}Sezione INFN, Bari, Italy
\item \Idef{org1133}Sezione INFN, Bologna, Italy
\item \Idef{org1146}Sezione INFN, Cagliari, Italy
\item \Idef{org1155}Sezione INFN, Catania, Italy
\item \Idef{org1271}Sezione INFN, Padova, Italy
\item \Idef{org1286}Sezione INFN, Rome, Italy
\item \Idef{org1316}Sezione INFN, Trieste, Italy
\item \Idef{org36377}Nuclear Physics Group, STFC Daresbury Laboratory, Daresbury, United Kingdom
\item \Idef{org1258}SUBATECH, Ecole des Mines de Nantes, Universit\'{e} de Nantes, CNRS-IN2P3, Nantes, France
\item \Idef{org35706}Suranaree University of Technology, Nakhon Ratchasima, Thailand
\item \Idef{org1304}Technical University of Split FESB, Split, Croatia
\item \Idef{org1017659}Technische Universit\"{a}t M\"{u}nchen, Munich, Germany
\item \Idef{org1168}The Henryk Niewodniczanski Institute of Nuclear Physics, Polish Academy of Sciences, Cracow, Poland
\item \Idef{org17361}The University of Texas at Austin, Physics Department, Austin, TX, United States
\item \Idef{org1173}Universidad Aut\'{o}noma de Sinaloa, Culiac\'{a}n, Mexico
\item \Idef{org1296}Universidade de S\~{a}o Paulo (USP), S\~{a}o Paulo, Brazil
\item \Idef{org1149}Universidade Estadual de Campinas (UNICAMP), Campinas, Brazil
\item \Idef{org1239}Universit\'{e} de Lyon, Universit\'{e} Lyon 1, CNRS/IN2P3, IPN-Lyon, Villeurbanne, France
\item \Idef{org1205}University of Houston, Houston, Texas, United States
\item \Idef{org20371}University of Technology and Austrian Academy of Sciences, Vienna, Austria
\item \Idef{org1222}University of Tennessee, Knoxville, Tennessee, United States
\item \Idef{org1310}University of Tokyo, Tokyo, Japan
\item \Idef{org1318}University of Tsukuba, Tsukuba, Japan
\item \Idef{org21360}Eberhard Karls Universit\"{a}t T\"{u}bingen, T\"{u}bingen, Germany
\item \Idef{org1225}Variable Energy Cyclotron Centre, Kolkata, India
\item \Idef{org1306}V.~Fock Institute for Physics, St. Petersburg State University, St. Petersburg, Russia
\item \Idef{org1323}Warsaw University of Technology, Warsaw, Poland
\item \Idef{org1179}Wayne State University, Detroit, Michigan, United States
\item \Idef{org1143}Wigner Research Centre for Physics, Hungarian Academy of Sciences, Budapest, Hungary
\item \Idef{org1260}Yale University, New Haven, Connecticut, United States
\item \Idef{org15649}Yildiz Technical University, Istanbul, Turkey
\item \Idef{org1301}Yonsei University, Seoul, South Korea
\item \Idef{org1327}Zentrum f\"{u}r Technologietransfer und Telekommunikation (ZTT), Fachhochschule Worms, Worms, Germany
\end{Authlist}
\endgroup

\else
\input{authors-paper.tex}            
\fi
\author{(ALICE Collaboration)}
\else
\ifdraft
\author{ALICE Collaboration \\ \vspace{0.3cm} 
\today\\ \color{red}DRAFT \dvers\ \hspace{0.3cm} \$Revision: 1002 $\color{white}:$\$\color{black}}
\else
\author{ALICE Collaboration}
\fi
\fi
\fi
\begin{abstract}
In this paper measurements are presented of $\pi^{\pm}$, K$^{\pm}$, p
and $\bar{\rm p}$ production at mid-rapidity ($\left|y\right|<0.5$),
in \PbPb\ collisions at \snn~=~2.76~TeV as a function of
centrality. The measurement covers the transverse momentum (\pt) range
from 100, 200, 300 MeV/$c$ up to 3, 3, 4.6~GeV/$c$, for $\pi$, K, and
p respectively.  The measured \pt\ distributions and yields are
compared to expectations based on hydrodynamic, thermal and
recombination models.  The spectral shapes of central collisions show
a stronger radial flow than measured at lower energies, which can be
described in hydrodynamic models. In peripheral collisions, the \pt\
distributions are not well reproduced by hydrodynamic models.  Ratios
of integrated particle yields are found to be nearly independent of
centrality. The yield of protons normalized to pions is a factor $\sim
1.5$ lower than the expectation from thermal models.


\ifdraft 
\ifpreprint
\end{abstract}
\end{titlepage}
\else
\end{abstract}
\end{frontmatter}
\newpage
\fi
\fi
\ifdraft
\thispagestyle{fancyplain}
\else
\end{abstract}
\ifpreprint
\end{titlepage}
\else
\end{frontmatter}
\fi
\fi
\setcounter{page}{2}

\svnidlong 
{$HeadURL: svn+ssh://svn.cern.ch/reps/mfPapers/2010SpectraID/trunk/PRCPbPb/papercontent.tex $} 
{$LastChangedDate: 2013-02-08 15:50:01 +0100 (Fri, 08 Feb 2013) $} 
{$LastChangedRevision: 817 $} 
{$LastChangedBy: mfloris $}
\svnid
{$Id: prcSpectraMain.tex 698 2012-09-18 15:40:35Z mfloris $}
\section{Introduction}
\label{sec:introduction}

The ultimate goal of heavy-ion collisions is the study of the properties of a deconfined and chirally-restored state of matter: the Quark--Gluon Plasma (QGP). Indications of its existence have been already provided by previous studies at the SPS~\cite{Heinz:2000bk} and at RHIC~\cite{Arsene:2004fa,Adcox:2004mh,Back:2004je,Adams:2005dq,Shuryak:2004cy}.
With the advent of the LHC a new energy regime is being studied, aiming at a precise characterization of the QGP properties.

The matter created in heavy-ion collisions exhibits strong collectivity, behaving as a nearly-perfect liquid as observed at RHIC~\cite{Huovinen:2006jp,Muller:2006ee}. Its collective properties can be studied through transverse momentum (\pt) distributions of identified particles. A solid understanding of the bulk properties of the expanding fireball is necessary for the interpretation of many observables. Any signal produced in the QGP phase has to be folded with the space-time evolution of the whole system, which has to be taken into account for comparison of theory and data.

Models based on relativistic hydrodynamics have been very successful
in describing observables such as the transverse momentum
distributions of identified particles, up to a few GeV/$c$. These
distributions contain information about the transverse expansion and
the temperature at the moment when the hadrons decouple from the
system~\cite{Schnedermann:1993ws,Heinz:2004qz}. It is commonly assumed
that a significant fraction of the collective flow builds up in the
expansion of the fireball in the initial partonic
phase~\cite{Muller:2006ee}. In this picture, the system would cool
down as a consequence of the expansion and undergo a phase transition
from a partonic to a hadronic phase. The hadrons continue to interact,
building up additional collective flow and potentially changing the
relative abundances. The hadronic yields are fixed at the moment when
inelastic collisions no longer play a role in the
system~\cite{Andronic:2011yq,Andronic:2008gu,Becattini:2010lb,Cleymans:1998fq}. However,
it is usually assumed that the hadronic phase does not affect particle
abundances~\cite{Heinz:2004qz,Rapp:2000gy}.
It was also suggested that the temperature of the hadronic
(``chemical'') freeze-out can be related to the phase transition
temperature~\cite{Andronic:2008gu,BraunMunzinger:2003zz,Heinz:2006ur}. Abundances
of particles have been fitted very successfully over a wide range of
energies (from \snn~=~2~GeV to
\snn~=~200~GeV~\cite{Andronic:2011yq,Andronic:2008gu,Cleymans:1998fq})
with thermal (or ``statistical hadronization'') models.  From these
fits, one can extract the thermal properties of the system at the
moment when the particle abundances are fixed, the key parameters
being the ``chemical freeze-out'' temperature \Tch\ and the
baryochemical potential \muB\ (determined by the net baryon content of
the system). As will be discussed, the new data presented in this
work 
seem to question part of the assumptions in these
models, as also reported in~\cite{prl-spectra}.  The system eventually decouples when elastic
interactions cease, at the ``kinetic freeze-out'' temperature
\Tfo. This temperature, together with the expansion velocity at the
moment of decoupling can be inferred from the \pt\ distributions of
identified particles.

In the intermediate \pt\ region ($2 \lesssim \pt \lesssim 8$~GeV/$c$) the baryon-to-meson ratios have been shown to reach values $\gtrsim 1$ for $\pt \sim 3$~GeV/$c$, much larger than in pp collisions~\cite{Abelev:2006jr}. 
It was suggested that this could be a consequence of hadronization via recombination of quarks from the plasma (in the coalescence models~\cite{Fries:2003vb,Greco:2003xt}). 

In this paper, we present the measurement of \pt\ spectra of $\pi^{\pm}$, K$^{\pm}$, p and $\bar{\rm p}$ in \PbPb\ collisions at \snn~=~2.76 TeV, as a function of centrality and over a wide \pt\ range (from 100, 200, 300 MeV/$c$ up to 3, 3, 4.6~GeV/$c$, for $\pi$, K, and p respectively) . The ALICE experiment, thanks to its unique Particle IDentification (PID) capabilities, is well suited for these measurements. Previous results on identified particle production in pp collisions have been reported in~\cite{Aamodt:2011zj}. 
The paper is organized as follows. In Sec. \ref{sec:datasampleAndAnalysis} the data sample and the analysis technique are discussed. 
The systematic uncertainties are presented in Sec.~\ref{sec:systematics}, and the results in Sec.~\ref{sec:results}. These are discussed in the context of theoretical models in Sec.~\ref{sec:discussion}.
Finally, we come to our conclusions in Sec.~\ref{sec:conclusions}.


\section{Data Sample and Analysis Method}
\label{sec:datasampleAndAnalysis}

\begin{figure*}[tbp]
  \centering
  \includegraphics[width=0.49\linewidth]{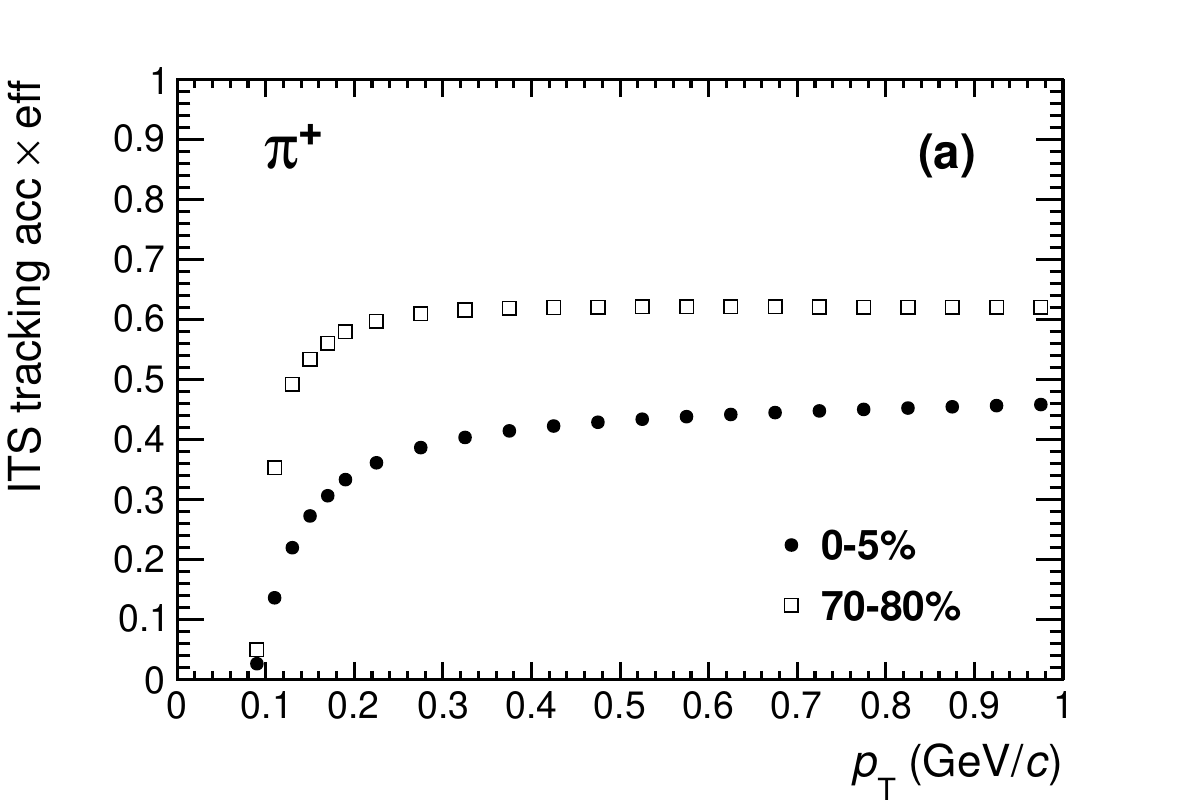}
  \includegraphics[width=0.49\linewidth]{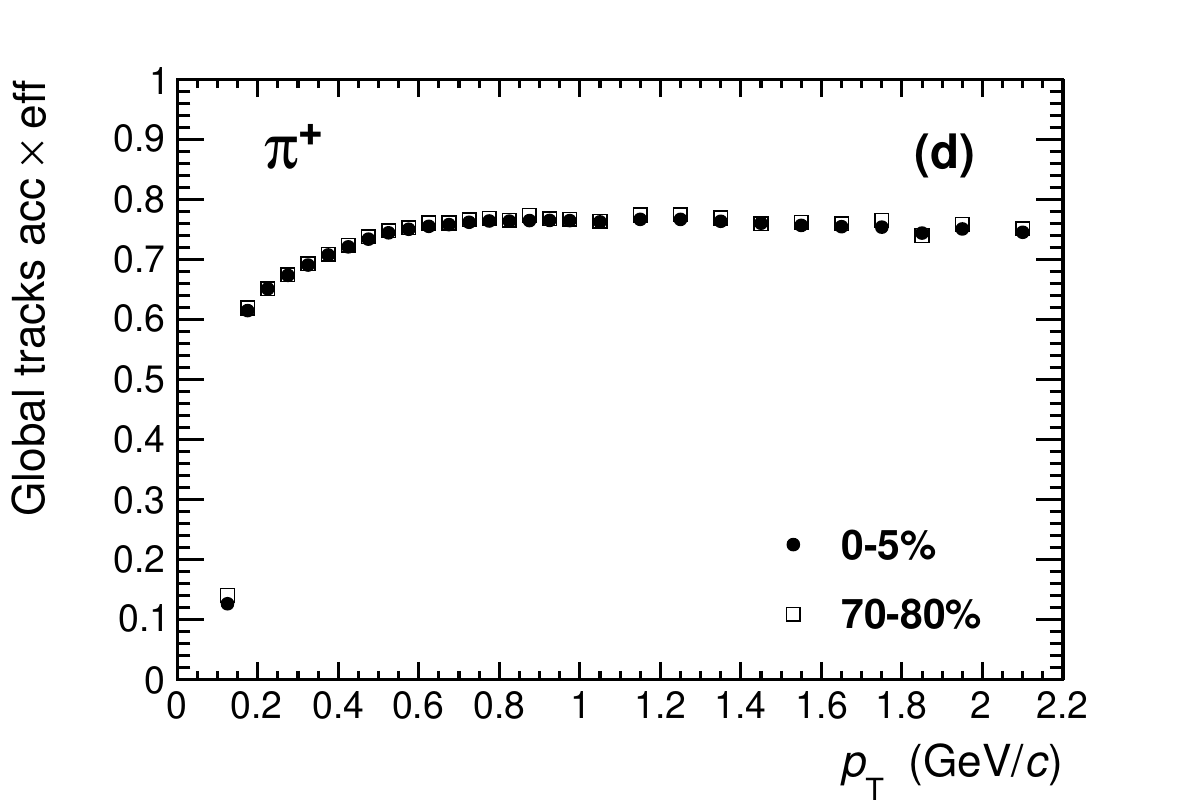}
  \vfill
  \includegraphics[width=0.49\linewidth]{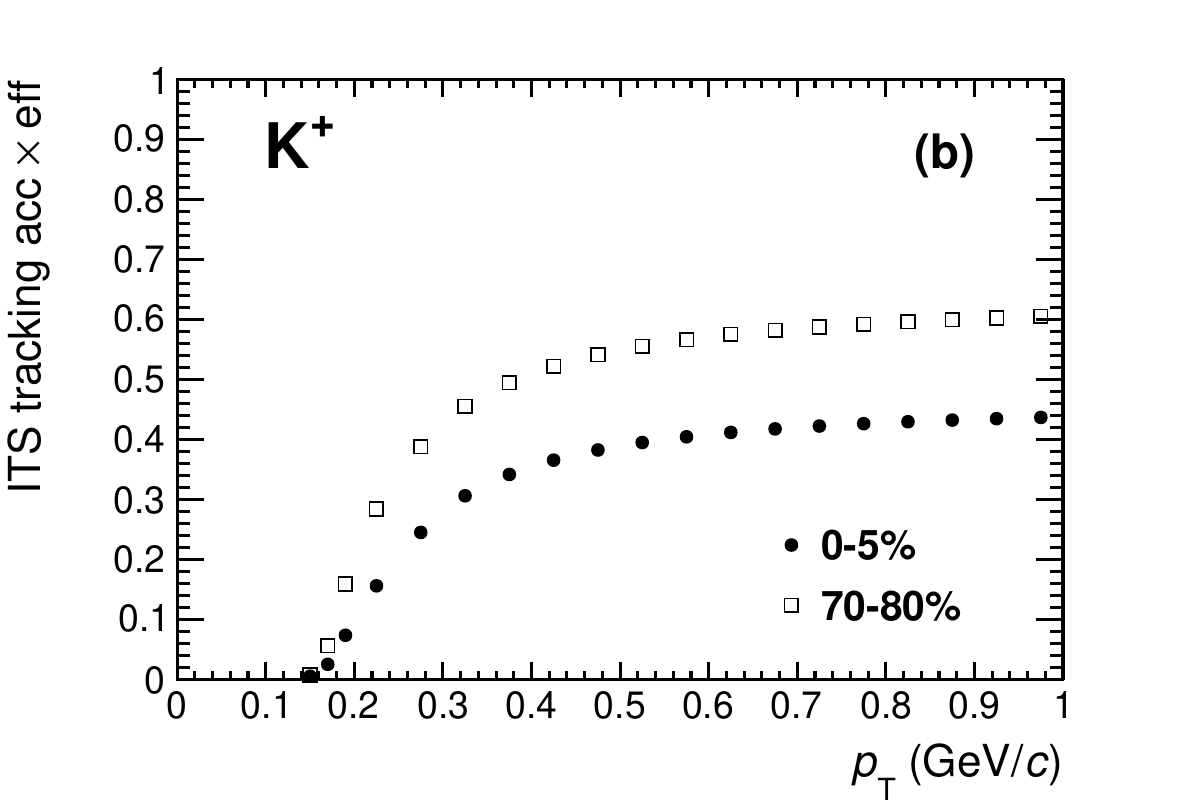}
  \includegraphics[width=0.49\linewidth]{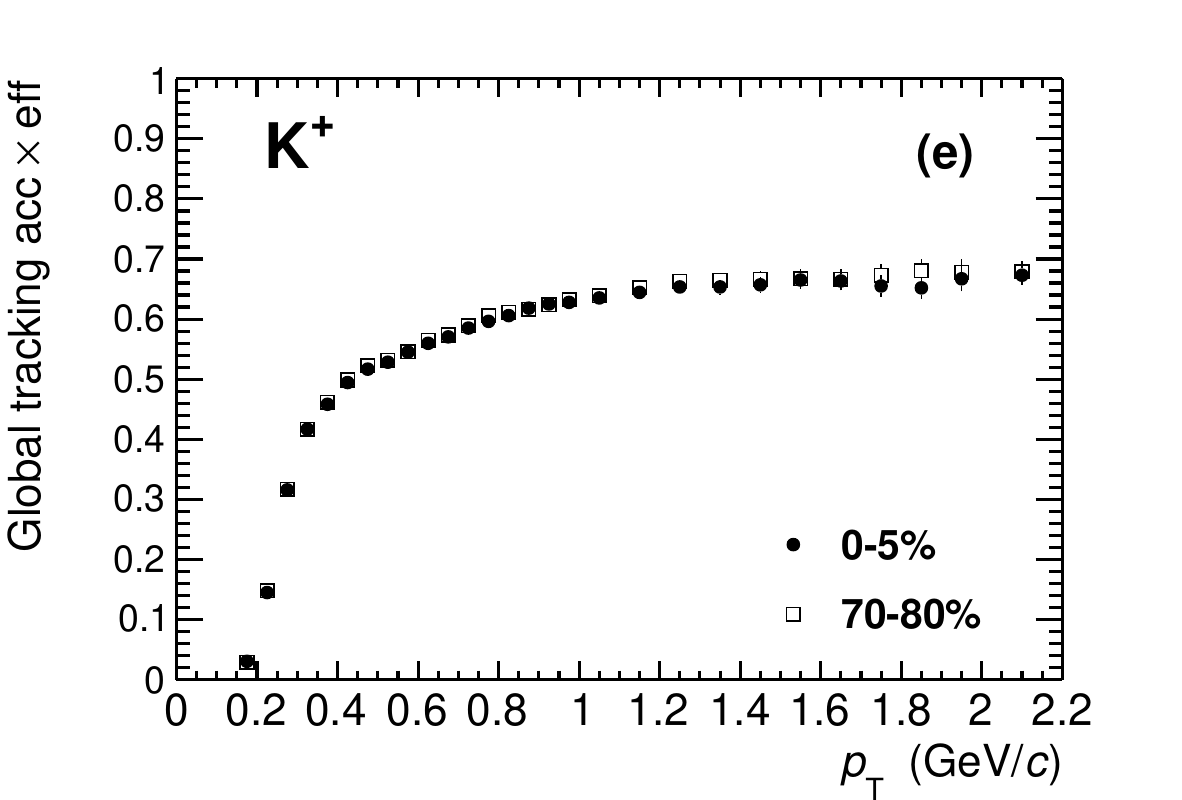}
  \vfill
  \includegraphics[width=0.49\linewidth]{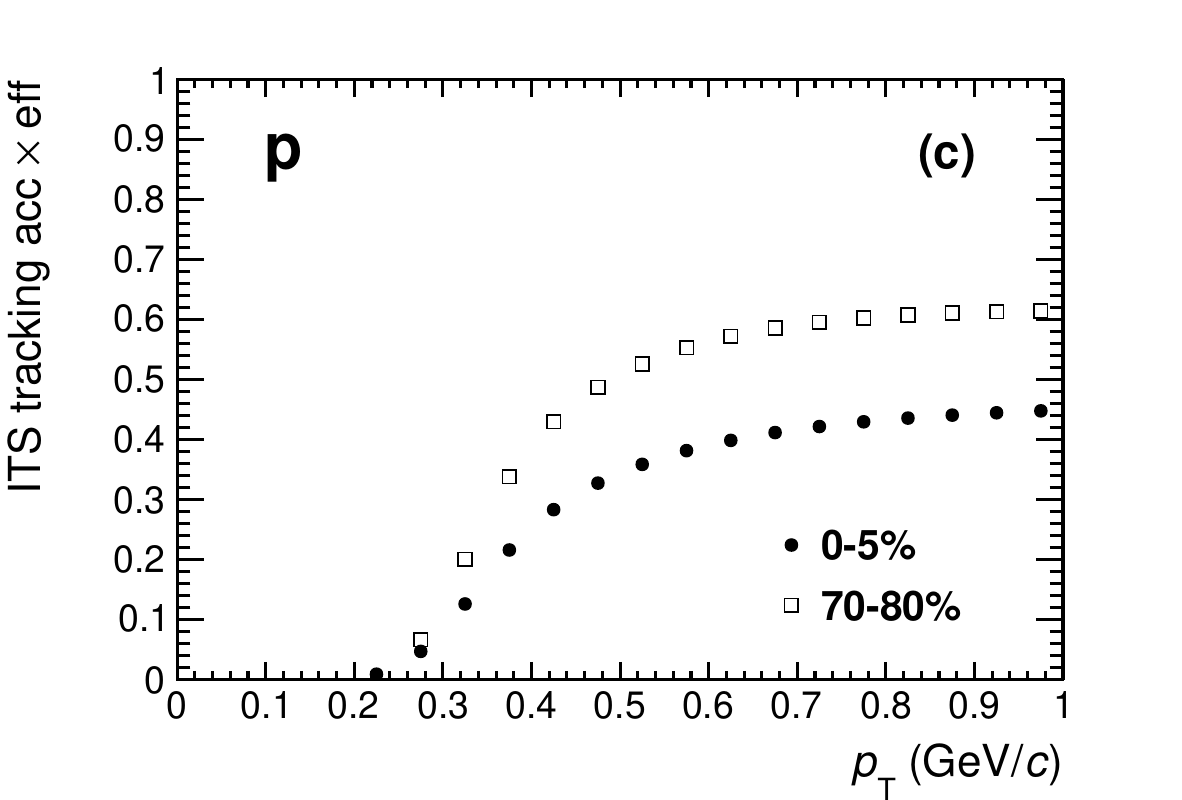}
  \includegraphics[width=0.49\linewidth]{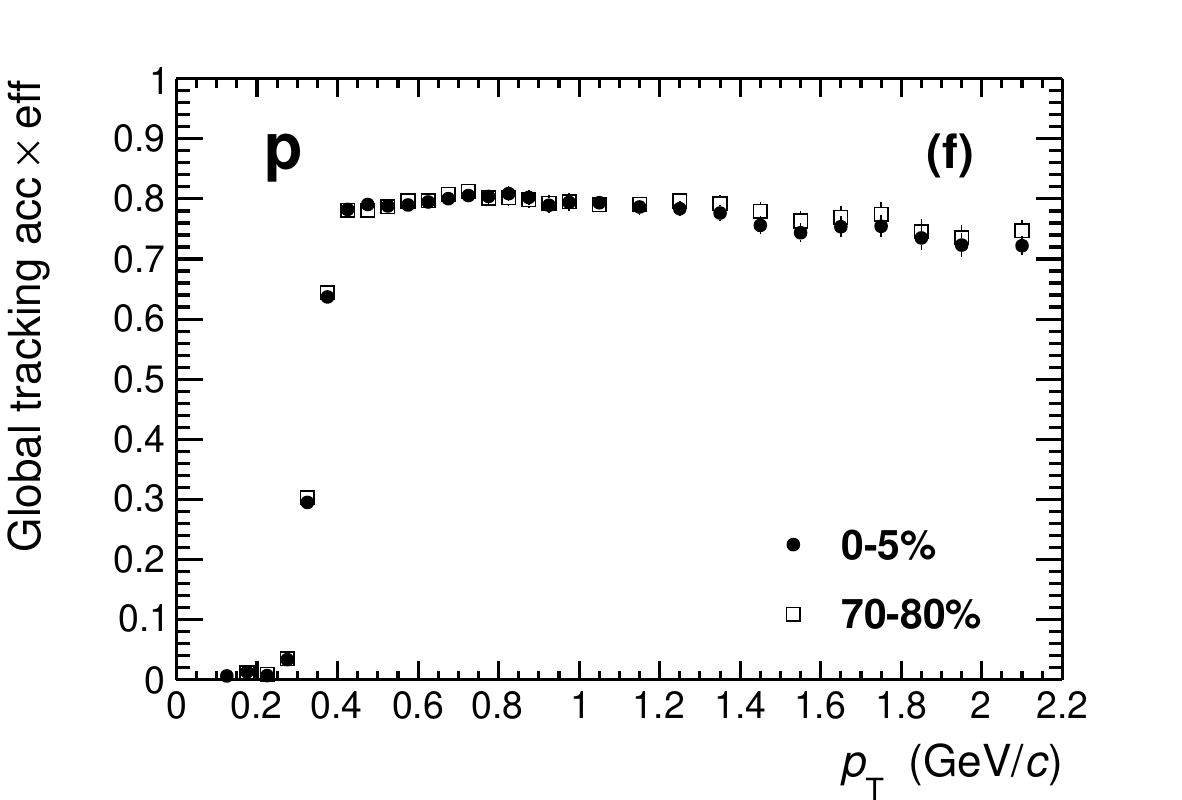}
  \vfill
  \caption{ITS standalone tracking efficiency for (a) pions, (b) kaons, (c) protons and global tracking efficiency for (c) pions, (d) kaons, (e) protons in central and peripheral collisions. }

  \label{fig:its-sa-global-trkeff}
\end{figure*}

\begin{figure}[hbt!]
  \centering
  \includegraphics[width=\mywidth]{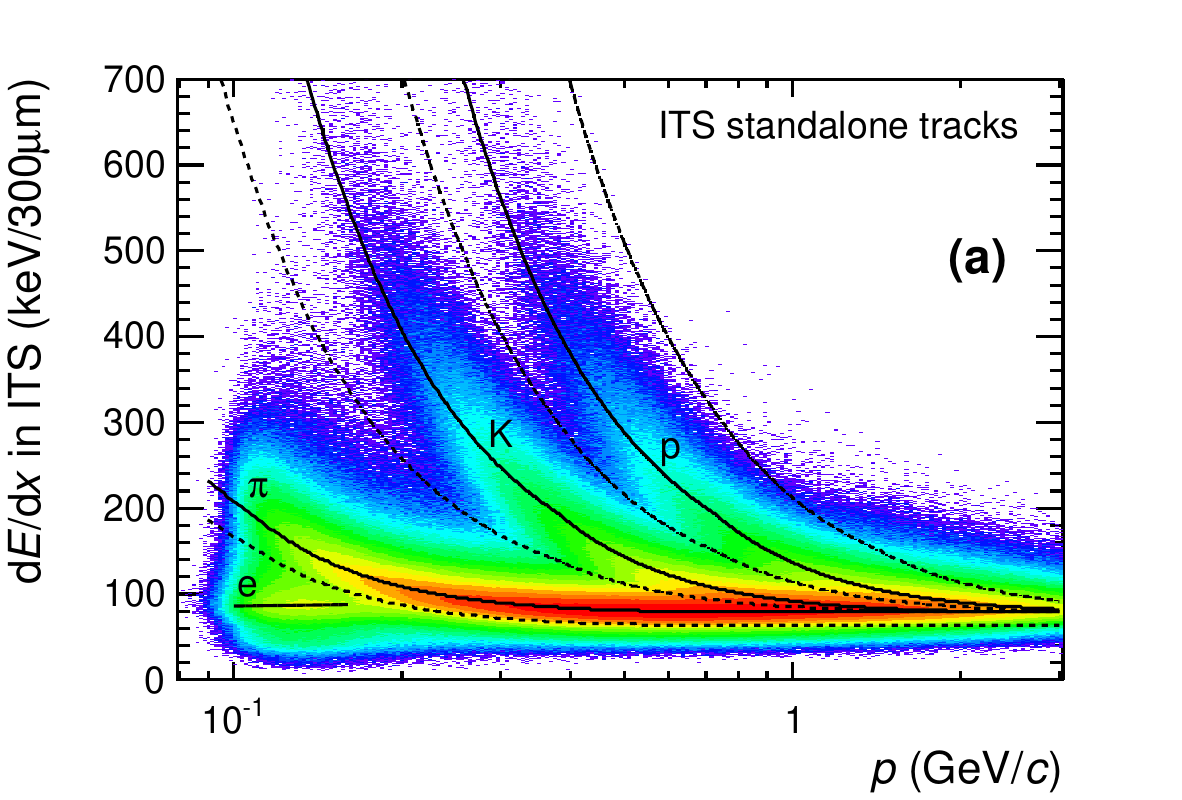}
  \includegraphics[width=\mywidth]{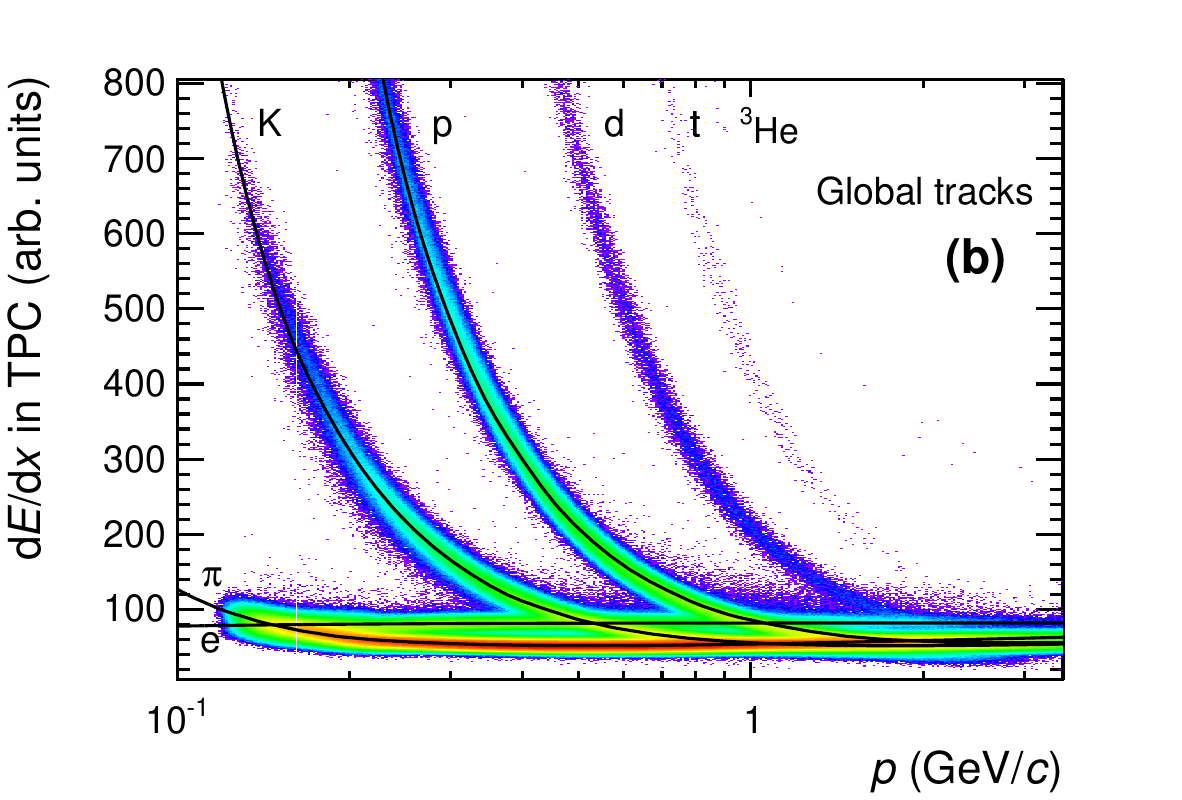}
  \includegraphics[width=\mywidth]{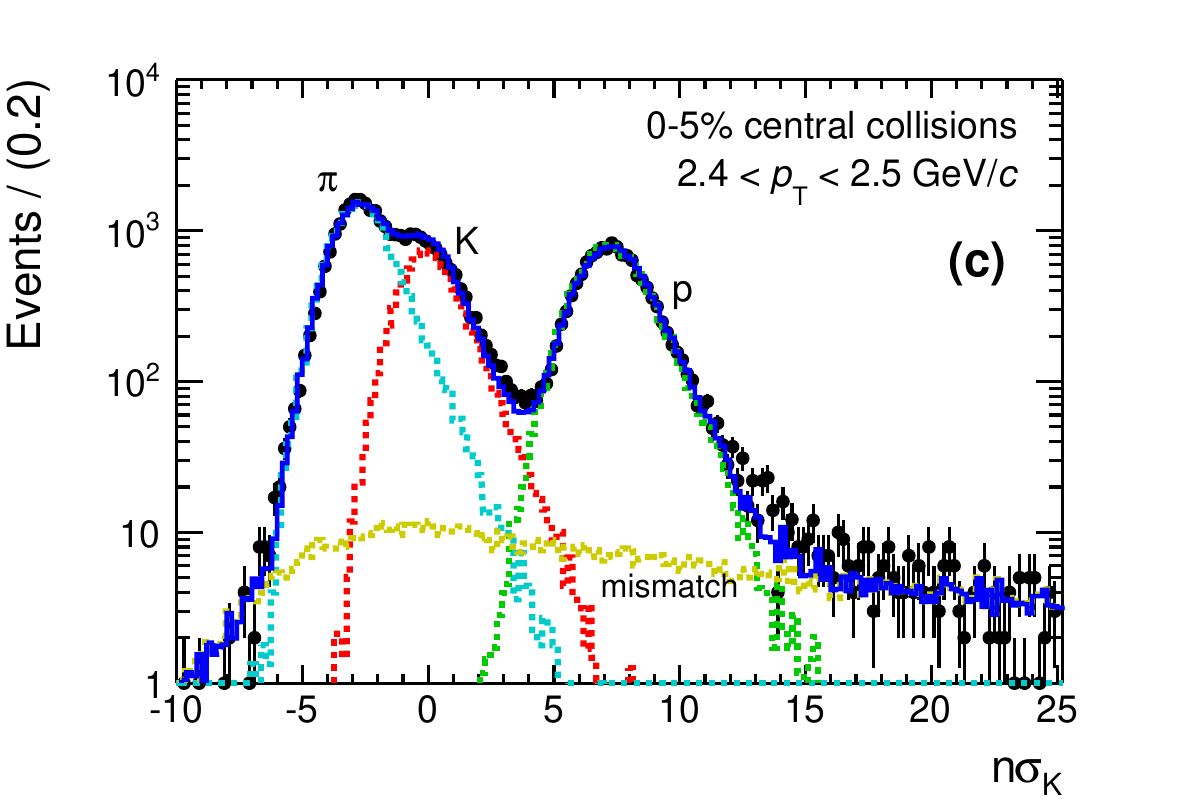}
 \caption{(color online) Performance of the PID detectors: (a) \dedx\ distribution measured in the ITS, the continuous curves represent the Bethe-Bloch parametrization,  the dashed curves the asymmetric bands used in the PID procedure; (b) \dedx\ measured in the TPC with global tracks (see text for the definition of global tracks), the continuous curves represent the Bethe-Bloch parametrization; (c) fit of the TOF time distribution with the expected contributions for negative tracks and for the kaon mass hypothesis, in the bin $2.4 < \pt < 2.5$~GeV/$c$.}
  \label{fig:pid-performance}
\end{figure}

The data used for this analysis were collected during the first \PbPb\ run at the Large Hadron Collider in the Autumn of 2010. The sample consists of 4 million events, after event selection.

A detailed description of the ALICE detector can be found in~\cite{Alessandro:2006yt,Aamodt:2008zz}.
The central tracking and PID detectors used in this work cover the pseudorapidity range $\left| \eta \right| < 0.9$ and are, from the innermost one outwards, the Inner Tracking System (ITS), the Time Projection Chamber (TPC), the Transition Radiation Detector (TRD), and the Time-Of-Flight array (TOF). The detector features a small material budget ($\sim 0.1~X/X_0$ for particles going through the TPC) and a low magnetic field, which allow for the reconstruction of low \pt\ particles. The central detectors are embedded in a 0.5~T solenoidal field, whose polarity was reversed to allow for systematic studies.

A pair of forward scintillator hodoscopes, the \VZERO\ detectors ($2.8 < \eta < 5.1$ and $-3.7 < \eta < -1.7$), measures the arrival time of particles with a resolution of one ns  and was used for triggering purposes and centrality determination. 
The data were collected using a minimum bias trigger requiring a combination of hits in the two innermost layers of the ITS (Silicon Pixel Detector, SPD, see below) and in the \VZERO. The trigger condition used during the data-taking has been changed as a function of time, to cope with the increasing luminosity delivered by the LHC. This time dependence was eliminated off--line by requiring two hits in the SPD and one hit in either of the \VZERO\ detectors. This condition was shown to be fully efficient for the 90\% most central events~\cite{centrality-paper}.

The signal in the \VZERO\ was required to lie in a narrow time window  of about 30~ns around the nominal collision time, in order to reject any contamination from beam-induced backgrounds. Only events produced in the vertex fiducial region of $\left|V_{z}\right|<10~$cm were considered in the analysis (where $V_{z}$ is the vertex position along the beam direction).
In the sample used for this analysis, a non-negligible fraction of the ions were located outside of their nominal RF bucket in the bunch, giving rise to ``satellite'' collisions. These events are produced well outside the vertex fiducial region, but could give rise to ``fake'' vertices due to the combinatorial algorithm which reconstructs the vertices assuming that particles are coming from the area around the nominal region. These events were rejected cutting on the correlation of arrival times of beam fragments to a pair of ``Zero Degree Calorimeters'' (ZDCs), placed close to the beam pipe, 114~m away from the interaction point on either side of the detector.  For details, see~\cite{Aamodt:2010cz,Aamodt:2010pb}.

The \VZERO\ amplitude distribution was also used to determine the centrality of the events. In a first step, it was fitted with a Glauber Monte Carlo model to compute the fraction of the hadronic cross section corresponding to any given range of the \VZERO\ amplitude. Based on these studies, the data were divided in several centrality percentiles, selecting on signal amplitudes measured in the \VZERO~\cite{centrality-paper}. 
The results in this paper are reported in ten centrality bins, ranging from 0-5\% to 80-90\%,
The centrality intervals and the corresponding charged particle multiplicity measured in $\left|\eta\right|<0.5$ (called \dNdeta\ in the following) are summarized in sec.~\ref{sec:results}. The \dNdeta\ in the centrality bin 80-90\%, shown here for the first time, was computed with the same analysis as described in~\cite{Aamodt:2010cz}. 
The contamination from electromagnetic processes is negligible down to 80\% centrality. In the bin 80-90\% an upper limit for this contribution  was estimated as 6\%, using the energy distributions of the ZDCs and looking for
 the single (or few) neutron peaks on top of the distribution which would be expected for hadronic interactions~\cite{ALICE:2012aa,centrality-paper}. 

The production of $\pi^{\pm}$, K$^{\pm}$, p and $\bar{\rm p}$ was measured at mid-rapidity ($\left|y\right|<0.5$) via three independent analyses, each one focusing on a sub-range of the total \pt~distribution, with emphasis on the individual detectors and specific techniques to optimize the signal extraction. The ranges covered by the three analyses are summarized in Table~\ref{tab:range_analyses}. 

The ITS is composed of six layers of silicon detectors using three different technologies. The two innermost layers, based on a silicon pixel technology (SPD), are also used in the trigger logic, as they provide online the number of pixel chips hit by the produced particles, as mentioned above. The four outer layers, made of drift (SDD) and strip (SSD) detectors, provide identification via the specific energy loss. Moreover, using the ITS as a standalone tracker enables the reconstruction and identification of particles that do not reach the TPC (at low momentum) or cross its dead sectors.
This makes the identification of $\pi$, K, and p possible down to respectively 0.1, 0.2, 0.3~GeV/$c$ in \pt. 
In the first analysis, ``ITS standalone'' tracks and \dedx\ were used. At least 4 points are required to form a track, out of which at least 1 must be in the SPD and 3 in the drift or strip detectors. With such a small number of tracking points and the high multiplicity in central heavy ion collisions, the probability of having tracks with wrongly associated clusters is not negligible. This contribution is strongly suppressed by applying a cut on the $\chi^2$ per cluster $<2.5$ and not allowing tracks to share clusters. These cuts, however, introduce a strong centrality dependence of the efficiency, as shown in Fig.~\ref{fig:its-sa-global-trkeff} (a), (b), (c). The efficiency saturates at $\sim 0.6$ mostly due to the requirement of an SPD point: this detector was only $\sim 80\%$ operational in 2010~\cite{Aamodt:2010ft}.
For each track, the \dedx\ is calculated using a truncated mean: the average of the lowest two points in case four points are measured, or a weighted sum of the lowest (weight 1) and the second lowest point (weight 1/2), in case only three points are measured. The final \dedx\ resolution is about $10\%$. 
The particle identity was assigned according to the distance, measured in units of the resolution, from the expected energy loss curves. While no upper limit on this distance was in general used, a 2$\sigma$ lower bound was applied in the case of pions, to remove contamination from electrons at low \pt. The procedure results in asymmetric ranges around the curves for $\pi$, K, and p, to reflect the asymmetric nature of the energy loss (Fig.~\ref{fig:pid-performance}, top). The range of this analysis is determined at low \pt\ by the ITS standalone tracking efficiency, at high \pt\ by the contamination from other particle species: the analysis is stopped when the systematic uncertainty coming from this is no longer negligible.

The other two analyses were based on global tracks, which combine the
information from the ITS, the TPC and the
TRD. 
This provides good resolution in the transverse distance of closest
approach to the vertex, \dcaxy, and hence good separation of primary
and secondary particles.  The track selection required at least 70
clusters in the TPC and at least 2 points in the ITS, out of which at
least one must be from the SPD, to improve the \dcaxy\ resolution.
The tracking efficiency, shown in Fig.~\ref{fig:its-sa-global-trkeff}
(c), (d), (e), depends only mildly on centrality. It saturates at $\sim
70\%$ because of the inactive channels in the SPD. Simulations with a
fully operational ITS show that the intrinsic efficiency of the
detector is $>90\%$. The rise of the efficiency at low \pt\ is due to
interactions with the detector material, and it is thus sharper for
protons. The efficiency reaches a maximum when the curvature is big
enough so that a track can cross the TPC readout chamber boundaries
within a relatively small area so that the two track parts can be
easily connected (at $\pt \sim 0.6$~GeV/$c$). The straighter tracks at
higher \pt~are affected by the geometrical acceptance.  This effect is
more pronounced for protons than for pions, because the efficiency is
folded with the decay probability for pions. While protons are stable
particles, there is a non-zero decay probability for pions which
decreases towards higher \pt. The shape of the efficiency for kaons,
on the other hand, is dominated by the decay probability.

The TPC identifies particles via the specific energy loss in the  gas: up to 159 samples can be measured. A truncated mean, utilizing only 60\% of the available samples, is employed in the analysis
 (Fig.~\ref{fig:pid-performance}, b). This leads to a Gaussian (and hence symmetric) response function, in contrast to the ITS. The resolution is  $\sim$~5\% in peripheral and $\sim$~6.5\% in central collisions.
Further outwards, the TOF measures the time-of-flight of the particles, allowing identification at higher \pt. The TRD tracking information, if present, is used to constrain the extrapolation to the TOF. The total time resolution is about 85~ps and it is determined by three contributions: the intrinsic timing resolution of the detector and associated electronics, the tracking and the start time. This makes the identification possible out to \pt~=~3~GeV/$c$ for pions and kaons, and 4.6~GeV/$c$ for protons.

\begin{table}[tp]
  \centering
  \caption{\pt\ range (GeV/$c$) covered by the different analyses.}
  \begin{tabular}{c|ccc}
    \hline
    Analysis & $\pi$ & K & p \\
    \hline
    \hline
    ITS standalone & 0.10--0.60 & 0.20--0.50 & 0.30--0.60 \\
    TPC/TOF & 0.20--1.20 & 0.25--1.20 & 0.45--1.80 \\
    TOF fits  & 0.50--3.00 & 0.45--3.00 & 0.50--4.60\\
    \hline
 \end{tabular}
  \label{tab:range_analyses}
\end{table}

\begin{figure}[tbp]
  \centering
  \includegraphics[width=\mywidth]{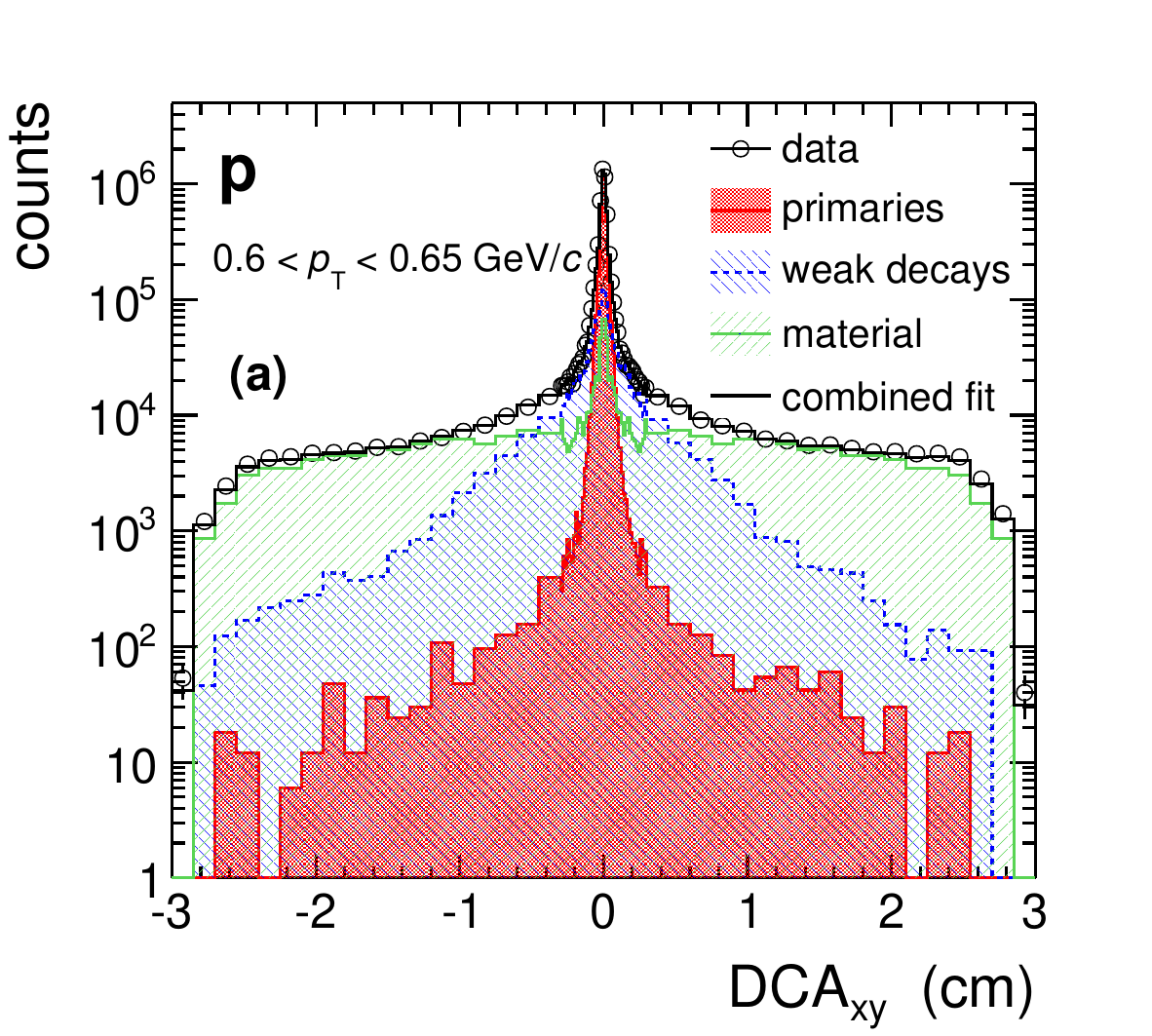} 
  \includegraphics[width=\mywidth]{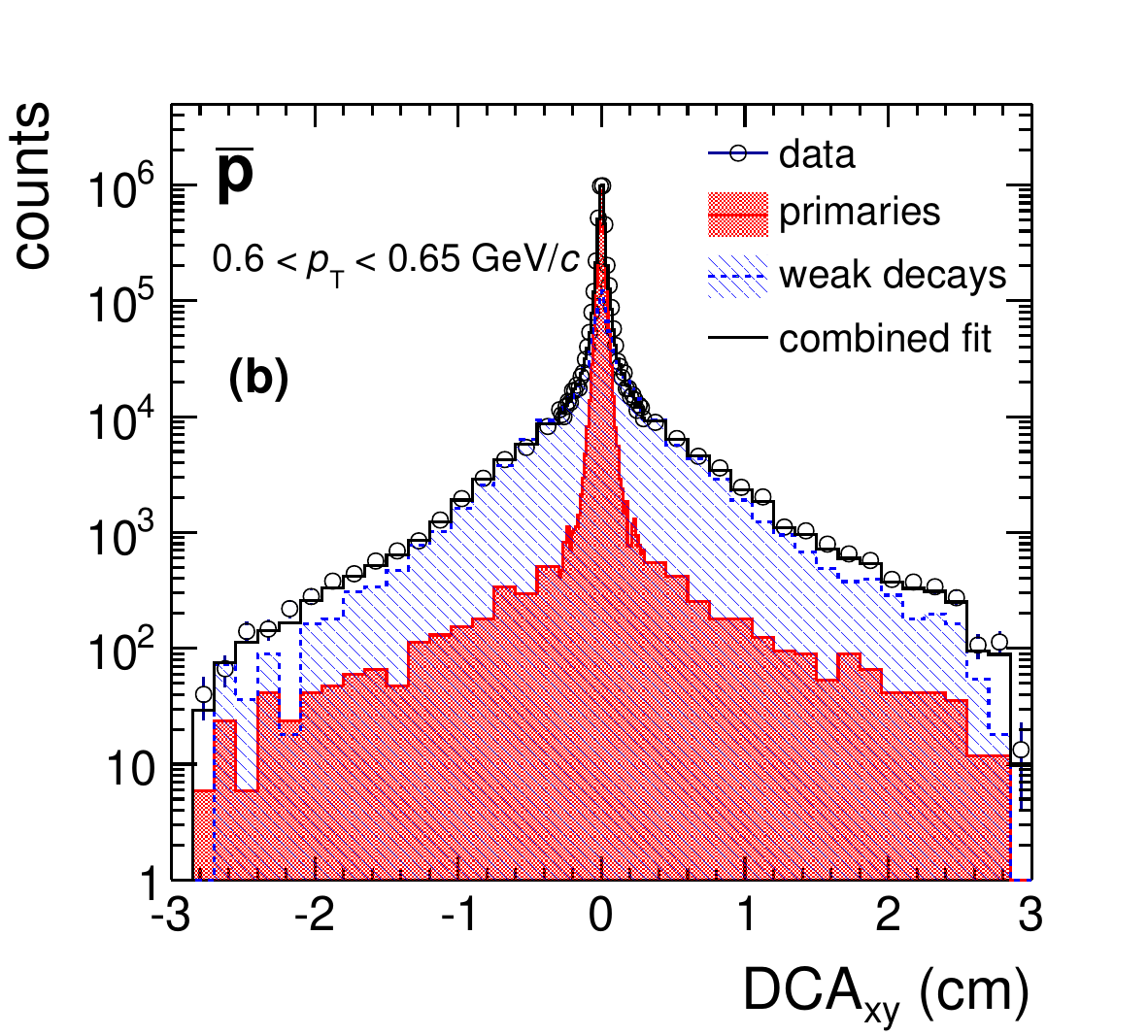}
  \caption{(color online) \dcaxy\  of (a) protons and (b) antiprotons in the \pt-range between 0.6 GeV/$c$ and 0.65 GeV/$c$ together with the Monte Carlo templates which are fitted to the data (0-5\% most central collisions). The dashed areas represent the individual templates, the continuous curve the combined fit.}
  \label{fig:dcaDistribution}
\end{figure}

In the intermediate \pt\ range, track-by-track identification is possible, based on the combined TPC and TOF signals (``TPC/TOF''). It was required that the particles are within 3$\sigma$ of the expected values measured in the TPC and/or TOF. The TOF information was used starting at \pt~=~0.65, 0.6, 0.8~GeV/$c$ for $\pi$, K, and p respectively, where the 3$\sigma$ compatibility cut was required for both TPC and TOF. The additional requirement of a matching TOF hit reduces the overall efficiency shown in Fig.~\ref{fig:its-sa-global-trkeff} (c), (d), (e) by about 30\%,
 due to the TOF geometrical acceptance and to the additional material (see also Sec.~\ref{sec:systematics}). The range of this analysis is determined at low \pt\ by the global tracking efficiency, at high \pt\ by the contamination from other particle species.
Finally, in the third analysis,  a statistical identification based  on the TOF signal alone was used (``TOF fits''). In order to extend the measurement beyond the region of clear separation, the difference between the measured time-of-flight and the expected one for the particle species under study (normalized to the resolution) was examined. For each \pt\ bin, this distribution was fitted  with the expected shapes (called ``templates'' in the following) for $\pi$, K, and p, allowing the three particles to be distinguished when the separation is as low as $\sim$2$\sigma$. An additional template is needed to account for the tracks which are associated to a wrong hit in the TOF (mismatch). The templates are built from data, using the measured TOF time response function (described by a Gaussian with an exponential tail) and sampling real tracks to get a realistic track-length distribution.
This fit was repeated for each mass hypothesis, to allow for the calculation of the correct rapidity interval. A worst-case example, for 0-5\% central collisions, in the bin $2.4 < \pt < 2.5$~GeV/$c$ and for the kaon mass hypothesis, can be seen in Fig.~\ref{fig:pid-performance} (bottom).

In this paper, results for ``primary'' particles are presented, defined as prompt particles produced in the collision, including decay products, except those from weak decays of strange particles. The contamination from secondary particles produced by weak decays or interaction with the material is mostly relevant for (anti)protons. Since strangeness production is typically underestimated in current event generators, and low \pt\ interactions with the material are not modeled perfectly in transport codes, the contamination was extracted from data. The transverse distance-of-closest-approach (\dcaxy) distribution for selected tracks was fitted with three distributions (``templates'' in the following) corresponding to the expected shapes for primary particles, secondaries from material and secondaries from weak decays, as extracted from Monte Carlo. This contribution can reach ~35\% at \pt~=~300 MeV/$c$ for protons. An example of \dcaxy\ fits for the bin $0.6 < \pt < 0.65$~GeV/$c$ is shown in Fig.~\ref{fig:dcaDistribution}: the shapes of the three contributions are very different. The distribution corresponding to primary particles reflects the resolution of the \dcaxy. The small non-Gaussian tails seen in the figure are mostly due to large angle scattering~\cite{Beringer:1900zz}, tracks with wrongly assigned ITS clusters or to the combination of tracks with hits on the first and/or second ITS layer which have slightly different \dcaxy\ resolution.   The secondary particles from weak decays show a wider distribution which reflects the large $c\tau$ of weakly decaying particles (of the order of several centimeters). Finally, the secondaries from material show very flat tails at high \dcaxy. The last contribution is negligible in the case of antiprotons. The \dcaxy\ distribution for pions is similar to the one for protons, but with much suppressed contribution of secondaries. The distribution for kaons is almost entirely composed of primary particles.

The efficiency correction and the templates used in the secondary correction procedure were computed with about 1 million Monte Carlo events, generated using the HIJING~\cite{hijing} event generator, tuned to reproduce approximately the \dNdeta\ as measured for central collisions~\cite{Aamodt:2010pb}. The transport of particles through the detector was simulated using GEANT~3~\cite{Geant:1994zzo}.
The results of the three analyses were combined using the (largely independent) systematic uncertainties as weights in the overlapping ranges, after checking for their compatibility.


\section{Systematic Uncertainties}
\label{sec:systematics}

Several effects related to the transport of particles through the detector and support materials contribute to the systematic uncertainties, namely: (i) the amount of secondary protons which are produced in the material, (ii) corrections for energy loss through electromagnetic interactions, (iii) absorption of antimatter due to hadronic interactions. The first effect is taken into account in the data-driven \dcaxy\ fit procedure, and its uncertainties are estimated with the \dcaxy\ fit variations discussed below.
 The uncertainties due to the second effect were estimated varying the material budget in the simulation by $\pm$ 7\% and are of the order of 3\% and 1\%, for pions and kaons, 5\% and 2\% for protons in the two lowest \pt-bins respectively and then quickly decrease further, becoming negligible towards higher momenta. In order to account for the last effect, different transport codes (GEANT 3, GEANT 4~\cite{Agostinelli:2002hh} and FLUKA~\cite{Battistoni:2007zzb}) were compared.  GEANT 3 (the default transport code used in this work) is known not to reproduce the cross sections relevant for the interactions with the material at low \pt~\cite{Aamodt:2010dx}. The efficiencies were scaled with a factor computed with a dedicated FLUKA~\cite{Aamodt:2010dx} simulation. This allows the uncertainty for anti-protons to be limited to $< 6\%$ despite the large absorption cross-section, as there are sufficient existing data on $\bar{p}-$nucleus collisions to validate the transport codes. A detailed comparison of GEANT~3 with the few existing measurements of  hadronic interaction cross sections of low momentum kaons and pions~\cite{Lee:2002eq,Friedman:1997eq,Ashery:1981tq,Carlson199693,Bendiscioli:1994uv} reveals differences of about 20-30\%. After folding with the relevant percentage of particles which are lost due to hadronic interaction before entering the TPC or the TOF, the resulting uncertainty is of the order of 2-3\% for K$^{-}$ and $\pi$ and below 1\% for K$^{+}$.

Uncertainties in the estimate of contamination from secondary particles can arise from differences in the \dcaxy\ distributions between data and Monte Carlo. In particular, the \dcaxy\ distribution for secondary particles coming from weak decays is affected by the $c\tau$ of the decaying mother, and the actual template used in the fits is a mixture of contributions from different particles.
The uncertainties due to the secondary subtraction procedure were estimated for all analyses by varying the range of the \dcaxy\ fit, by using different track selections (for instance using TPC-only tracks), by applying different cuts on the (longitudinal) \dcaz\ and by varying the particle composition of the Monte Carlo templates used in the fit. Overall, the effect is of the order of 3\% to 1\% for pions and 4\% to 1\% for protons. Such an uncertainty is not relevant for kaons, which have a negligible contamination from secondary particles. 
Moreover the agreement between the ITS standalone and the TPC analyses provides an additional crosscheck of this procedure, because different tracking methods have different sensitivity to contamination from secondaries.

At the lowest \pt, the main contribution to the systematic uncertainties comes from the ITS standalone tracking efficiency, due to the small number of tracking points and the strong centrality dependence. 
The uncertainty was estimated using the global tracking as a reference. For each reconstructed global track, a corresponding ITS standalone track is sought in a narrow window defined as $\Delta \eta < 0.03$, $\Delta \phi < 0.03$, and $\Delta \pt < 0.1\times\pt$. If a track is found, the ITS standalone tracking is considered efficient. This pseudo-efficiency is then compared in data and Monte Carlo to extract the corresponding systematic uncertainty, estimated to be about 10\% for central and 3\% for peripheral collisions. 
The efficiency of the ITS standalone reconstruction depends strongly on centrality (Fig.~\ref{fig:its-sa-global-trkeff}, left), so that a careful matching between the multiplicity in data and Monte Carlo is important. As mentioned, the simulation was adjusted to reproduce the \dNdeta\ measured for central collisions. The occupancy of all ITS layers and the reconstructed track multiplicity were compared between data and Monte Carlo. Residual differences, due to the different dependence of multiplicity on centrality in data and Monte Carlo, contribute to the systematic uncertainty on the tracking mentioned above and are of the order 2\% . 
 The sensitivity to the variation of the track selection cuts (\dcaxy, $\chi^2$, number of clusters) was found to be of the order of 7\% for all particles. Moreover, the Lorentz force causes shifts of the cluster position in the ITS, pushing the charge in opposite directions when switching the polarity of the magnetic field of the experiment (\EcrossB\ effect). This effect is not fully reproduced in the Monte Carlo simulation and was estimated analyzing data samples collected with different magnetic field polarities. This uncertainty is only relevant at the lowest \pt\ and is 3\% for pions, and 1\% for kaons and protons.

\ifpreprint
\begin{table}
\else
\begin{table*}
\fi

\centering \caption{Main sources of systematic uncertainty.} \label{tab:syst}
\ifpreprint
\footnotesize
\fi

\begin{tabular}{l|cc|cc|cc}
\hline
\hline
effect & \multicolumn{2}{c|}{$\pi^{\pm}$} & \multicolumn{2}{c|}{K$^{\pm}$} & \multicolumn{2}{c}{p and \pbar} \\ [3pt]
\hline
\pt\ range (GeV/$c$) & 0.1 & 3 & 0.2 & 3 & 0.35 & 4.5\\
\hline

correction for &  \multirow{2}{*}{1.5\%} & \multirow{2}{*}{1\%} & \multicolumn{2}{c|}{\multirow{2}{*}{negl.}} & \multirow{2}{*}{4\%} & \multirow{2}{*}{1\%} \\ 
secondaries & & & & & & \\[3pt]

material & \multirow{2}{*}{5\%} & \multirow{2}{*}{negl.} & \multirow{2}{*}{3\%} & \multirow{2}{*}{negl.} & \multirow{2}{*}{3\%} & \multirow{2}{*}{negl.} \\
budget & & & & & & \\ [3pt]

hadronic      &  \multirow{2}{*}{2\%} & \multirow{2}{*}{1\%} &   \multirow{2}{*}{4\%} & \multirow{2}{*}{1\%} &  
6\% & 1\% (\pbar) \\
interactions & & & & &  4\% & negl. (p) \\  [3pt]  

\hline
\pt\ range (GeV/$c$) & 0.1 & 0.5 & 0.2 & 0.5 & 0.35 & 0.65\\
\hline

ITS tracking & \multicolumn{2}{c|}{\multirow{2}{*}{10\%}} & \multicolumn{2}{c|}{\multirow{2}{*}{10\%}} & \multicolumn{2}{c}{\multirow{2}{*}{10\%}} \\
efficiency (Central 0-5\%)  &&&&&&\\ [3pt]

ITS tracking & \multicolumn{2}{c|}{\multirow{2}{*}{4\%}} & \multicolumn{2}{c|}{\multirow{2}{*}{4\%}} & \multicolumn{2}{c}{\multirow{2}{*}{4\%}} \\
efficiency (Peripheral 80-90\%)   &&&&&&\\[3pt]

ITS PID  &  \multicolumn{2}{c|}{2\%} & \multicolumn{2}{c|}{4\%} & \multicolumn{2}{c}{4.5\%} \\ [3pt]

\hline
\pt\ range (GeV/$c$) & 0.3 & 0.65 & 0.3 & 0.6 & 0.5 & 0.8\\
\hline

global tracking&  \multicolumn{2}{c|}{\multirow{2}{*}{4\%}} & \multicolumn{2}{c|}{\multirow{2}{*}{4\%}} & \multicolumn{2}{c}{\multirow{2}{*}{4\%}} \\
efficiency &&&&&& \\ [3pt]

TPC PID (Central, 0-5\%) & \multicolumn{2}{c|}{3\%} & \multicolumn{2}{c|}{5\%} & \multicolumn{2}{c}{1.5\%} \\ [3pt]

TPC PID (Peripheral 80-90\%)& \multicolumn{2}{c|}{1.5\%} & \multicolumn{2}{c|}{3.5\%} & \multicolumn{2}{c}{1\%} \\ [3pt]

\hline
\pt\ range (GeV/$c$) & 0.5 & 3 & 0.5 & 3 & 0.5 & 4.5\\
\hline

TOF matching  & \multicolumn{2}{c|}{\multirow{2}{*}{3\%}} & \multicolumn{2}{c|}{\multirow{2}{*}{6\%}} & \multicolumn{2}{c}{\multirow{2}{*}{3\%}} \\ 
efficiency & & & & & & \\ [3pt]
TOF PID & 2\% & 7\% & 3\% & 15\% & 5\% & 25\% \\ [3pt]
\hline\noalign{\smallskip}
\hline\noalign{\smallskip}

\end{tabular}
\ifpreprint
\end{table}
\else
\end{table*}
\fi

\begin{figure*}[p]
  \centering
  \includegraphics[width=0.45\linewidth]{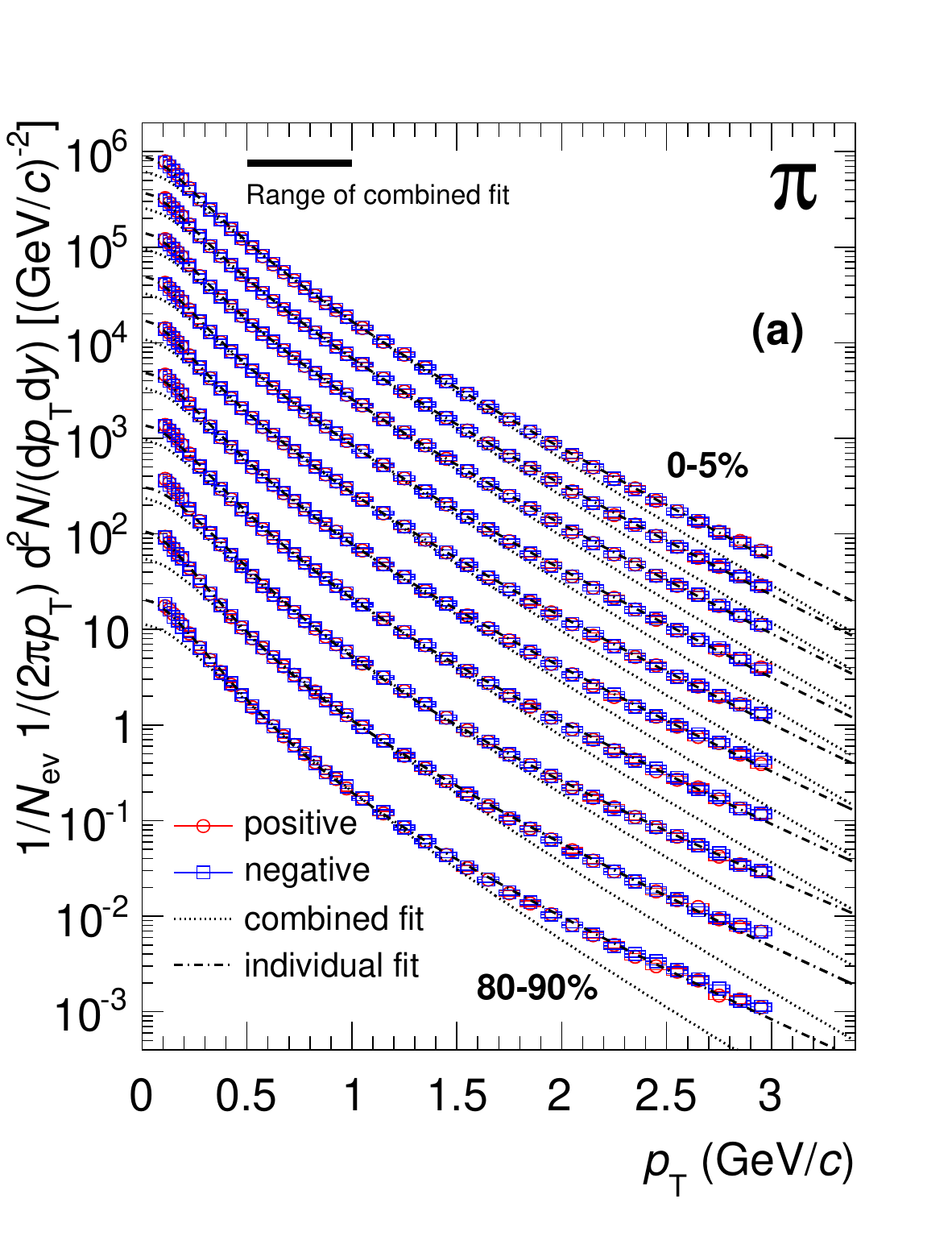}
  \includegraphics[width=0.45\linewidth]{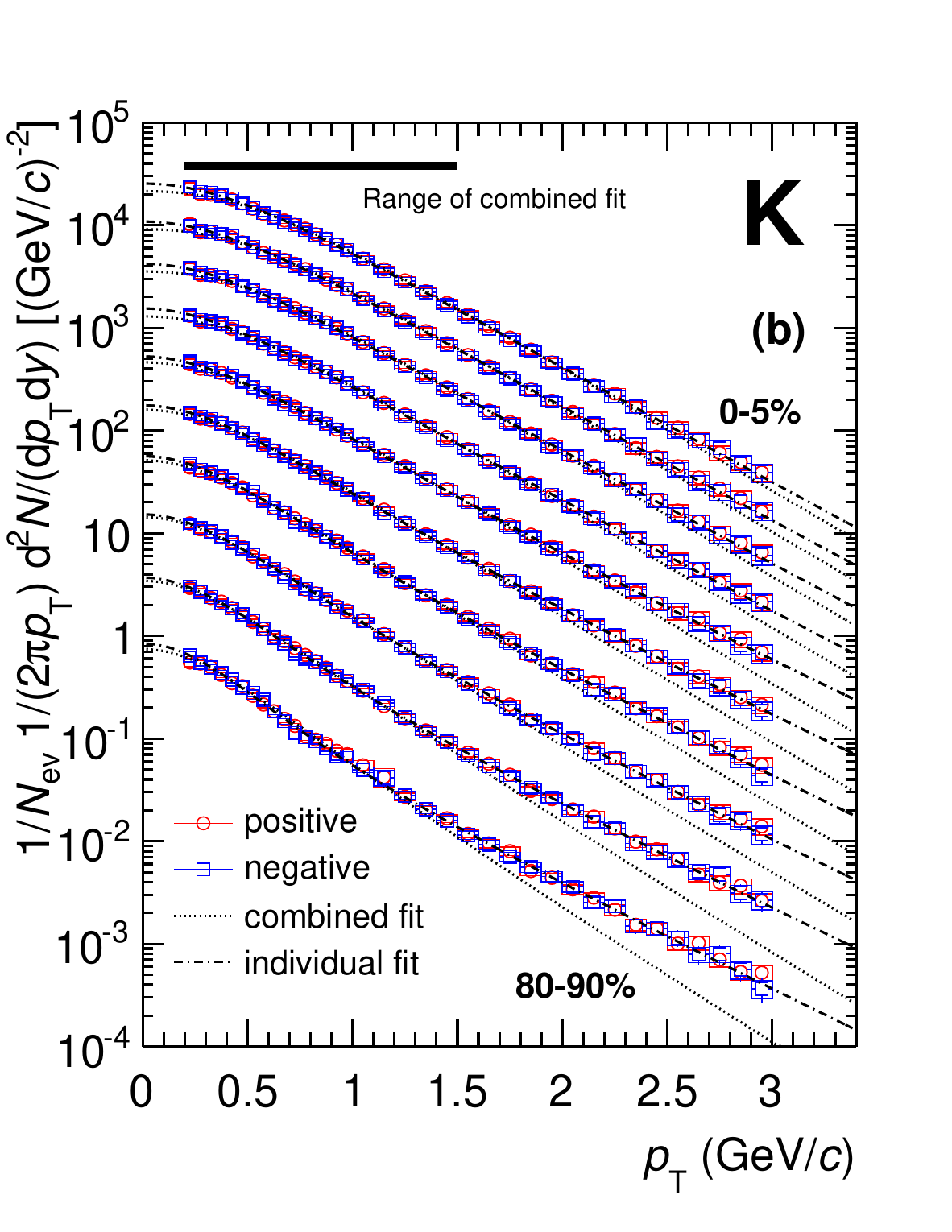}
  \includegraphics[width=0.45\linewidth]{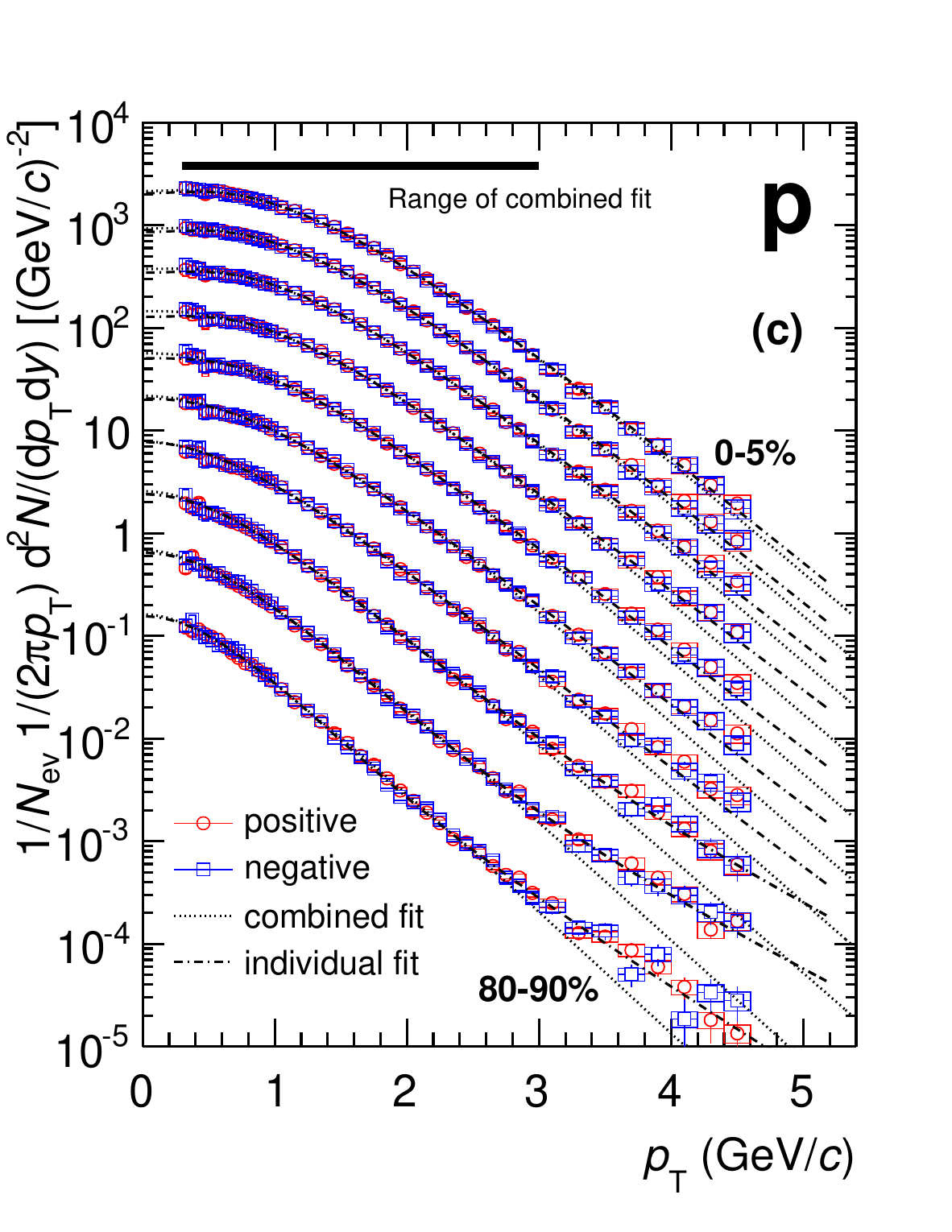}
  \caption{(color online) Transverse momentum (\pt) distribution of (a) $\pi$, (b) K, and (c) p as a function of centrality, for positive (circles) and  negative (squares) hadrons. Each panel shows central to peripheral data; spectra scaled by  factors $2^n$ (peripheral data not scaled). Dashed curves: blast-wave fits to individual particles; dotted curves: combined blast-wave fits (see text for details). Statistical (error bars) and systematic (boxes) uncertainties plotted. An additional normalization uncertainty (Table~\ref{tab:yields}) has to be added in quadrature.}
  \label{fig:spectra-results}
\end{figure*}

The uncertainties related to the ITS PID  method were estimated by using different techniques for the identification.
A 3$\sigma$ cut on the maximum difference in energy loss of the particles was applied (instead of using all particles as in the default strategy). Alternatively, an unfolding method was used, in which for each \pt\ bin, the \dedx\ distribution is fitted with Monte Carlo templates for each one of the species, similar to what is done in the TOF fits. The effect is found to be 2\%, 4\%, 7\% for $\pi$, K, and p respectively.
Close to the high \pt\ boundaries of the analyses (Table~\ref{tab:range_analyses}), the possibility of misidentifying other species is not completely negligible. This is corrected for with Monte Carlo, and the corresponding uncertainty, due to residual differences in particle ratios and \dedx\ between data and Monte Carlo is estimated to be 2\% for kaons and 1\% for protons.

The uncertainty in the global tracking efficiency was estimated comparing the track matching efficiency from TPC to ITS and from ITS to TPC in data and Monte Carlo. The effect is found to be $\sim$4\%.
The uncertainties related to the track selection were investigated by a variation of the track cuts. They are estimated (for most central collisions) to be 3\% for pions and kaons and  4\% for protons. This uncertainty decreases slightly for peripheral collisions.
The uncertainties related to the TPC/TOF PID procedure were estimated varying the PID cut between 2 and 5 $\sigma$. 
As pions are the most abundant particles, their corresponding systematic uncertainty does not exceed 3\% even in regions where the mean pion energy loss crosses the kaon and proton curves. On the other hand, it can reach up to 4-5\% for kaons. The corresponding systematics of the better separated protons are below 1.5\%. As the \dedx\ resolution becomes slightly worse with increasing multiplicity, the systematic uncertainty shows a similar slight increase. The values quoted here represent  upper bounds for all centralities as they correspond to the 0-5\% most central collisions.

The uncertainties on the response functions of the PID detectors were  found to have a negligible effect: they were carefully tuned using high statistics data and this analysis is limited to regions were the separation of different particles is $\gtrsim$2$\sigma$.

As mentioned above, the tracks reaching the TOF detector have to cross a substantial amount of additional material budget (about $23\%~X/X_0$), mostly from the Transition Radiation Detector (TRD)~\cite{Aamodt:2008zz}. The systematic uncertainties on the TOF matching were estimated comparing the matching efficiency evaluated in Monte Carlo and from data using samples of cleanly identified particles in TPC. Good agreement is observed in case of pions and kaons, with deviations at the level of at most 3\% and 6\%, respectively, over the full \pt\ range.

 In the case of protons and antiprotons good agreement is also observed, for $\pt> 1$~ GeV/$c$, with maximal deviations of 7.5\%. 
Since the TRD was not fully installed in 2010, the analysis was repeated for regions in azimuth with and without installed TRD modules, allowing to cross-check the uncertainty on the material. The effect is found to be 3\%, 6\%, 3\% for $\pi$, K, and p respectively.  The uncertainties related to the identification in the ``TOF fits'' analysis were estimated varying the parameters of the expected sources in the fit by $\pm$ 10\%. This is one of the main sources of uncertainty and increases with \pt, as the separation between different species decreases. It is $< 5$\% at $\pt \sim 0.5$~GeV/$c$ and $\lesssim 7\%, 15\%, 25\%$ at high \pt, for $\pi$, K, and p respectively.

The main sources of systematic uncertainties are summarized in Table~\ref{tab:syst}.

The measured spectra are further affected by an additional normalization uncertainty, coming from the centrality estimation. The centrality percentiles are determined with sharp cuts on the VZERO amplitude distribution, which are affected by a 1\% uncertainty~\cite{centrality-paper}. This translates into a normalization uncertainty on the spectra. The total normalization uncertainty is about +12\% -8.5\% for peripheral events (also including a 6\% contribution due to contamination from electromagnetic processes) and  negligible for central events, see Table~\ref{tab:yields}.


\section{Results}
\label{sec:results}

\begin{figure}[tp]
  \centering
  \includegraphics[width=\mywidth]{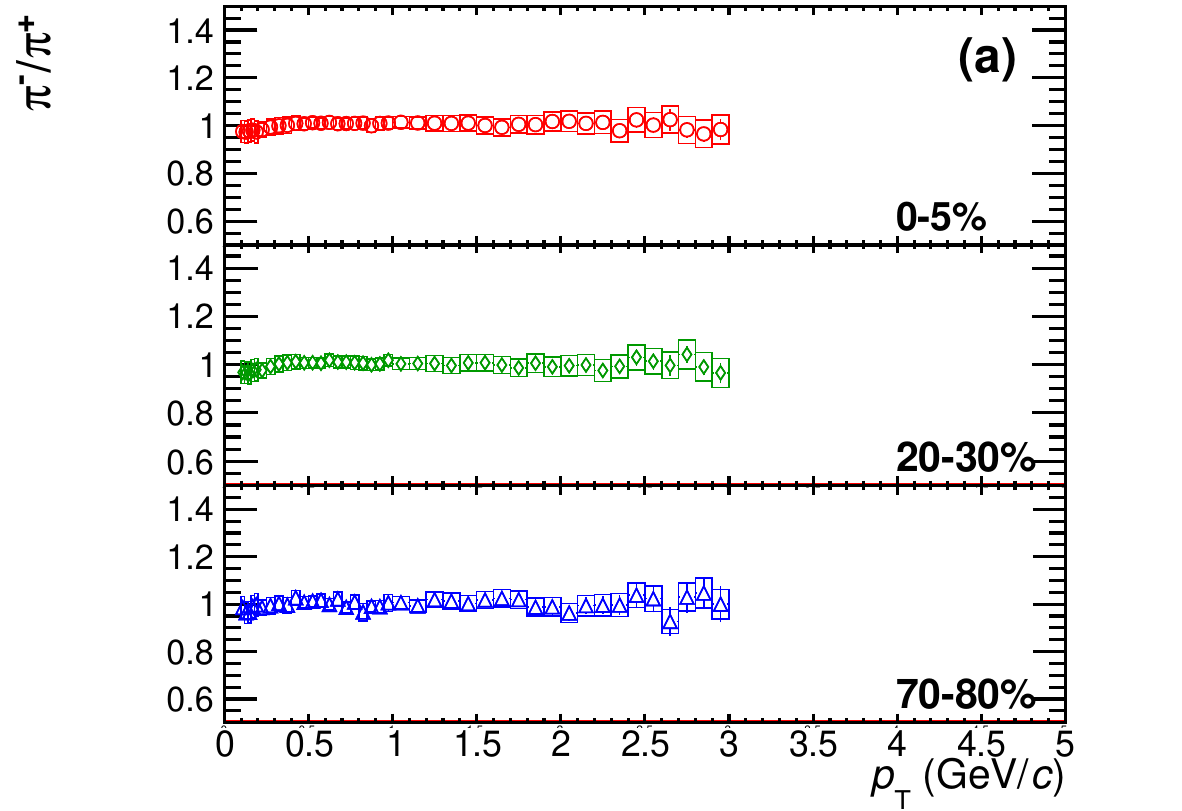}
  \includegraphics[width=\mywidth]{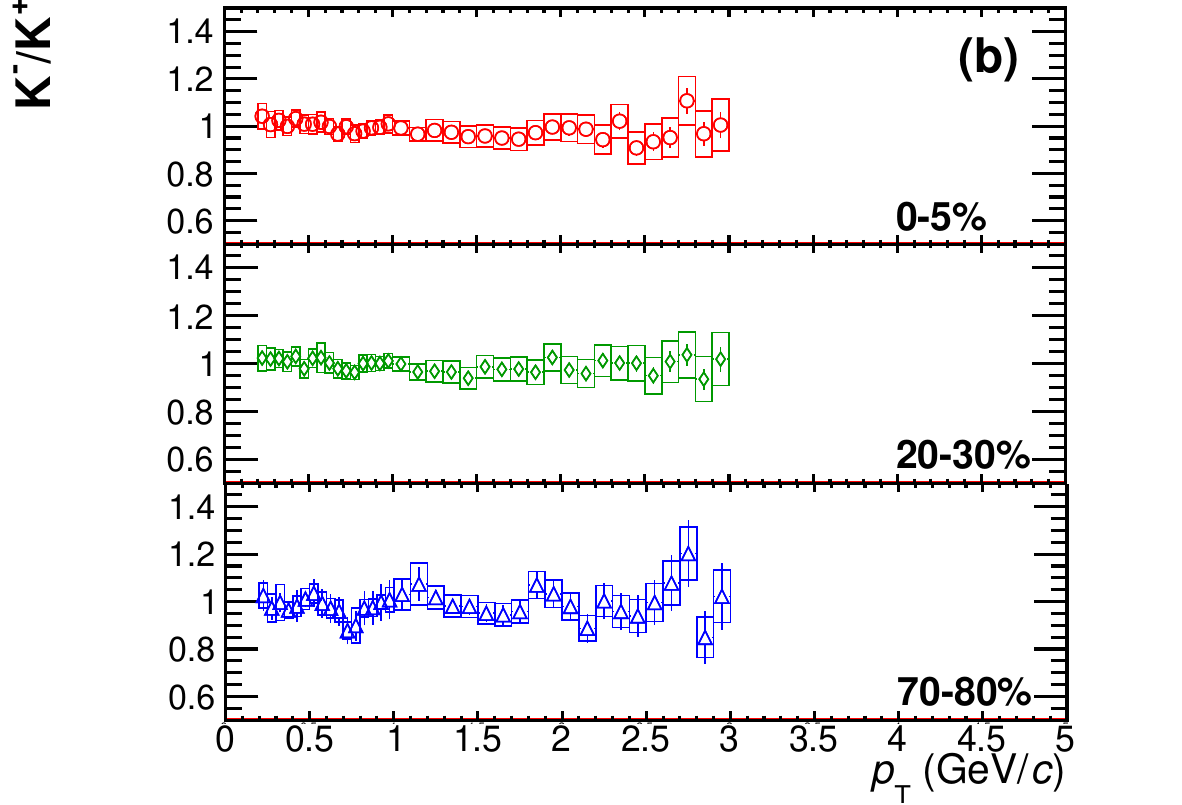}
  \includegraphics[width=\mywidth]{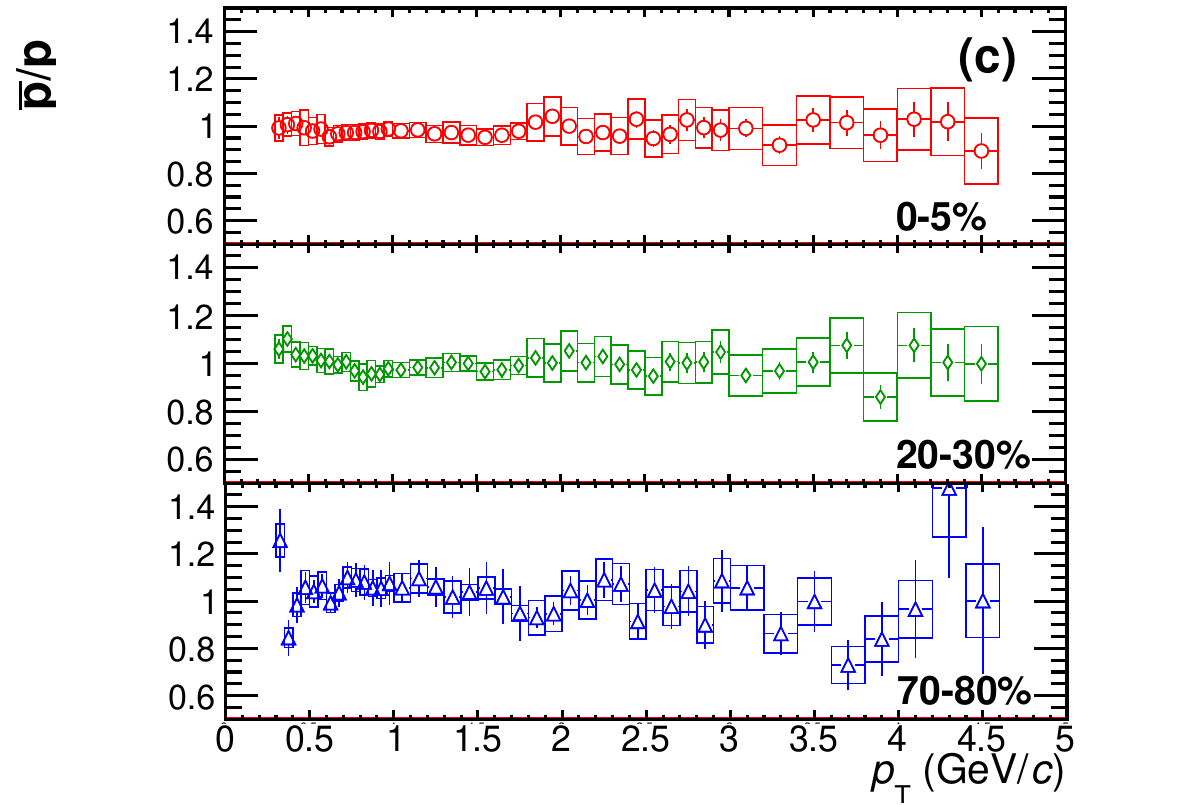}
  \caption{(color online) (a) $\pi^-/\pi^+$  (b) $K^-/K^+$  (c) $\bar{\rm p}/{\rm p}$ ratios as a function of \pt\ for central, semi-central and peripheral events.}
  \label{fig:ratio-neg-pos-vs-pt}
\end{figure}

\begin{figure}[htb!]
  \centering
  \includegraphics[width=\mywidth]{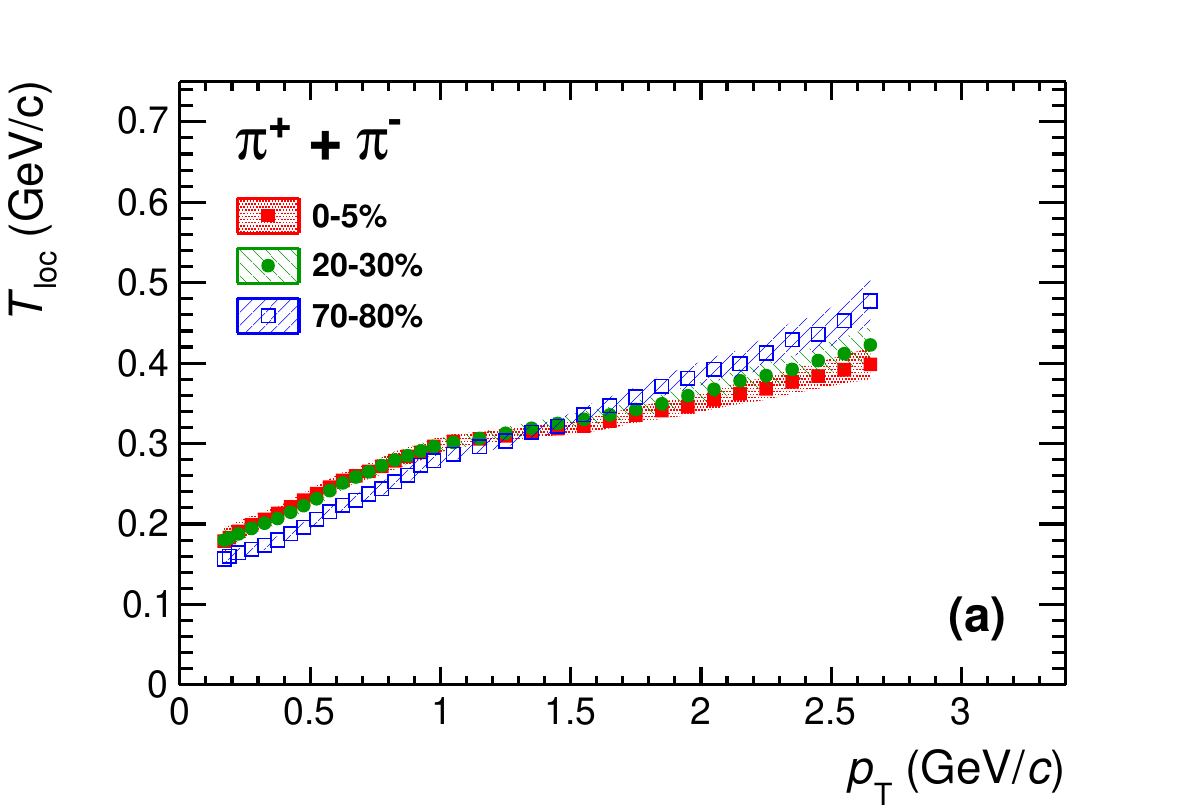}
  \includegraphics[width=\mywidth]{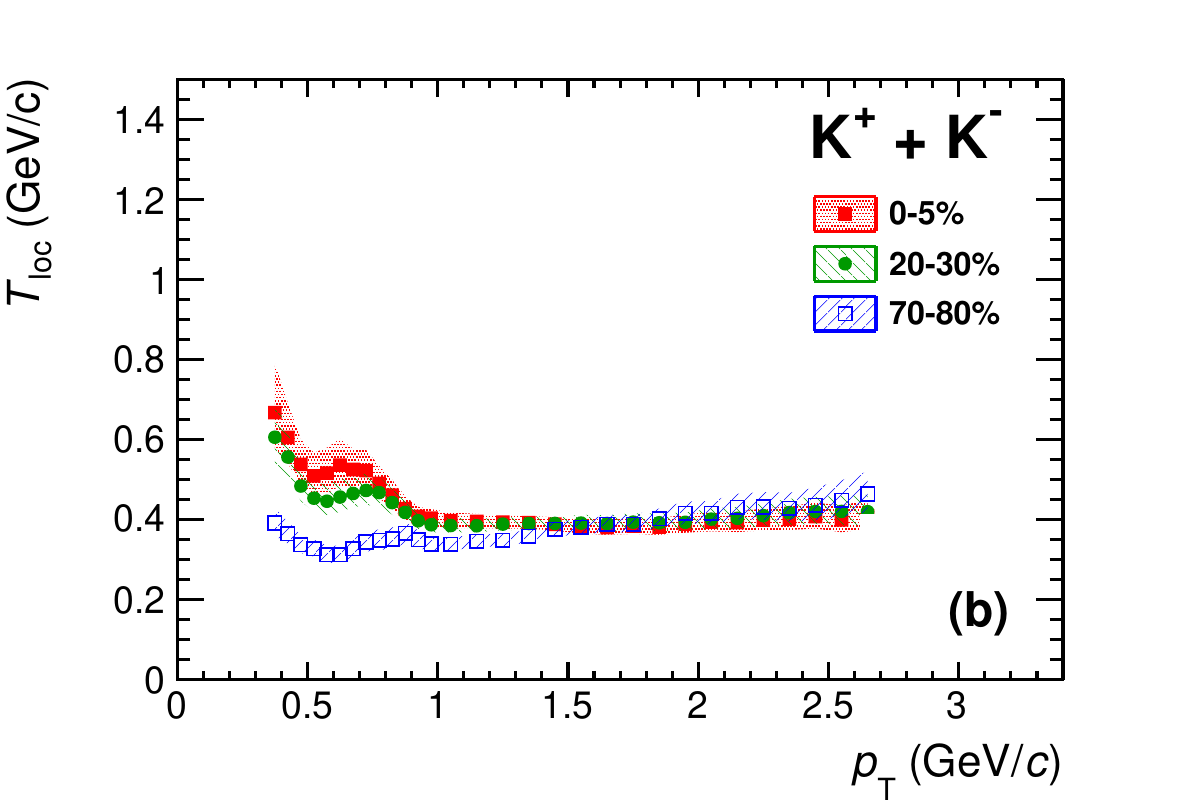}
  \includegraphics[width=\mywidth]{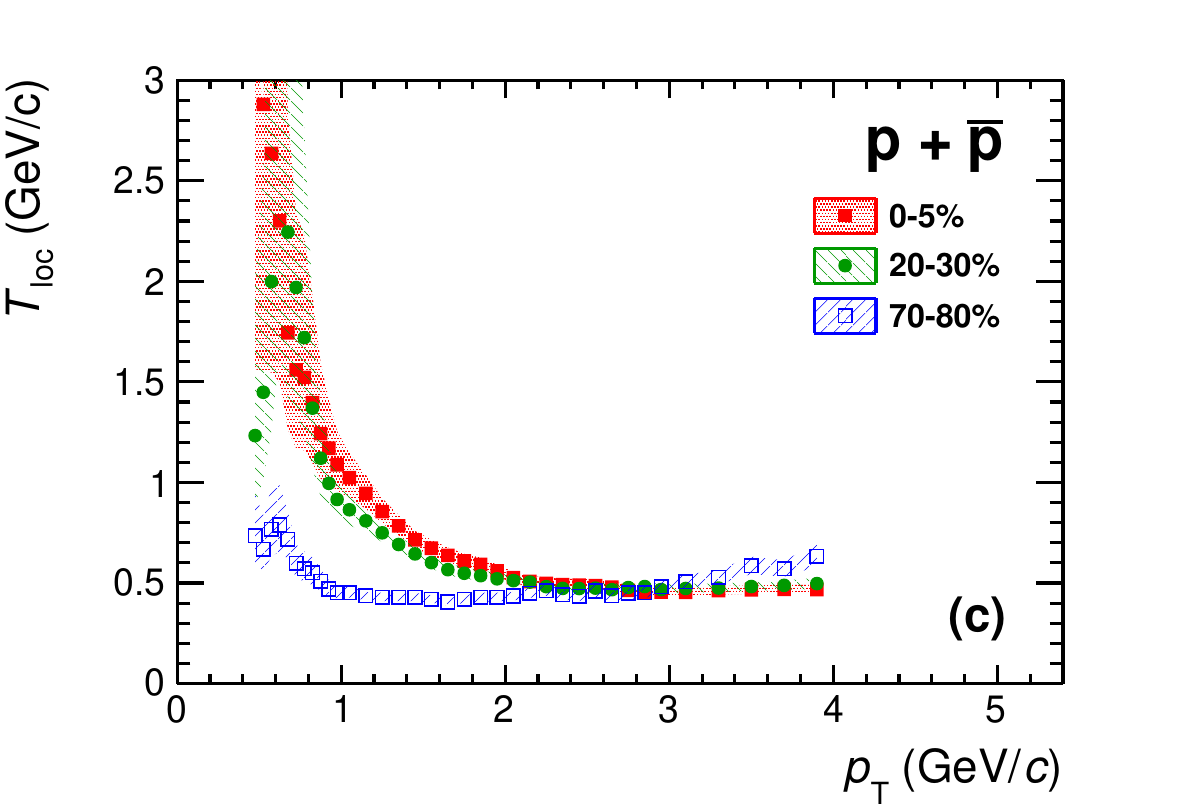}
  \caption{(color online) Local slopes of the \pt\ distributions of
    (a) $\pi$, (b) K,  (c) p (summed charge states) as a function
    of \pt\ and centrality. See text for details.}
  \label{fig:spectra-local-slope}
\end{figure}

\subsection {Transverse Momentum Distributions}

The combined spectra in the centrality bins reported in
Table~\ref{tab:yields} are shown in
Fig.~\ref{fig:spectra-results}. The distributions of positive and
negative particles are compatible within uncertainties at all \pt, as
expected at LHC energies. This is clearly seen in the ratio of
negative to positive \pt\ spectra shown in
Fig.~\ref{fig:ratio-neg-pos-vs-pt}. For this reason, we focus in the following mostly on results for combined charges. 

The change of shapes with centrality is apparent in Fig.~\ref{fig:spectra-results}: the spectra get
harder with increasing centrality. The spectra for all particle
species have an almost exponential shape at high \pt\ in central
collisions: contrary to what is observed in pp collisions~\cite{Aamodt:2011zj} there is no
obvious onset of a high-\pt\ power-law tail. In more peripheral collisions, the onset of the power-law is visible at high \pt. This is quantified in Fig.~\ref{fig:spectra-local-slope} which shows the local inverse slope $T_{\rm loc}$ of the spectra, as a function of \pt, computed with a fit using five bins in the vicinity of each \pt\ bin with the function
\begin{equation}
  \label{eq:1}
  \frac{1}{\pt} \frac{\mathrm{d}N}{\mathrm{d}\pt} \propto e^{-\pt/T_{\rm loc}}.
\end{equation}

This approach guarantees a numerically stable extraction of the local
inverse slope parameter $T_{\rm loc}$, but leads to a significant
correlation of the depicted uncertainties for neighboring points. Residual
correlations in the bin-by-bin systematics artificially increase the
uncertainties on $T_{\rm loc}$ shown in Fig.~\ref{fig:spectra-local-slope}.
As seen in the figure, the inverse slopes of K and p become larger
with decreasing \pt\ (the spectra are flatter at low \pt), and this
change is more pronounced for central collisions. The proton spectrum
at low \pt\ is nearly flat, making $T_{\rm loc}$ unconstrained (for a
flat distribution, $T_{\rm loc} \rightarrow \infty$).  Above a certain
\pt, roughly 1 GeV/$c$ for K and 2 GeV/$c$ for p, the slopes do not
change with \pt\ for central and semi-central collisions, consistent
with an exponential shape. Protons and kaons converge to similar
values of $T_{\rm loc} \sim 0.45$~GeV/$c$. For peripheral collisions,
a modest increase of $T_{\rm loc}$ is seen at the highest \pt,
suggesting the onset of a power-law behavior. The $T_{\rm loc}$ trend
is different for pions. The $\pi$ spectra are not purely
exponential. At high \pt\ the power-law rise is more suppressed in
central collisions compared to peripheral ones. At low \pt\ $T_{\rm
  loc}$ increases with \pt, opposite to the trend observed for protons
and kaons. This steepening of the pion spectra is a general feature of
heavy-ion collisions, and is due to the large contribution of
resonance decays to the pion spectrum, as already noted
in~\cite{Sollfrank:1990qz,Schnedermann:1993ws}. Above $\pt \simeq
1$~GeV/$c$, the rate of increase of $T_{\rm loc}$ is slower than at
lower \pt, and is less pronounced for central collisions.

\subsection{\pt-Integrated Yields and Mean Transverse Momentum}
\label{sec:pt-integ-yields}

In order to extrapolate to zero \pt\ for the extraction of \pt-integrated yields and \avpT, the spectra were fitted individually with a blast-wave function~\cite{Schnedermann:1993ws}:

\begin{equation}
  \frac{1}{\pt} \frac{\mathrm{d}N}{\mathrm{d}\pt} \propto \int_0^R r \mathrm{d}r\, m_{\rm T}\, I_0 \left( \frac{p_{\rm T}\sinh \rho}{T_{kin}} \right) K_1 \left( \frac{m_{\rm T}\cosh \rho}{T_{kin}} \right),
  \label{eq:blast-wave}
\end{equation}

\noindent where the velocity profile $\rho$ is described by
\begin{equation}
 \rho = \tanh^{-1} \beta_{\rm T} = \tanh^{-1} \Biggl(\left(\frac{r}{R}\right)^{n} \beta_{s} \Biggr) \; .
 \label{eq:rhoBWdefintion}
\end{equation}

Here, $\mt = \sqrt{\pt^2+m^2}$ is the transverse mass, $I_0$ and $K_1$
the modified Bessel functions, $r$ is the radial distance in the
transverse plane, $R$ is the radius of the fireball, $\beta_{\rm T}$
is the transverse expansion velocity and $\beta_{s}$ is the transverse
expansion velocity at the surface. From these equations one can also
derive the average transverse expansion velocity \avbT.  The free
parameters in the fit are the freeze-out temperature \Tfo, the average
transverse velocity \avbT\ and the exponent of the velocity profile
$n$. This function describes very well all particle species over the
whole measured \pt\ range (as individual fits). It should be noted,
however, that from fits to a single particle species no physics
meaning can be attached to those parameters.  A combined fit to
different particle species can provide insight on the freeze-out
parameters, and this is discussed in detail in the next section.  The
fraction of extrapolated yield is small: about 7\%, 6\%, 4\% for
$\pi$, K, p, respectively. The systematic uncertainties due to the
extrapolation amounts to 2.5\%, 3\%, 3\% for the yields and to 2\%,
1\%, 1\% for \avpT (independent of centrality). This was estimated
using different fit functions~\cite{Abelev:2008ez} (Boltzmann, \mt\
exponential, \pt\ exponential, Tsallis-Levy, Fermi-Dirac,
Bose-Einstein), restricting the fit range to low \pt\ for those
functions not giving a satisfactory description of the spectra over
the full range. The \avpT\ is computed extrapolating with the blast
wave function to 100~GeV/$c$ (infinity, effectively). The difference
between the \avpT\ computed with and without the extrapolation at high
\pt\ is $<1\%$, $\sim1.5$\%, $<1\%$ for $\pi$, K, and p respectively.
The extracted particle yields and \avpT\ as a function of centrality
are summarized in Table~\ref{tab:yields} and Table~\ref{tab:meanpt}.

\begin{figure}[tp]
  \centering
  \includegraphics[width=\mywidth]{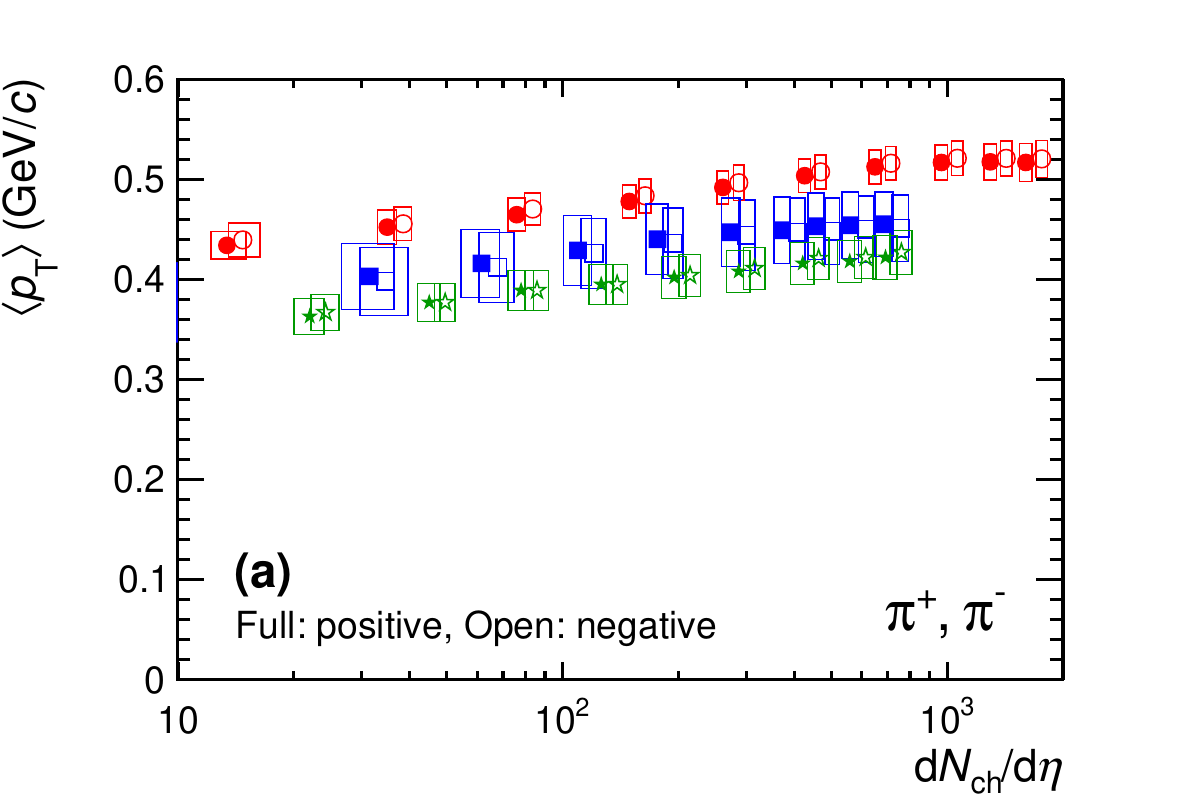}
  \includegraphics[width=\mywidth]{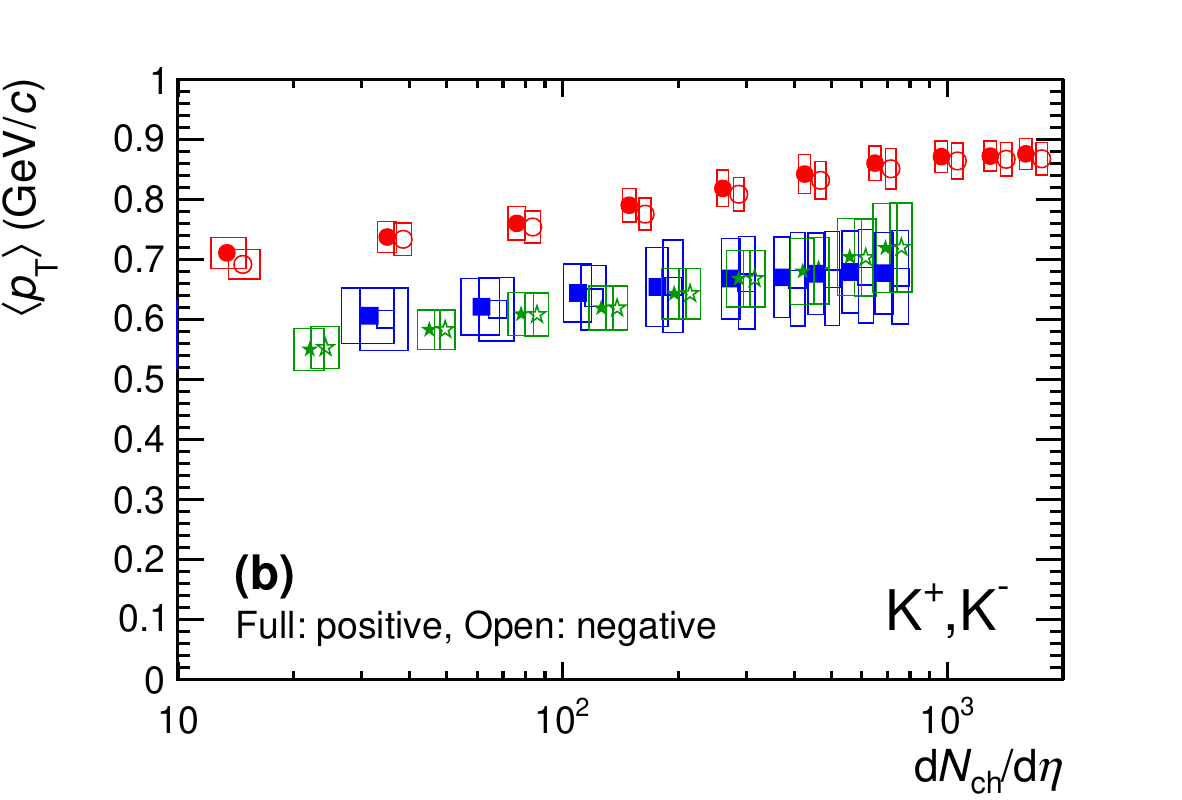}
  \includegraphics[width=\mywidth]{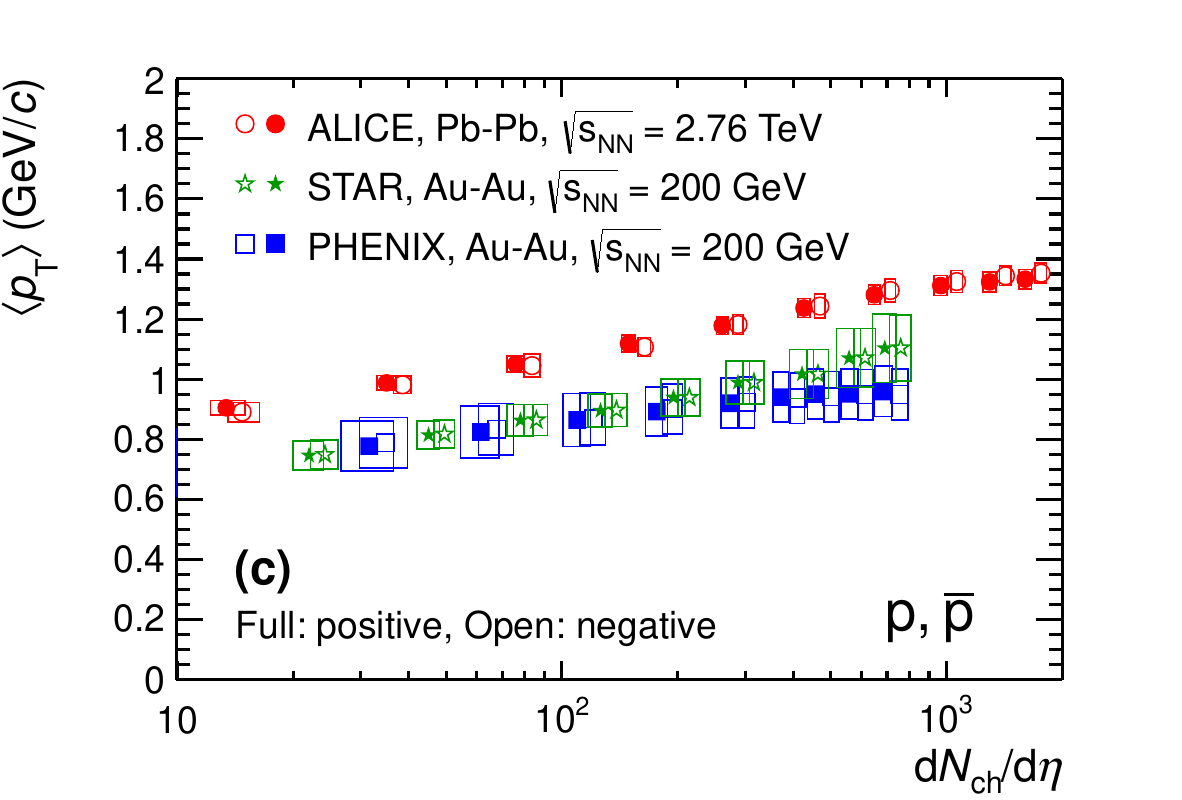}
  \vfill
  \caption{(color online) Mean transverse momentum \avpT\ as a function of \dNdeta\ for (a) $\pi$, (b) K,  (c) p, compared to RHIC results at \snn~=~200~GeV~\cite{Abelev:2008ez,Adler:2003cb}. Negative charge results displaced for better readability. Boxes: systematic uncertainties.}
  \label{fig:meanpt}
\end{figure}

Figure~\ref{fig:meanpt} shows the mean transverse momentum \avpT\ as a
function of \dNdeta, compared to previous results at RHIC\footnote{The
  RHIC data are plotted as a function of \dNdeta\ using the measured
  \dNdeta\ of each individual
  experiment~\cite{Adler:2004zn,Bearden:2001qq,Abelev:2008ez}. In the
  case of PHENIX, the \dNdeta\ are published in 5\% percentiles, while
  the spectra are published mostly in 10\% percentiles: whenever
  needed, the value of the \dNdeta\ used in the figure is a linear
  interpolation of the 5\% percentiles; the \dNdeta\ measurement,
  moreover, is only available up to 70\% centrality in PHENIX.  Since
  there is some disagreement in the \dNdeta\ measurements from
  different RHIC experiments for peripheral event, we decided not to
  plot in Fig.~\ref{fig:meanpt} the PHENIX results below 70\%
  centrality. The discrepancy is also visible in
  Fig.~\ref{fig:spectra-vs-hydro-2}, where the difference in
  normalization between STAR and PHENIX is apparent. The STAR (anti)proton
  measurement is inclusive of products from weak decays of strange
  particles, and therefore not included in all comparisons shown in
  this
  section.}~\cite{Abelev:2008ez,Adler:2004zn,Adler:2003cb}. The
\avpT\ increases with centrality and is higher than the
lower energy results for comparable charged particle densities.  The
\avpT\ at RHIC was found to be compatible with a scaling as a function
of \dNdeta\
for different energies~\cite{Abelev:2008ez}. This scaling is clearly excluded at the
LHC.

\begin{sidewaystable*}[!ph]

  \ifpreprint
  \vspace{1cm}
  \else
  \vspace{10cm}
  \fi
  \begin{minipage}[b]{\linewidth}
    \ifpreprint
    \footnotesize
    \fi

  \centering
  
  \caption{\label{tab:yields}Charged particle multiplicity density~\cite{Aamodt:2010cz}
    (total uncertainties) and mid-rapidity particle yields
    $\displaystyle{\frac{\mathrm{d}N_{i}}{\mathrm{d}y}\bigl\vert_{|y|<0.5}}$
    (statistical uncertainties and systematic uncertainties including extrapolation
    uncertainty). The last column indicates the additional
    normalization uncertainty coming from the centrality definition.}

  \begin{tabular*}{\linewidth}{@{} c@{\extracolsep{\fill}}c@{\extracolsep{\fill}}c@{\extracolsep{\fill}}c@{\extracolsep{\fill}}c@{\extracolsep{\fill}}c@{\extracolsep{\fill}}c@{\extracolsep{\fill}}c@{\extracolsep{\fill}}c@{\extracolsep{\fill}}c @{}} 
    \hline
    Centrality & $\dNdeta$ & $\pi^{+}$ & $\pi^{-}$ & K$^{+}$ & K$^{-}$
    & p & \pbar & Norm. Uncertainty \\
    \hline
    \hline

    0--5\%   &   $1601 \pm 60$  & 
    $733 \pm 54$   & 
    $732 \pm 52$  & 
    $109 \pm 9$   & 
    $109 \pm 9$   & 
    $34  \pm 3$   & 
    $33  \pm 3$    &
    0.5\% \\

    5--10\%  &   $1294 \pm 49$  & 
    $606  \pm 42$   & 
    $604  \pm 42$  & 
    $91   \pm 7$   & 
    $90   \pm 8$   & 
    $28   \pm 2$   & 
    $28   \pm 2$     &
    0.5\% \\

    10--20\% &   $966  \pm 37$  & 
    $455  \pm 31$   & 
    $453  \pm 31$  & 
    $68   \pm 5$   & 
    $68   \pm 6$   & 
    $21.0 \pm 1.7$   & 
    $21.1 \pm 1.8$     &
    0.7\% \\

    20--30\% &   $649  \pm 23$  & 
    $307  \pm 20$   & 
    $306  \pm 20$  & 
    $46   \pm 4$   & 
    $46   \pm 4$   & 
    $14.4 \pm 1.2$   & 
    $14.5 \pm 1.2$     &
    1\% \\

    30--40\% &   $426  \pm 15$  & 
    $201 \pm 13$   & 
    $200 \pm 13$  & 
    $30  \pm 2$   & 
    $30  \pm 2$   & 
    $9.6 \pm 0.8$   & 
    $9.7 \pm 0.8$     &
    2\% \\

    40--50\% &   $261  \pm 9 $  & 
    $124    \pm 8$   & 
    $123    \pm 8$  & 
    $18.3   \pm 1.4$   & 
    $18.1   \pm 1.5$   & 
    $6.1    \pm 0.5$   & 
    $6.2    \pm 0.5$     &
    2.4\% \\

    50--60\% &   $149  \pm 6 $  & 
    $71   \pm 5$   & 
    $71   \pm 4$  & 
    $10.2 \pm 0.8$   & 
    $10.2 \pm 0.8$   & 
    $3.6  \pm 0.3$   & 
    $3.7  \pm 0.3$     &
    3.5\% \\

    60--70\% &   $76   \pm 4 $  & 
    $37  \pm 2$   & 
    $37  \pm 2$  & 
    $5.1 \pm 0.4$   & 
    $5.1 \pm 0.4$   & 
    $1.9 \pm 0.2$   & 
    $2.0 \pm 0.2$     &
    5\% \\

    70--80\% &   $35   \pm 2 $  & 
    $17.1  \pm 1.1$   & 
    $17.0  \pm 1.1$  & 
    $2.3   \pm 0.2$   & 
    $2.3   \pm 0.2$   & 
    $0.90  \pm 0.08$   & 
    $0.93  \pm 0.09$     &
    6.7\% \\

    80--90\% & 13.4 +1.6 -1.2     & 
    $6.6  \pm 0.4$   & 
    $6.6  \pm 0.4$  & 
    $0.85 \pm 0.08$   & 
    $0.86 \pm 0.09$   & 
    $0.36 \pm 0.04$   & 
    $0.36 \pm 0.04$     &
    +12\% -8.5\% \\

    \hline
  \end{tabular*}

\ifpreprint
  \vspace{2cm}
\else
  \vspace{5cm}
\fi
\end{minipage}

\begin{minipage}[b]{\linewidth}
    \ifpreprint
    \footnotesize
    \fi

  \centering
  \caption{  \label{tab:meanpt}
\avpT\ as a function of centrality (GeV/$c$), statistical uncertainties and
    systematic uncertainties including extrapolation uncertainty summed in quadrature (systematic uncertainties dominate).}
  \begin{tabular*}{\linewidth}{@{} c@{\extracolsep{\fill}}c@{\extracolsep{\fill}}c@{\extracolsep{\fill}}c@{\extracolsep{\fill}}c@{\extracolsep{\fill}}c@{\extracolsep{\fill}}c @{}} 
    \hline
    Centrality & $\pi^{+}$ & $\pi^{-}$ & K$^{+}$ & K$^{-}$ & p & \pbar \\
    \hline
    \hline
    0--5\%   & 0.517 $\pm$ 0.019 & 0.520 $\pm$ 0.018 & 0.876 $\pm$ 0.026 & 0.867 $\pm$ 0.027 & 1.333 $\pm$ 0.033 & 1.353 $\pm$ 0.034 \\[0.3em]
    5--10\%  & 0.517 $\pm$ 0.018 & 0.521 $\pm$ 0.017 & 0.872 $\pm$ 0.025 & 0.866 $\pm$ 0.028 & 1.324 $\pm$ 0.033 & 1.344 $\pm$ 0.033 \\[0.3em]
    10--20\% & 0.517 $\pm$ 0.017 & 0.521 $\pm$ 0.017 & 0.871 $\pm$ 0.027 & 0.864 $\pm$ 0.030 & 1.311 $\pm$ 0.034 & 1.325 $\pm$ 0.036 \\[0.3em]
    20--30\% & 0.512 $\pm$ 0.017 & 0.516 $\pm$ 0.017 & 0.860 $\pm$ 0.029 & 0.851 $\pm$ 0.034 & 1.281 $\pm$ 0.033 & 1.295 $\pm$ 0.039 \\[0.3em]
    30--40\% & 0.504 $\pm$ 0.017 & 0.507 $\pm$ 0.017 & 0.842 $\pm$ 0.032 & 0.831 $\pm$ 0.031 & 1.237 $\pm$ 0.032 & 1.243 $\pm$ 0.041 \\[0.3em]
    40--50\% & 0.492 $\pm$ 0.017 & 0.497 $\pm$ 0.018 & 0.818 $\pm$ 0.030 & 0.808 $\pm$ 0.028 & 1.178 $\pm$ 0.030 & 1.182 $\pm$ 0.033 \\[0.3em]
    50--60\% & 0.478 $\pm$ 0.017 & 0.483 $\pm$ 0.017 & 0.790 $\pm$ 0.028 & 0.775 $\pm$ 0.027 & 1.118 $\pm$ 0.028 & 1.107 $\pm$ 0.032 \\[0.3em]
    60--70\% & 0.465 $\pm$ 0.017 & 0.470 $\pm$ 0.016 & 0.760 $\pm$ 0.028 & 0.754 $\pm$ 0.027 & 1.050 $\pm$ 0.027 & 1.045 $\pm$ 0.039 \\[0.3em]
    70--80\% & 0.452 $\pm$ 0.017 & 0.455 $\pm$ 0.017 & 0.737 $\pm$ 0.027 & 0.733 $\pm$ 0.028 & 0.987 $\pm$ 0.025 & 0.981 $\pm$ 0.031 \\[0.3em]
    80--90\% & 0.434 $\pm$ 0.014 & 0.439 $\pm$ 0.017 & 0.711 $\pm$ 0.027 & 0.692 $\pm$ 0.026 & 0.905 $\pm$ 0.026 & 0.890 $\pm$ 0.035 \\[0.3em]
    \hline
  \end{tabular*}
\end{minipage}
\end{sidewaystable*}

\begin{figure}[tp]
  \centering
  \includegraphics[width=\mywidth]{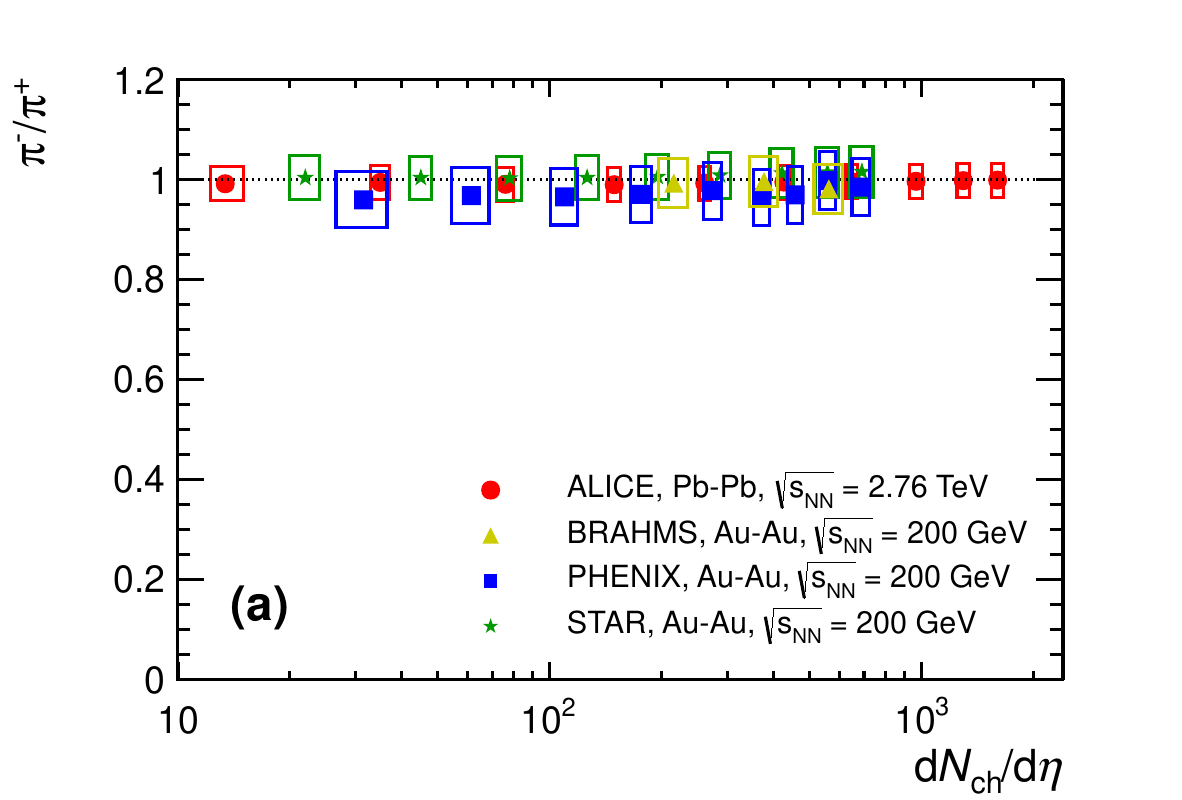}
  \includegraphics[width=\mywidth]{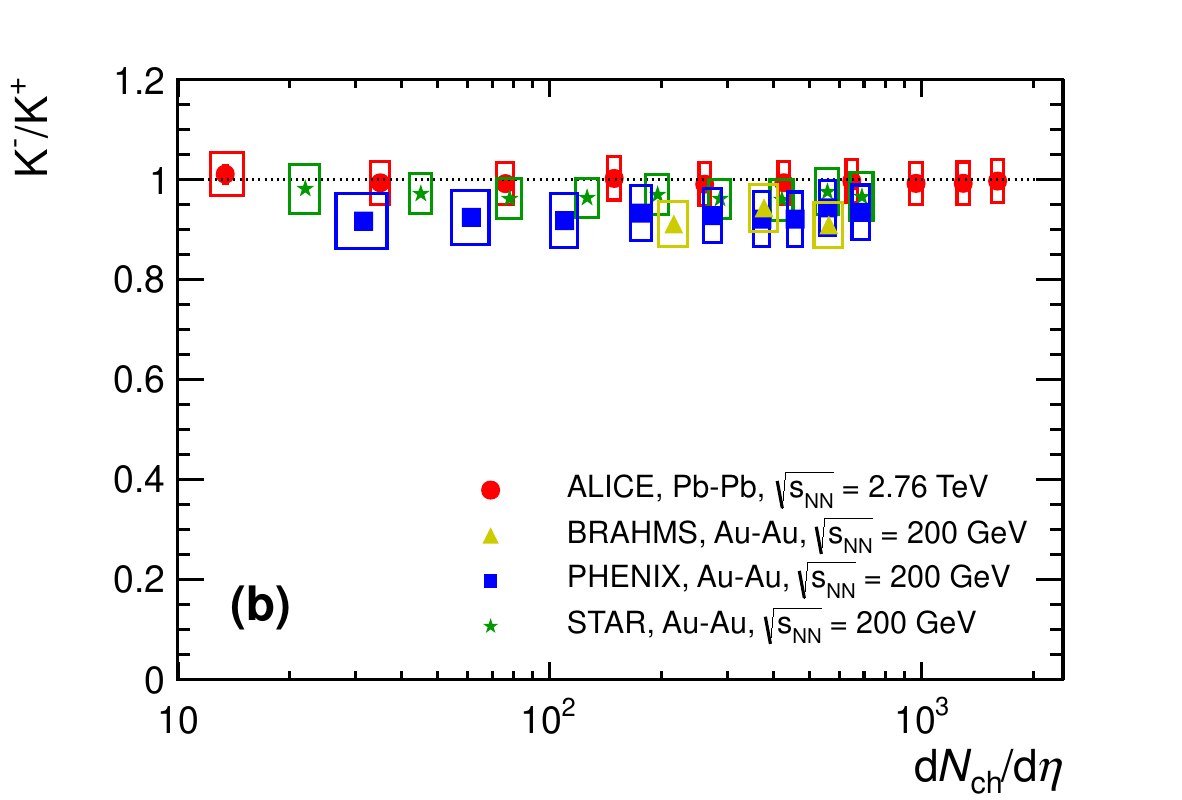}
  \includegraphics[width=\mywidth]{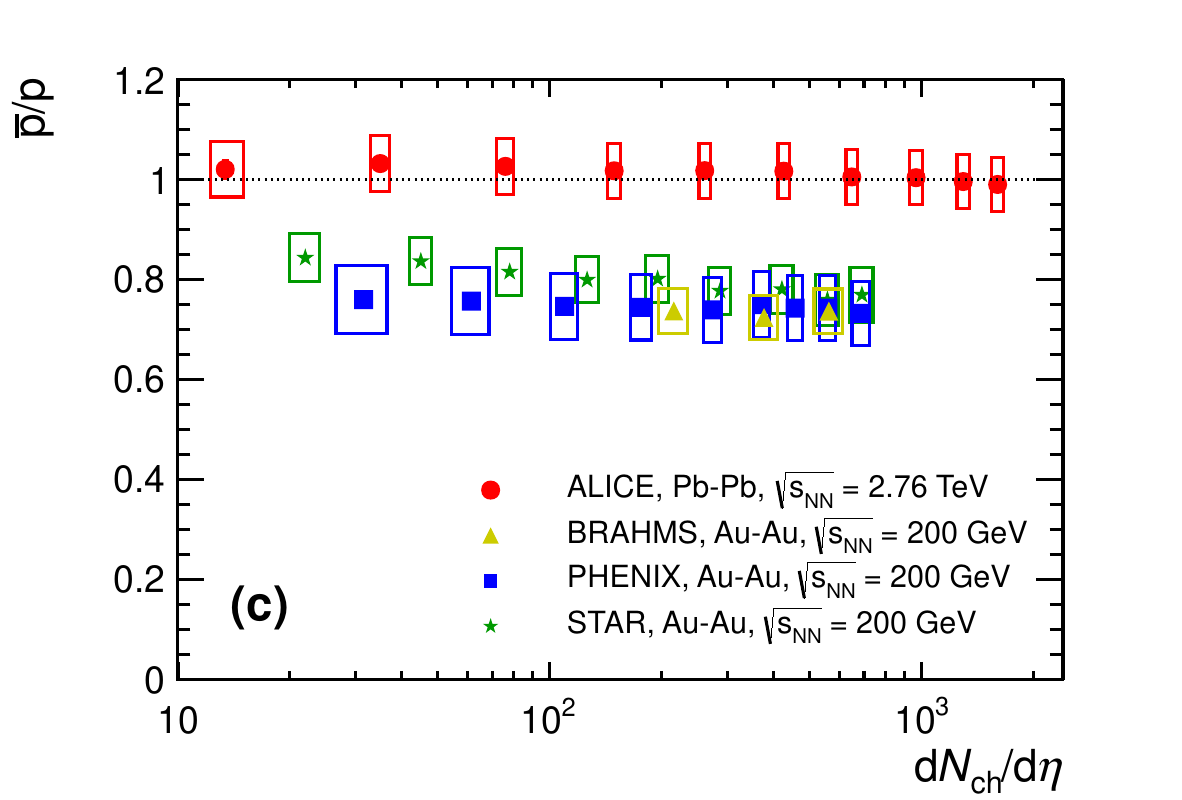}
  \caption{(color online) (a) $\pi^-/\pi^+$, (b) $K^-/K^+$  (c) $\bar{\rm p}/{\rm p}$ ratios as a function of \dNdeta,  compared to previous results at \snn~=~200~GeV.}
  \label{fig:ratiosposneg_vs_centr}
\end{figure}

\begin{figure}[tp]
  \centering
  \includegraphics[width=\mywidth]{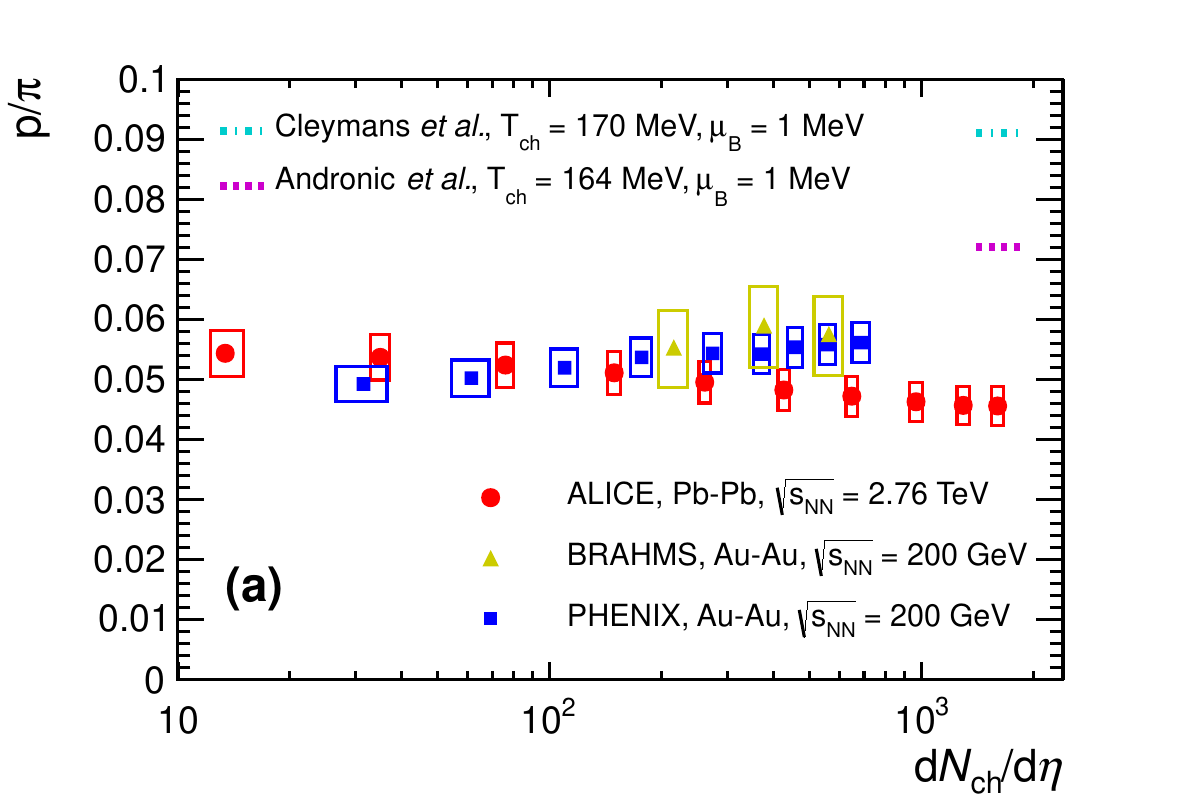}
  \includegraphics[width=\mywidth]{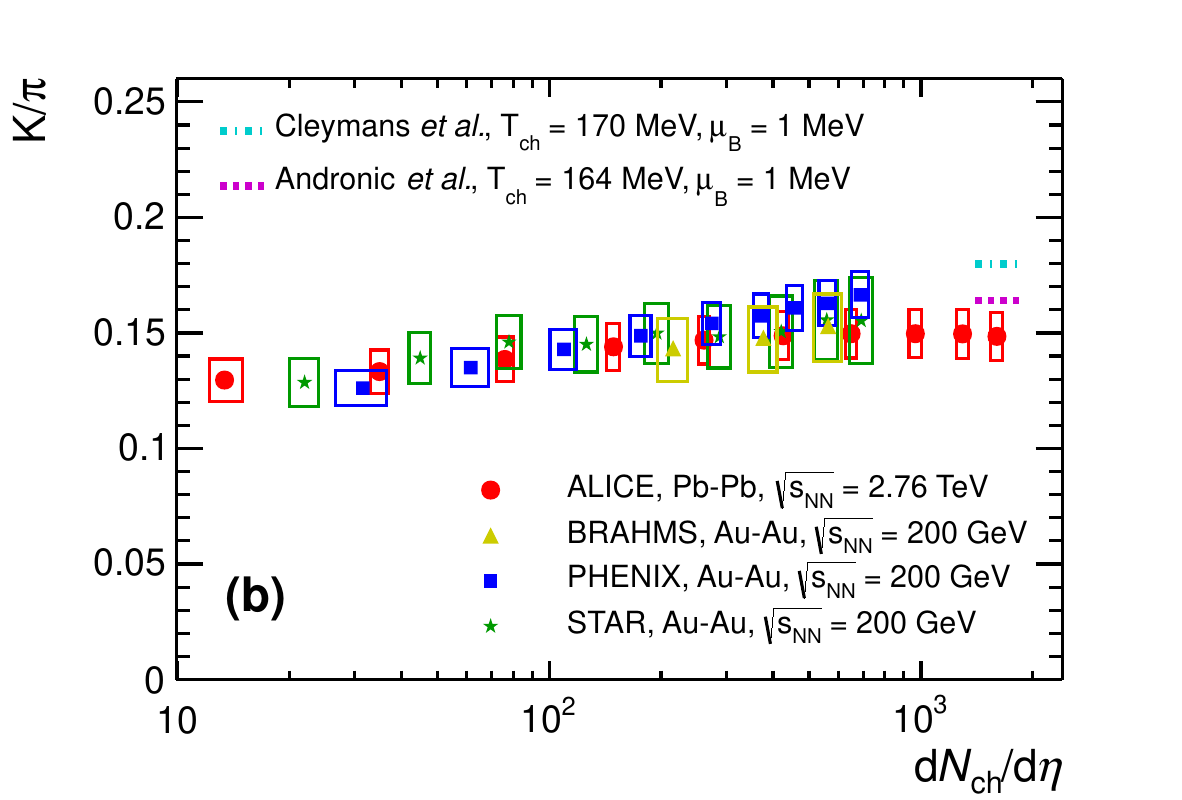}
  \caption{(color online) (a) $\ppi =
    (\mathrm{p} +\mathrm{\bar{p}})/(\pi^++\pi^-)$  and (b) $\kpi = (\mathrm{K^+} +
    \mathrm{K^-}))/(\pi^++\pi^-)$  ratios as a function
    of \dNdeta, compared to previous results at \snn~=~200~GeV.}
  \label{fig:ratios_vs_centr}
\end{figure}

The ratios of negative to positive particle yields (Fig.~\ref{fig:ratiosposneg_vs_centr}) are compatible with unity for all centralities. The $\bar{\mathrm{p}}/\mathrm{p}$ ratio, in particular, confirms the expectation of a vanishing baryon transport to mid-rapidity at the LHC, in contrast to to the RHIC energy regime, where the $\bar{\mathrm{p}}/\mathrm{p}$ ratio was found to be about 0.8 at \snn~=~200~GeV~\cite{Abelev:2008ez}. The effect of the different antibaryon to baryon asymmetry between the two energies is almost absent in the sum of the positive and negative charges. Therefore, the ratios $\kpi=(\mathrm{K}^+ + \mathrm{K}^-)/(\pi^+ + \pi^-)$\ and $\ppi=(\mathrm{p} + \mathrm{\bar p})/(\pi^+ + \pi^-)$ are compared to  RHIC~\cite{Bearden:2001qq,Abelev:2008ez,Adler:2003cb} in Fig.~\ref{fig:ratios_vs_centr} as a function of charged particle multiplicity. The \kpi\ ratio hints at a small increase with centrality following the trend from lower energy data. The \ppi\ ratio suggests a small decrease with centrality and is slightly lower than the RHIC measurements.

\section{Discussion}
\label{sec:discussion}
\begin{figure*}[tp]
  \centering
  \includegraphics[width=0.48\linewidth]{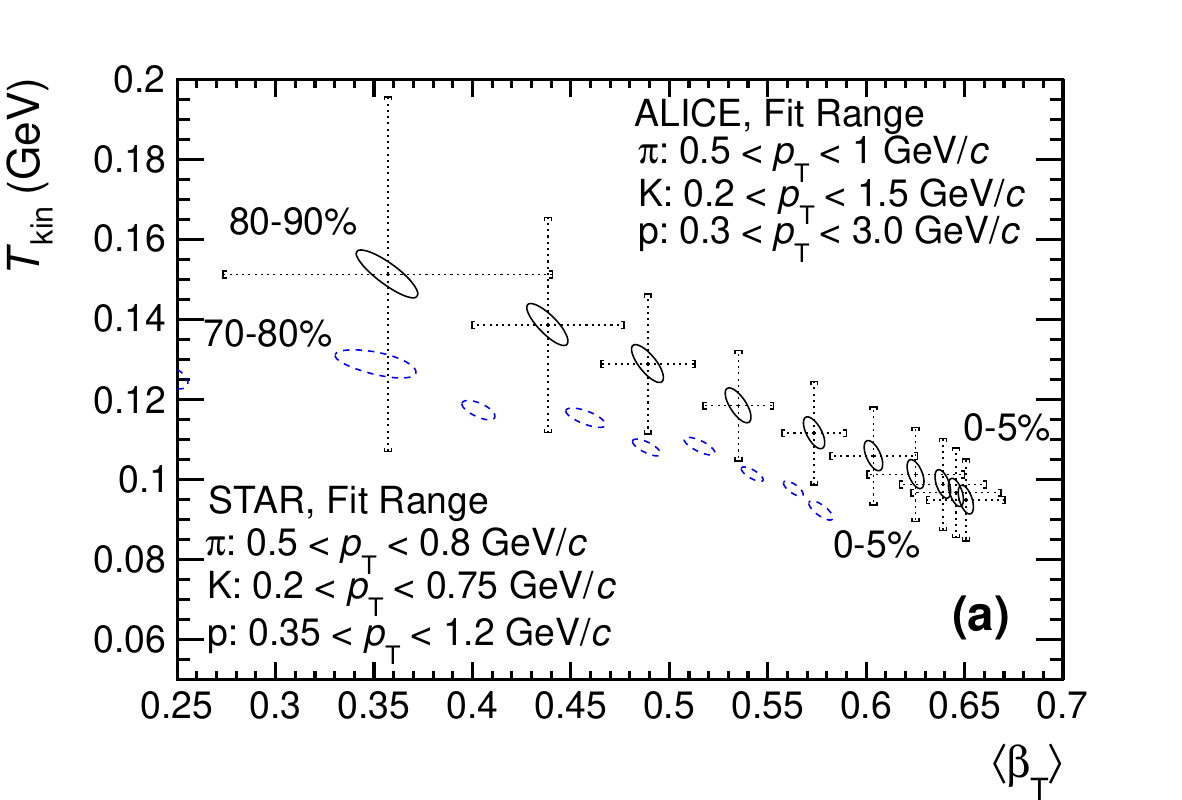}
  \includegraphics[width=0.48\linewidth]{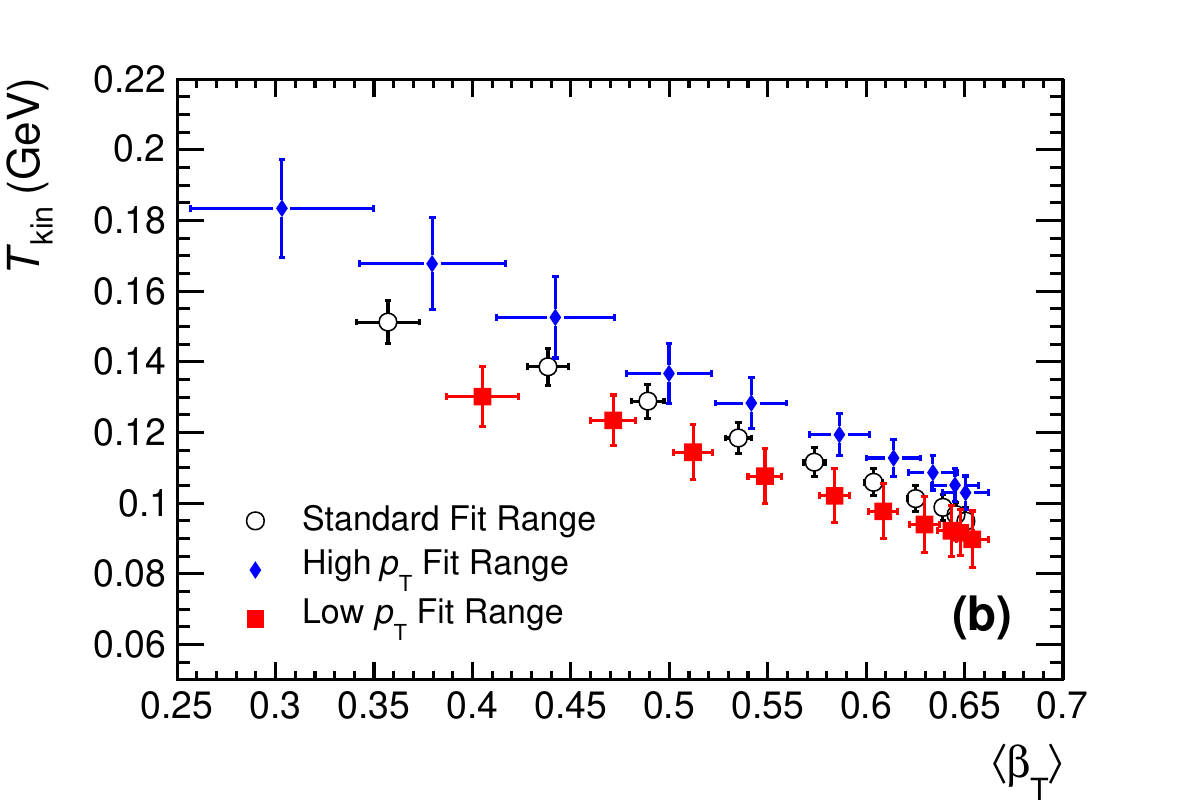}
  \caption{(color online) (a) Results of blast-wave fits, compared to similar fits at RHIC energies~\cite{Adams:2005dq}; the uncertainty contours include the effect of the bin-by-bin systematic uncertainties, the dashed error bars represents the full systematic uncertainty (see text for details), the STAR contours include only statistical uncertainties. (b) Comparison of fit results for different fit ranges; the error bars include only the effect of the bin-by-bin systematics (see text for details). }
  \label{fig:blast-wave-ellipses}
\end{figure*}

\begin{figure*}[tp]
  \centering
  \includegraphics[width=0.48\linewidth]{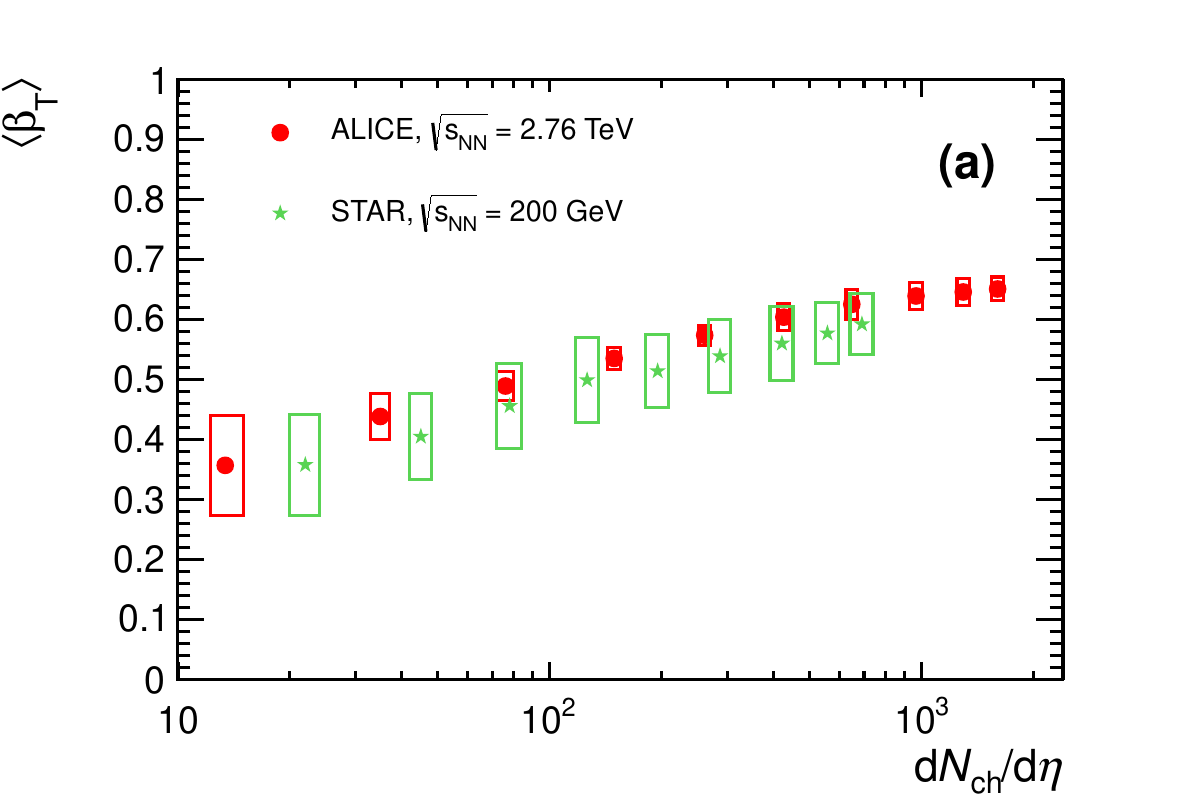}
  \includegraphics[width=0.48\linewidth]{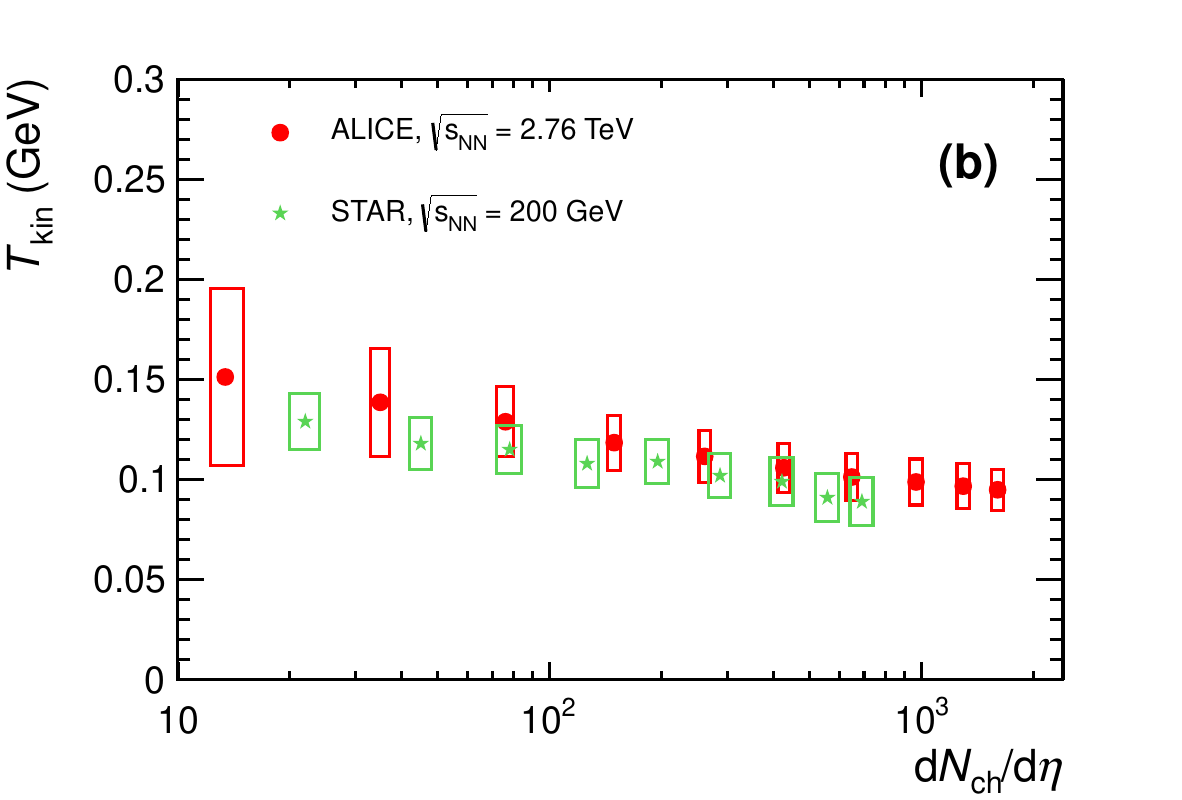}
  \caption{(color online)  Blast-wave parameters (a) \avbT\ and (b) \Tfo\ as a function of \dNdeta, compared to previous results at \snn~=~200~GeV~\cite{Adams:2005dq} (full systematic uncertainties for both experiments).}
  \label{fig:blast-wave-vs-dndeta}
\end{figure*}

\begin{figure}[tp]
  \centering
  \includegraphics[width=\mywidth]{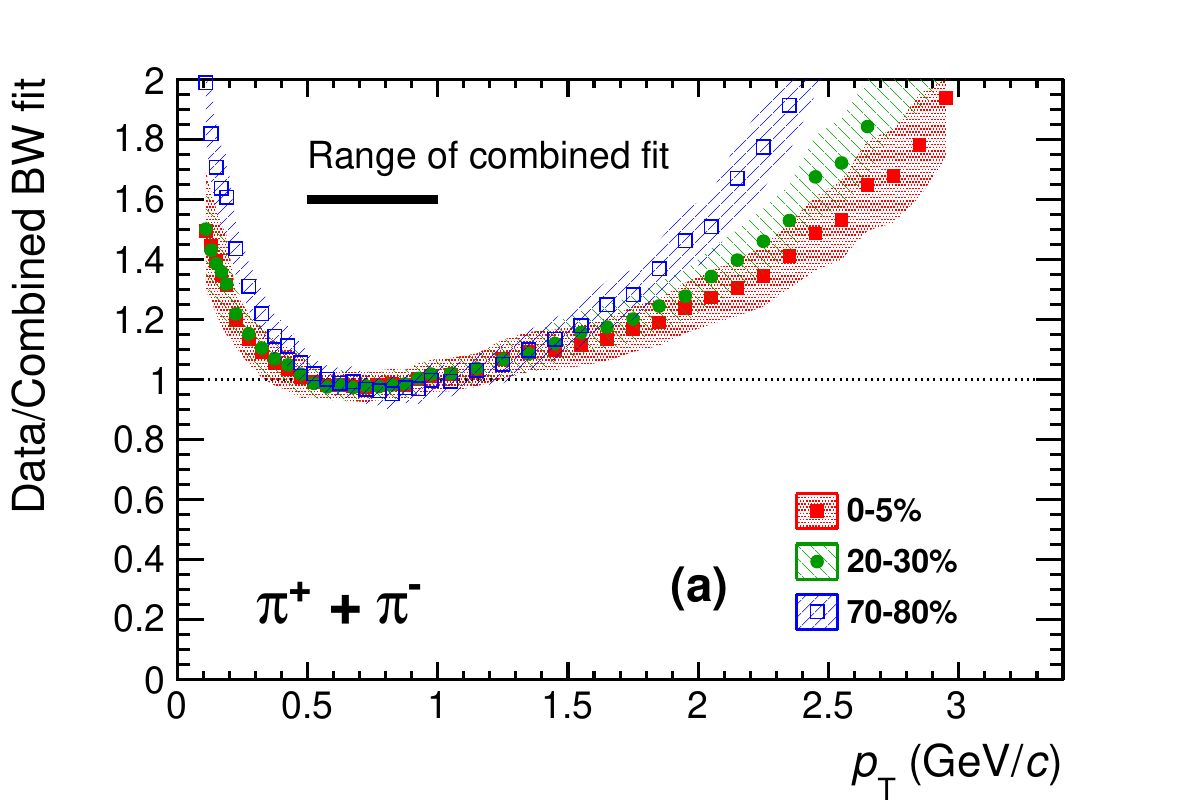}
  \includegraphics[width=\mywidth]{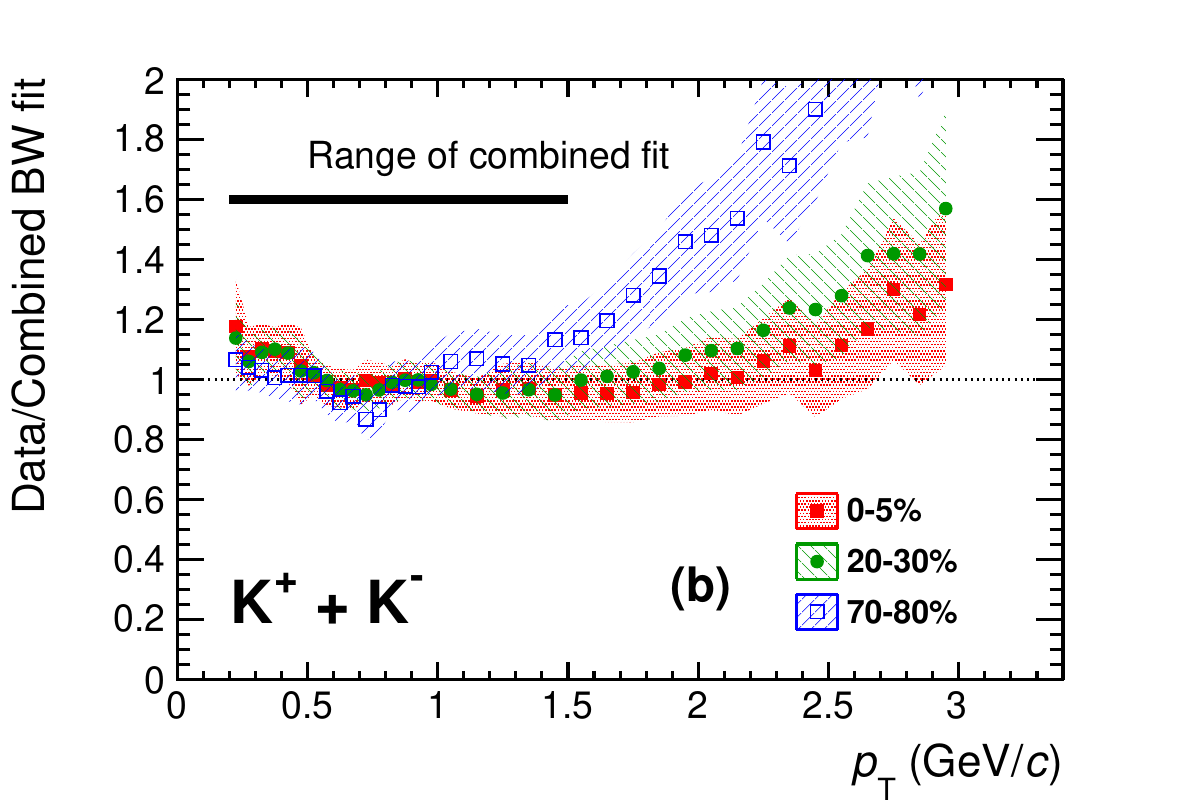}
  \includegraphics[width=\mywidth]{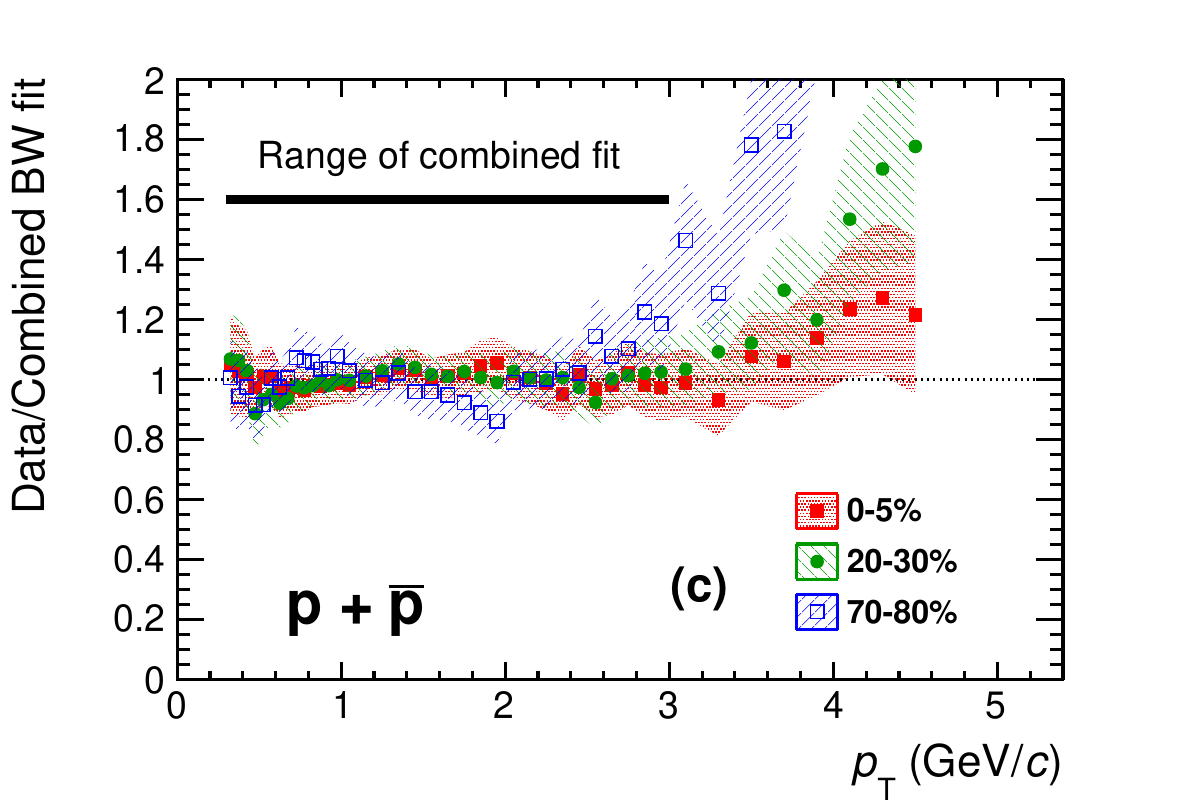}
  \caption{(color online) Ratio of the measured spectra to the combined blast-wave fit as a function of \pt\ for (a) $\pi$, (b) K,  (c) p.}
  \label{fig:RatioCombinedBW}
\end{figure}

\subsection{Transverse Momentum Distributions and Hydrodynamics}

 The change with centrality of the spectral shapes shown in Fig.~\ref{fig:spectra-results} and Fig.~\ref{fig:spectra-local-slope} can be interpreted in terms of hydrodynamics.
A flattening of the spectra, more pronounced at low \pt\ and for heavier particles, is expected in the hydrodynamical models as a consequence of the blue-shift induced by the collective expansion. The low-\pt\ change of the local slope shown in Fig.~\ref{fig:spectra-local-slope}, more pronounced for the proton spectra, thus suggests a progressively stronger radial flow with increasing centrality. The fact that the inverse slope converges to the same value for p and K at high \pt\ is also a generic feature of the blast wave parameterization.


\begin{figure}[tp]
  \centering
  \includegraphics[width=\mywidth]{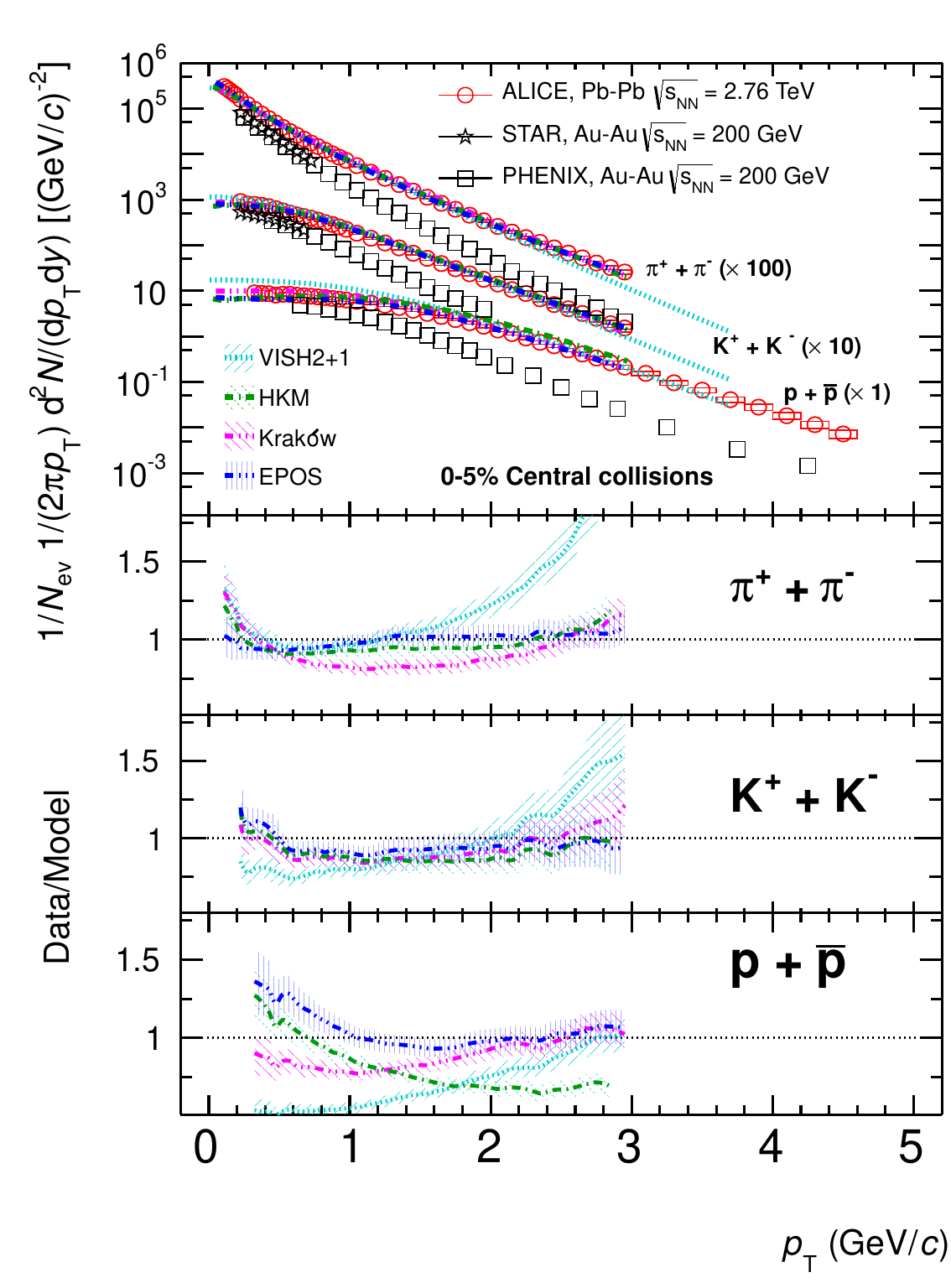}
  \caption{(color online) Spectra of particles for summed charge states in the centrality bin 0-5\%, compared to hydrodynamical models and results from RHIC at \snn~=~200~GeV~\cite{Abelev:2008ez,Adler:2003cb}. See text for details. Systematic uncertainties plotted (boxes); statistical uncertainties are smaller than the symbol size.}
  \label{fig:spectra-vs-hydro}
\end{figure}
\begin{figure}[tp]
  \centering
  \includegraphics[width=\mywidth]{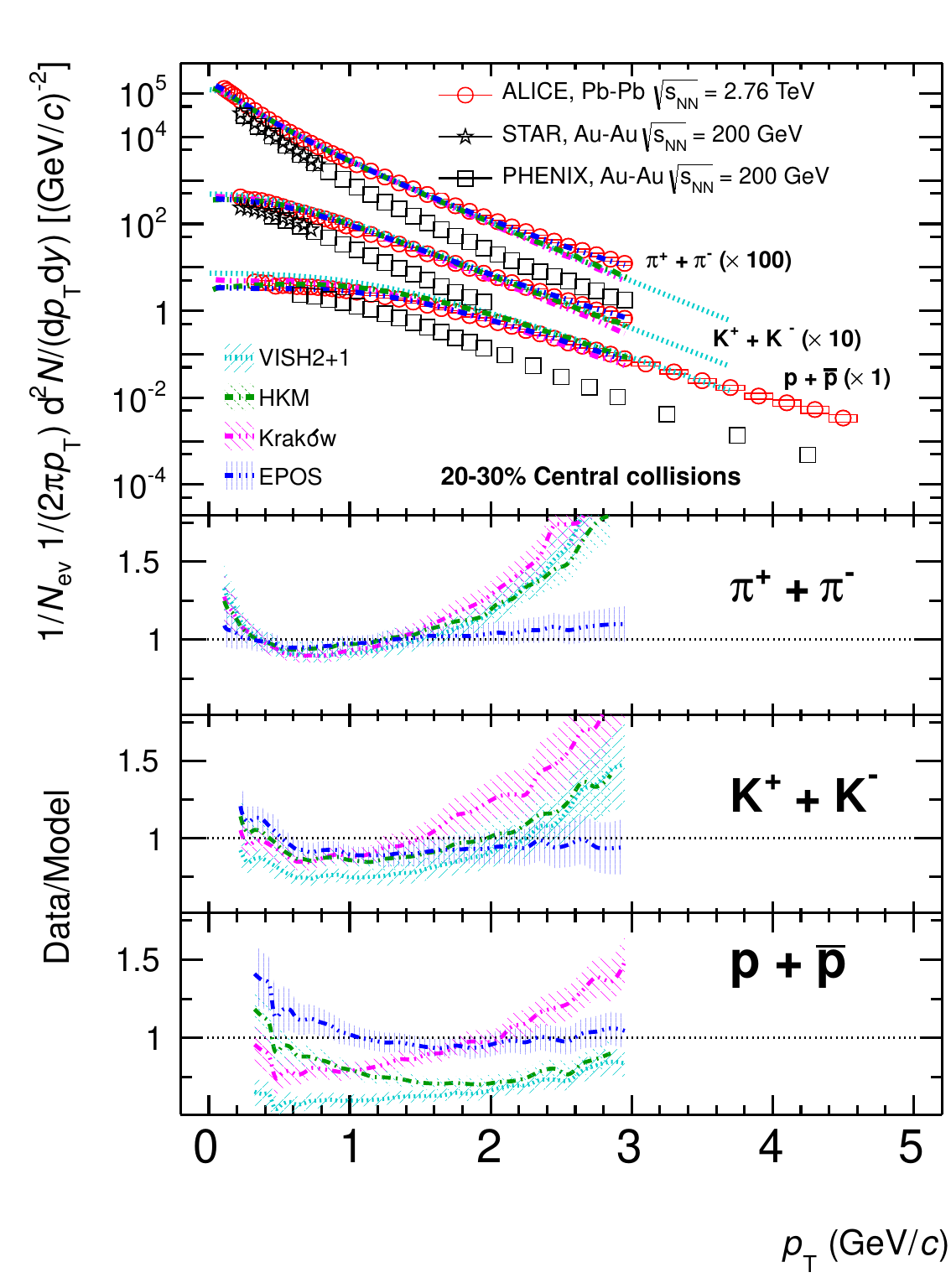}
  \caption{(color online) Spectra of particles for summed charge states in the centrality bin 20-30\% compared to hydrodynamical models and results from RHIC at \snn~=~200~GeV~\cite{Abelev:2008ez,Adler:2003cb}. See text for details. Systematic uncertainties plotted (boxes); statistical uncertainties are smaller than the symbol size.}
  \label{fig:spectra-vs-hydro-1}
\end{figure}

\begin{figure}[tp]
  \centering
  \includegraphics[width=\mywidth]{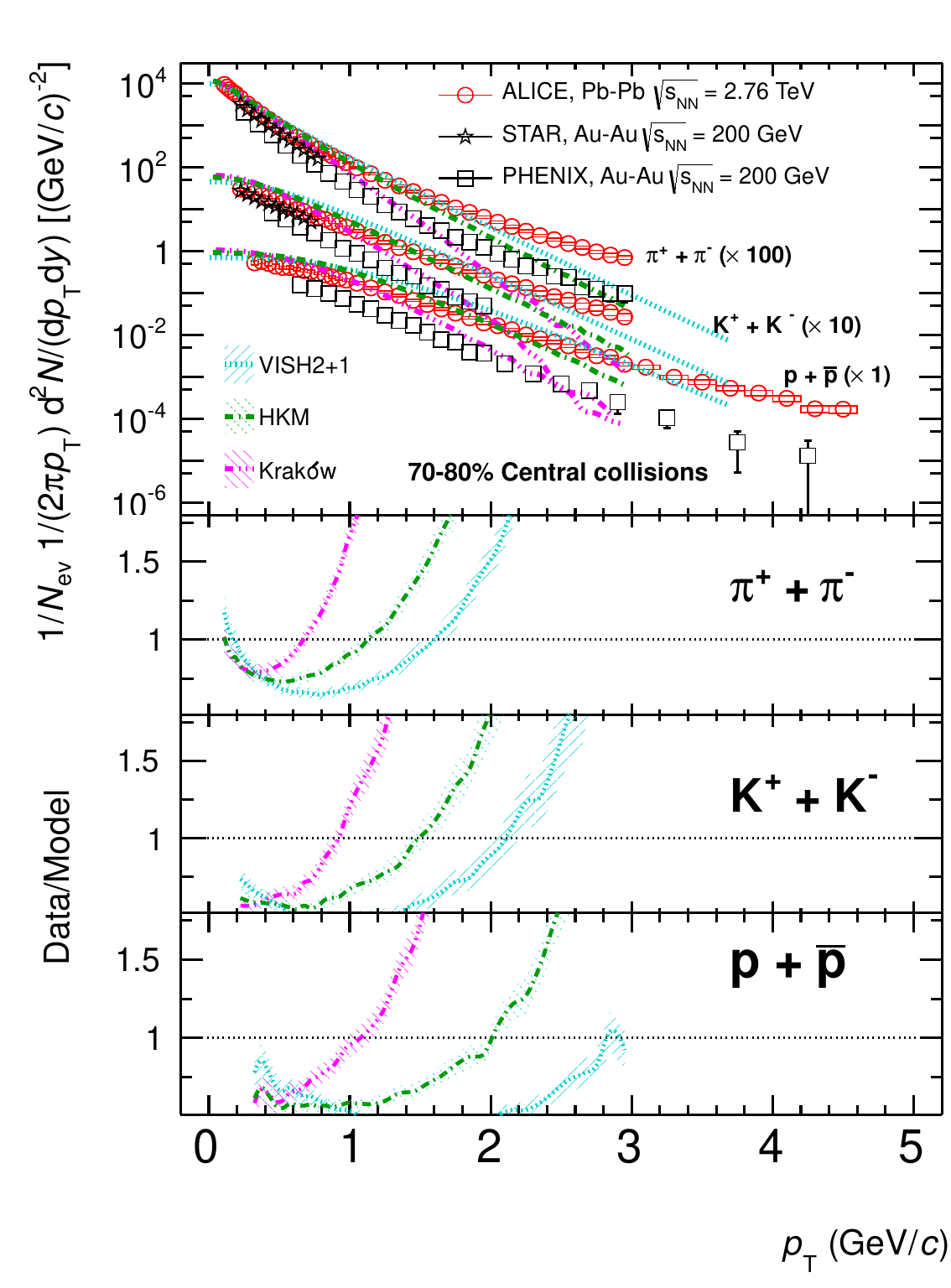}
  \caption{(color online) Spectra of particles for summed charge states in the centrality bin 70-80\% compared to hydrodynamical models and results from RHIC at \snn~=~200~GeV~\cite{Abelev:2008ez,Adler:2003cb}. See text for details. Systematic uncertainties plotted (boxes); statistical uncertainties are smaller than the symbol size.}
  \label{fig:spectra-vs-hydro-2}
\end{figure}

In order to quantify the freeze-out parameters at \snn~=~2.76~TeV, a  combined fit of the spectra with the blast-wave function Eq.~(\ref{eq:blast-wave}) was performed, in the ranges 0.5--1 GeV/$c$, 0.2--1.5 GeV/$c$, 0.3--3 GeV/$c$ for $\pi$, K, p, respectively. The pions at low \pt\ are known to have a large contribution from resonance decays, while at high \pt\ a hard contribution (not expected to be described by the blast-wave) may set in. Therefore, the values of the parameters extracted from the fit, and especially \Tfo, are sensitive to the fit range used for the pions. Forcing all species to decouple with the same parameters also makes the interpretation of the results arguable: different particles can in principle decouple at a different time, and hence with a different \avbT\ and \Tfo, from the hadronic medium, due to their different hadronic cross section.
These fits by no means replace a full hydrodynamical calculation: their usefulness lies in the ability to compare with a few simple parameters the measurements at different \snn. As will be discussed, the parameters extracted from such a combined fit depend on the range used for the different particles. Our standard fit ranges were therefore chosen to be similar to the ones used by the STAR collaboration at \snn~=~200~GeV/$c$ at the low \pt\ end. The high \pt\ boundaries were extended to higher \pt\ as compared to STAR, since at the LHC it is expected that the shapes are dominated by collective effects out to higher transverse momenta.
The results of the fit are summarized in Table~\ref{tab:fit-results} and shown in Fig.~\ref{fig:blast-wave-ellipses} (a) and in Fig.~\ref{fig:blast-wave-vs-dndeta}. The 1-sigma uncertainty ellipses shown in the figure reflect the bin-to-bin systematic uncertainties. The uncertainties shown as dashed bars and reported in Table~\ref{tab:fit-results} also include systematic uncertainties related to the stability of the fit: the effect of the variation of the lower fit bound for pions (to test the effect of resonance feed-down), the sensitivity to different particle species (i.e. excluding pions or kaons or protons) and to fits to the individual analyses. The value of \avbT\ extracted from the fit increases with centrality, while \Tfo\ decreases, similar to what was observed at lower energies (Figs.~\ref{fig:blast-wave-ellipses} and \ref{fig:blast-wave-vs-dndeta}, Tab.~\ref{tab:fit-results}). This was interpreted as a possible indication of a more rapid expansion with increasing centrality~\cite{Adams:2005dq}. In peripheral collisions this is consistent with the expectation of a shorter lived fireball with stronger radial gradients~\cite{Heinz:2004qz}.

The value of the $n$ parameter, Eq.~(\ref{eq:rhoBWdefintion}), is about 0.7 in central collisions and it increases towards peripheral collisions. The large values in peripheral collisions are likely due to the spectrum not being thermal over the full range: the $n$ parameter increases to reproduce the power law tail.

In order to further test the stability of the fit, it was repeated in the ranges 0.7--1.3 GeV/$c$, 0.5--1.5 GeV/$c$, 1--3 GeV/$c$ (``high \pt'') and 0.5--0.8 GeV/$c$, 0.2--1 GeV/$c$, 0.3--1.5 GeV/$c$ (``low \pt'') for $\pi$, K, p respectively.
The effect of the fit ranges is demonstrated in Fig.~\ref{fig:blast-wave-ellipses} (b). As can be seen, while the value of \avbT\ is relatively stable, especially for the most central bins, the value of \Tfo\ is strongly affected by the fit range, with differences of order 15 MeV for the most central events. For most peripheral events, \avbT\ shows some instability. However, it should be noticed that this parameter is mostly fixed by the low-\pt\ protons and its uncertainty increases significantly when the fit range for the protons is reduced. 

The combined fits in the default range are shown in Fig.~\ref{fig:spectra-results}, compared to fits to individual particle spectra. The dotted curves represent the combined blast-wave fits, while the dashed curves are the individual fits to each species. The individual fits reproduce the spectra over the full \pt\ range in all centrality bins, thus allowing for a reliable estimate of the \avpT\ and integrated yields, as mentioned above. The extrapolation at low and high \pt\ of the combined fits gets progressively worse for decreasing centrality. The ratios of the spectra to the combined fits are shown in Fig.~\ref{fig:RatioCombinedBW}. If the behavior of the spectra is purely hydrodynamical over the full \pt\ range considered, one would expect that the fireball blast-wave parameters determined by a fit in a limited \pt\ range are able to predict the full shape. This is what is observed in the most central bin for protons and kaons. The same is not true for the more peripheral bins, and the \pt\ threshold at which the function deviates from the data decreases with centrality, indicating the limit of applicability of the hydrodynamical picture (as also discussed below). 

The spectra for summed charge states in central (0-5\%), semi-central (20-30\%) and peripheral (70-80\%) collisions are compared to hydrodynamical models and previous results in \AuAu\ collisions at \snn~=~200~GeV in Figs.~\ref{fig:spectra-vs-hydro}, \ref{fig:spectra-vs-hydro-1}, and \ref{fig:spectra-vs-hydro-2}.

A dramatic change in spectral shapes from RHIC to LHC energies is
observed, with the protons in particular showing a flatter
distribution.  A comparison between the two energies based on the
values of \avbT\ and \Tfo\ from the combined blast-wave
fits~\cite{Adams:2005dq} is shown in
Fig.~\ref{fig:blast-wave-ellipses} and
\ref{fig:blast-wave-vs-dndeta}. For central collisions, about 10\%
stronger radial flow than at RHIC is observed at the LHC\footnote{The
  full systematic uncertainties on the parameters, quoted by both STAR
  and ALICE, include a number of checks on the stability of the fit
  which have a similar effect at different energies. When comparing
  results of different experiments, only the statistical and systematic
  uncertainties of the \pt\ distributions should be considered, but not those related to the
  stability of the fit.}.

The models shown in Figs.~\ref{fig:spectra-vs-hydro}, \ref{fig:spectra-vs-hydro-1}, and \ref{fig:spectra-vs-hydro-2} give for central collisions a fair description of the data. In the region $\pt \lesssim 3$~GeV/$c$ (Krak\'ow~\cite{Bozek:2012qs}), $\pt \lesssim 1.5$~GeV/$c$ (HKM~\cite{Karpenko:2012yf}) and  $\pt \lesssim 3$~GeV/$c$ (EPOS~\cite{Werner:2012xh}, with the exception of protons which are underestimated by about 30\% at low \pt),  the models describe the experimental data within $\sim$20\%, supporting a hydrodynamic interpretation of the \pt\ spectra in central collisions at the LHC.  
VISH2+1~\cite{Shen:2011eg} is a viscous hydrodynamic model that reproduces fairly well the pion and kaon distributions up to $\pt\sim 2$~GeV/$c$, but it misses the protons, both in shape and absolute abundance in all centrality bins. In this version of the model the  yields are thermal, with a chemical  temperature $T_{ch} = 165$~MeV,  extrapolated from lower energies. The difference between VISH2+1 and the data are possibly due to the lack of an explicit description of the hadronic phase in the model. This idea is supported by the comparison with HKM~\cite{Karpenko:2011qn,Karpenko:2012yf}. HKM is an ideal hydrodynamics model, in which after the hydrodynamic phase particles are injected into a hadronic cascade model (UrQMD), which further transports them until final decoupling. The hadronic phase builds up additional radial flow and affects particle ratios due to the hadronic interactions.  As can be seen, this model yields a better description of the data. The protons at low \pt, and hence their total number, are rather well reproduced, even if the slope is significantly smaller than in the data. Antibaryon-baryon annihilation is an important ingredient for the description of particle yields in this model~\cite{Karpenko:2011qn,Karpenko:2012yf}.  The Krak\'ow~\cite{Bozek:2011ua,Bozek:2011gq} model, on the other hand, uses an ansatz to describe deviation from equilibrium due to bulk viscosity corrections at freeze-out, which seems successful in reproducing the data. A general feature of these models is that, going to more peripheral events, the theoretical curves deviate from the data at high \pt\ (Figs.~\ref{fig:spectra-vs-hydro-1} and \ref{fig:spectra-vs-hydro-2}). This is similar to what is observed in the comparison to the blast-wave fits, and shows the limits of the hydrodynamical models. As speculated in~\cite{Bozek:2012qs}, this could indicate the onset of a non-thermal (hard) component, which in more peripheral collisions is not dominated by the flow-boosted thermal component. This picture is further substantiated by the change in the local slopes as seen in Fig.~\ref{fig:spectra-local-slope}.

The EPOS (2.17v3) model~\cite{Werner:2012xh} aims at describing all \pt\ domains with the same dynamical picture. In this model, the initial hard scattering creates ``flux tubes'' which either escape the medium and hadronize as jets, or contribute to the bulk matter, described in terms of hydrodynamics. After hadronization, particles are transported with a hadronic cascade model (UrQMD). EPOS shows a good agreement with the data for central and semi-central collisions. A calculation done with the same model, but disabling the late hadronic phase, yields a significantly worse description~\cite{Werner:2012xh}, indicating the important role of the late hadronic interactions in this model. An EPOS calculation for peripheral collisions was not available at the time of writing, but it will be important to see how well the peripheral data can be described in this model, since it should include all relevant physics processes.
Several other models implementing similar ideas (hydrodynamics model coupled to a hadronic cascade code, possibly with a description of fluctuations in the initial condition) are available in the literature~\cite{Ryu:2012at,Petersen:2011sb} but not discussed in this paper. 
The simultaneous description of additional variables, such as the $v_n$ azimuthal flow coefficients within the same model, will help in differentiating different hydrodynamical model scenarios.

\begin{table*}[tp]   \centering

\ifpreprint
\footnotesize
\fi
  \caption{Results of the combined blast-wave fits, in the default ranges 0.5--1 GeV/$c$, 0.2--1.5 GeV/$c$, 0.3--3 GeV/$c$ for $\pi$, K, p respectively. The first uncertainty in the table includes the effect of the bin-by-bin systematic uncertainties, the second is the full systematic uncertainty. See text for details.}

  \begin{tabular*}{\linewidth}{@{} c@{\extracolsep{\fill}}c@{\extracolsep{\fill}}c@{\extracolsep{\fill}}c@{\extracolsep{\fill}}c @{}}

\hline
Centrality  & \avbT & $\Tfo\ (\mathrm{GeV}/c)$ & n & $\chi^{2}/n_{dof}$ \\
\hline
\hline
0--5\%    &  $0.651 \pm 0.004  \pm 0.020 $ &  $0.095 \pm 0.004  \pm 0.010 $ &  $0.712 \pm 0.019  \pm 0.086 $ & 0.15\\
5--10\%   &  $0.646 \pm 0.004  \pm 0.023 $ &  $0.097 \pm 0.003  \pm 0.011 $ &  $0.723 \pm 0.019  \pm 0.116 $ & 0.20\\
10--20\%  &  $0.639 \pm 0.004  \pm 0.022 $ &  $0.099 \pm 0.004  \pm 0.011 $ &  $0.738 \pm 0.020  \pm 0.118 $ & 0.19\\
20--30\%  &  $0.625 \pm 0.004  \pm 0.025 $ &  $0.101 \pm 0.004  \pm 0.012 $ &  $0.779 \pm 0.022  \pm 0.133 $ & 0.22\\
30--40\%  &  $0.604 \pm 0.005  \pm 0.022 $ &  $0.106 \pm 0.004  \pm 0.012 $ &  $0.841 \pm 0.025  \pm 0.168 $ & 0.22\\
40--50\%  &  $0.574 \pm 0.005  \pm 0.016 $ &  $0.112 \pm 0.004  \pm 0.013 $ &  $0.944 \pm 0.029  \pm 0.142 $ & 0.22\\
50--60\%  &  $0.535 \pm 0.007  \pm 0.018 $ &  $0.118 \pm 0.004  \pm 0.014 $ &  $1.099 \pm 0.038  \pm 0.187 $ & 0.28\\
60--70\%  &  $0.489 \pm 0.008  \pm 0.024 $ &  $0.129 \pm 0.005  \pm 0.017 $ &  $1.292 \pm 0.052  \pm 0.194 $ & 0.36\\
70--80\%  &  $0.438 \pm 0.011  \pm 0.039 $ &  $0.139 \pm 0.005  \pm 0.027 $ &  $1.578 \pm 0.081  \pm 0.205 $ & 0.40\\
80--90\%  &  $0.357 \pm 0.016  \pm 0.084 $ &  $0.151 \pm 0.006  \pm 0.044 $ &  $2.262 \pm 0.191  \pm 0.498 $ & 0.52\\
\hline

  \end{tabular*}
  \label{tab:fit-results}
\end{table*}

\begin{figure}[tp]
  \centering
  \includegraphics[width=\mywidth]{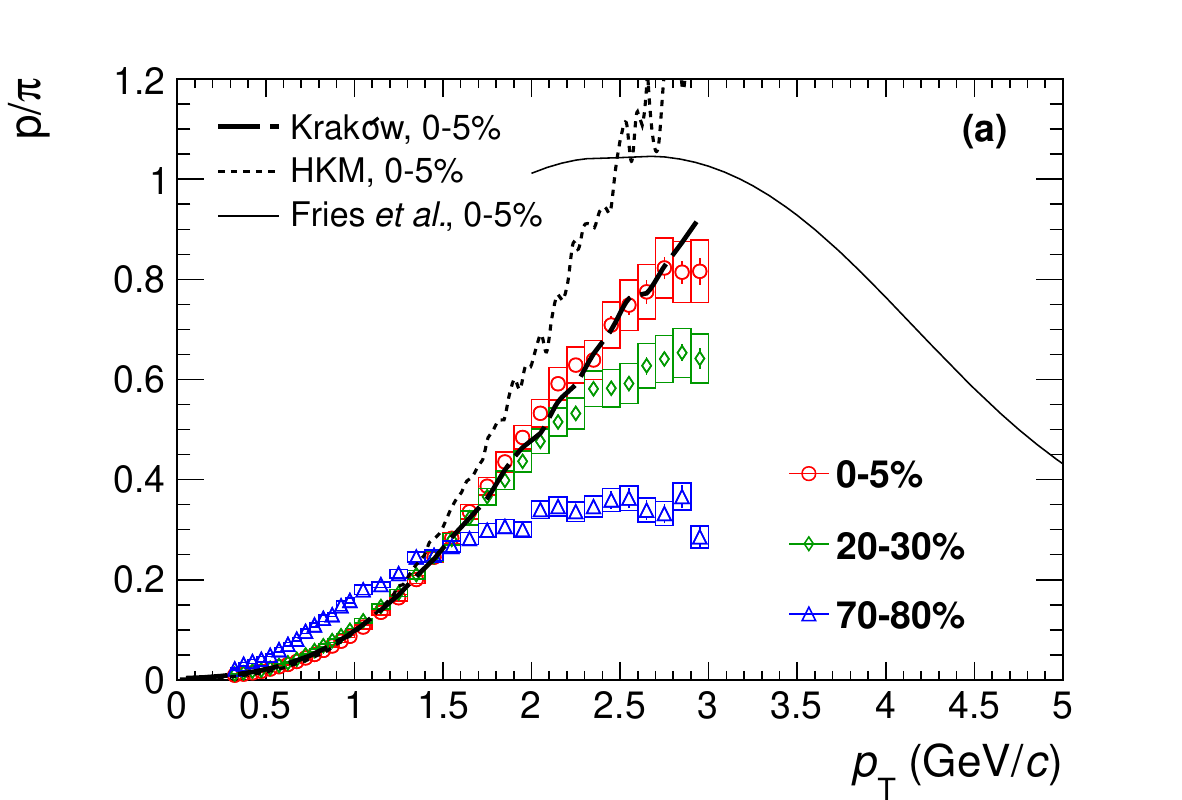}
  \hfill
  \includegraphics[width=\mywidth]{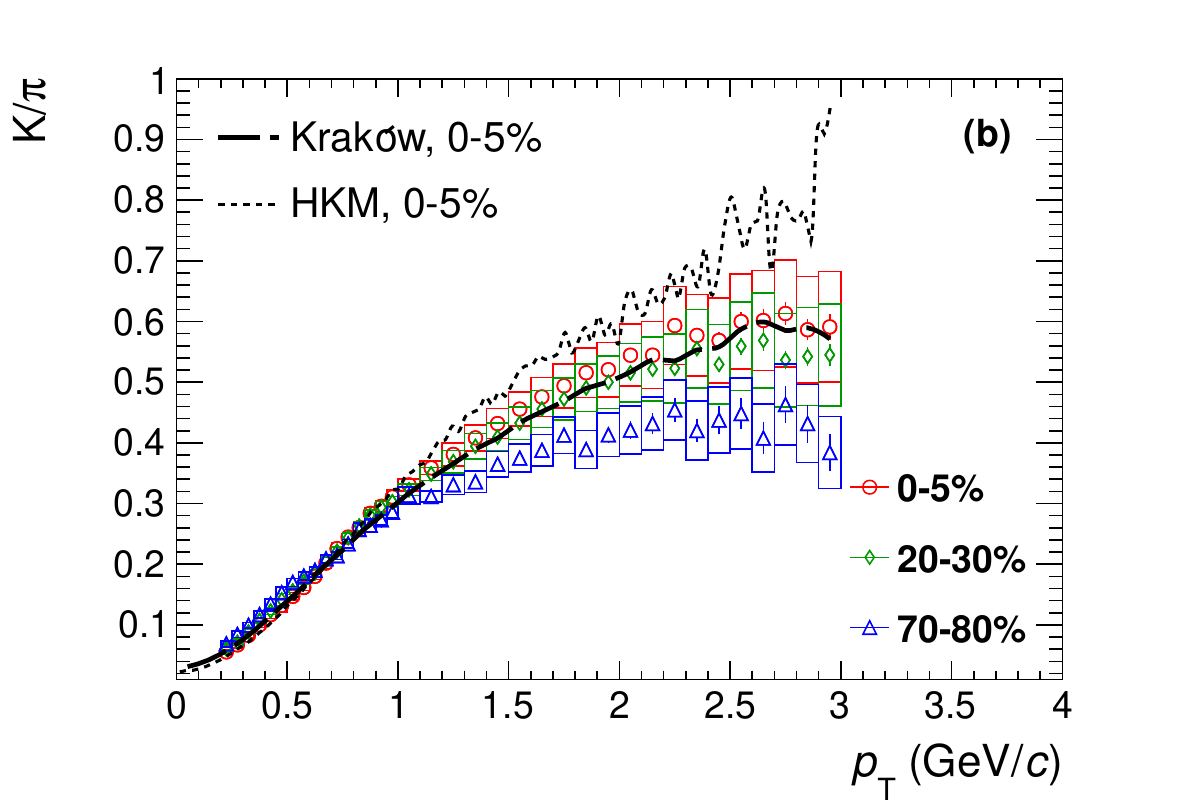}
  \caption{(color online) (a) $\ppi = (\mathrm{p}
    +\mathrm{\bar{p}})/(\pi^++\pi^-)$ as a function of \pt\ for
    different centrality bins compared to ratios from the
    Krak\'ow~\cite{Bozek:2012qs} and HKM~\cite{Karpenko:2012yf}
    hydrodynamic models and to a recombination
    model~\cite{Fries:2003vb,Fries:2003kq,FriesPrivateComm}; (b) $\kpi
    = (\mathrm{K^+} + \mathrm{K^-}))/(\pi^++\pi^-)$ ratio as a
    function of \pt\ for different centrality bins compared to ratios
    from the Krak\'ow~\cite{Bozek:2012qs} and
    HKM~\cite{Karpenko:2012yf} hydrodynamic models.}
  \label{fig:ratios-vs-pt}
\end{figure}

Figure~\ref{fig:ratios-vs-pt} shows the  $\ppi=(\mathrm{p} + \mathrm{\bar p})/(\pi^+ + \pi^-)$ and $\kpi=(\mathrm{K}^+ + \mathrm{K}^-)/(\pi^+ + \pi^-)$ ratios as a function of \pt. Both ratios are seen to increase as a function of centrality at intermediate \pt\, with a corresponding depletion at low \pt\ (the \pt\ integrated ratios show little dependence on centrality, Fig.~\ref{fig:ratios_vs_centr}). The \ppi\ ratio, in particular, shows a more pronounced increase, reaching a value of about 0.9 at \pt~=~3~\gevc\ . This is reminiscent of the increase in the baryon-to-meson ratio observed at RHIC in the intermediate \pt\ region~\cite{Lamont:2007ce,Abelev:2006jr}, which is suggestive of the recombination picture discussed in Sec.~\ref{sec:introduction}. It should be noted, however, that a rise of the ratio with \pt\ is an intrinsic feature of hydrodynamical models, where it is just due to the mass ordering induced by radial flow (heavier particles are pushed to higher \pt\ by the collective motion). In Fig.~\ref{fig:ratios-vs-pt} a prediction from two of the hydrodynamical models discussed above~\cite{Bozek:2011gq,Karpenko:2012yf} and a prediction from a recombination model~\cite{Fries:2003vb,Fries:2003kq,FriesPrivateComm} are shown.
 As can be seen, the ratio for central events is reasonably reproduced by the Krak\'ow hydrodynamical model, while HKM only reproduces the data up to $\pt\sim 1.5$~GeV/$c$ and the recombination prediction is higher than the data and predicts a flatter trend in the range 2-3 GeV/$c$.

This measurement is currently being extended  to higher \pt\  by ALICE using the HMPID detector (High Momentum Particle Identification~\cite{HMPIDTDR}, a ring imaging \u{C}erenkov), complemented by a statistical identification in the relativistic rise region of the TPC. A complementary study of the $\Lambda/\mathrm{K}_S^0$ ratio will also provide a good \pt\ coverage and additional constraints. 

\subsection{\pt-Integrated Yields}
\label{sec:pt-yields-discussion}

The integrated ratios can be interpreted in terms of the thermal models. Figure~\ref{fig:ratios_vs_centr} depicts the expectations from these models, which, based on lower energy data, used the values $\mathrm{T}\simeq 160-170~\mathrm{MeV}$ and $\mu_{B}\simeq 1~\mathrm{MeV}$ at the LHC~\cite{Cleymans:2006xj,Andronic:2005yp}. The \kpi\ ratio is consistent with these expectations, while the \ppi\ ratio is found to be lower by a factor of about 1.5. As discussed in \cite{prl-spectra}, this finding was one of the surprises of the first \PbPb\ run at the LHC and still needs to be understood. 
 
Some indication of a similar disagreement between data and the thermal model is also seen in the RHIC data,
with the proton measurements being 10-20\% lower than the thermal model predictions~\cite{Andronic:2011yq,Adams:2005dq}. This discrepancy was not considered to be significant, due to experimental uncertainties in the subtraction of secondary particles, differences between thermal model implementations and model uncertainties~\cite{Becattini:2010sk}. 

\subsection{Total proton spectrum}
\label{sec:total}

\begin{figure}[tb!]
  \centering
  \includegraphics[width=\mywidth]{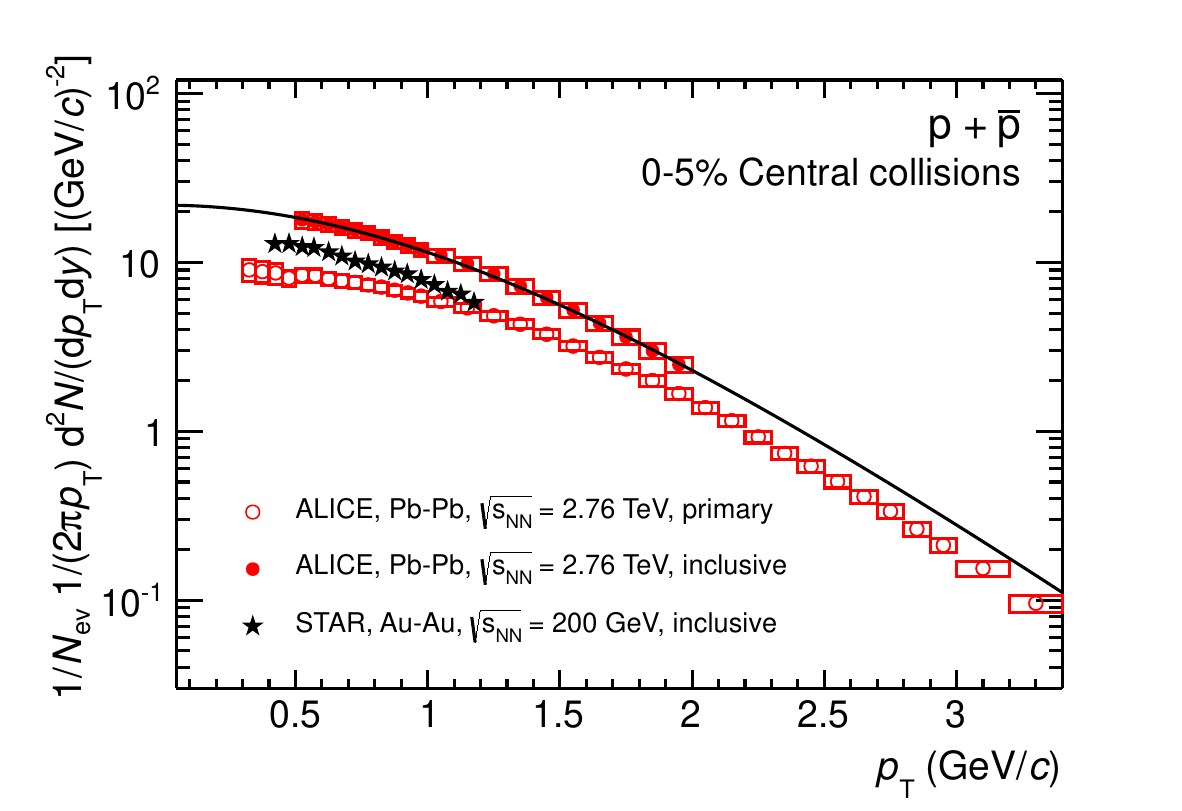}
  \caption{(color online) Transverse momentum \pt\ distribution of total protons based on tracks reconstructed using only TPC information (full circles) compared to primary protons (open circles) and corresponding total spectra measured by the STAR collaboration in Au-Au collisions at \snn~=~200~GeV/$c$ (full stars). 0-5\% most central events. Box: systematic uncertainty; Statistical uncertainties smaller than the symbols; curve: individual blast wave fit.}
  \label{fig:p-fully-inclusive}
\end{figure}

In order to compare directly with lower energy results, we performed a measurement of protons including feed-down from weak decays (``total proton spectrum''), based on tracks reconstructed using only TPC information, with no requirements on the ITS. These tracks have a similar efficiency for both primary and secondary particles, with the difference in the particle composition between data and Monte Carlo being of minor importance. The reconstructed sample thus already includes most of the secondary particles. The efficiency correction for the total proton measurement is about 25\% at 450 MeV/$c$ and 5\% at 1 GeV/$c$. In this analysis, the secondaries from detector material were subtracted using a \dcaxy\ fit procedure, similar to the one discussed in Sec.~\ref{sec:datasampleAndAnalysis}.
The total proton spectrum for central collisions (0-5\%) is compared in Fig.~\ref{fig:p-fully-inclusive} with the primary proton spectrum and with the previous STAR measurements of total protons at \snn~=~200~GeV. The comparison of the spectra confirms the change in shape and yield from RHIC to LHC, already discussed for primary protons. The total spectrum was fitted with a blast-wave function (also shown in Fig.~\ref{fig:p-fully-inclusive}) to compute the extrapolation for the extraction of the integrated yield.  Due to the limited $p_{T}$-coverage of the TPC-only analysis, the fraction of the extrapolated yield amounts to about 25\% resulting in a larger extrapolation uncertainty (about 12\%) as compared to the primary spectrum.  We found $\mathrm{d}N^{\rm total}_{\mathrm{p}+\mathrm{\bar{p}}}/\mathrm{d}y = 120 \pm 20~\mathrm{(syst+stat)}$ for the combined spectra of protons and antiprotons, resulting in the ratio p~(total)/$\pi$ = $0.082 \pm 0.010$. This is about 15\% lower than the ratio of $0.095 \pm 0.011$~\cite{Abelev:2008ez} measured by STAR at \snn~=~200~GeV (consistent with what was previously noted for primary protons), but the difference is not significant within uncertainties.

Subtraction of the total and primary proton yields  $\mathrm{d}N^{\rm sec}_{\rm p+\bar{p}}/\mathrm{d}y = 53 \pm 19~\mathrm{(syst+stat)}$ for secondary protons.  This number, normalized to the primary pion yield, can be compared to the thermal model prediction~\cite{Andronic:2008gu} with \Tch~=~164~MeV.  There are five main contributions to the secondary protons, summarized in Table~\ref{tab:thermal_model_secondaries}. 
Rescaling the ratios shown in Table~\ref{tab:thermal_model_secondaries} by the measured pion yield, we get $\mathrm{d}N^{\rm sec,model}_{\rm p+\bar{p}}/\mathrm{d}y = 62$, in good agreement with the measurement.  This result suggests that the disagreement between the data and the thermal model is most prominent for primary protons, while strange baryons contributing to the production of secondary protons are likely better described.  If the total protons measured in the data and in the thermal model are compared, the disagreement gets partially suppressed, because the number of secondaries is well reproduced in the thermal model, and the fraction of secondary protons is rather large ($\sim 50\%$). 

\begin{table}[tp]
\ifpreprint
\footnotesize
\fi
  \centering
  \caption{Contribution to secondary protons estimated in the thermal model~\cite{Andronic:2008gu} with \Tch~=~164~MeV.}
  \begin{tabular}{cccc}
    \hline
    Particle & Decay Channel & Branching Ratio & p (secondary)/$\pi$ \\
    \hline
    \hline
    $\Lambda    $&$ {\mathrm p} \pi $     & 63.9 \%  & $4.42\times10^{-2}$     \\
    $\Sigma^{+}$ &$ {\mathrm p} \pi^{0} $ &  51.6 \%  &  $1.27\times 10^{-2}$    \\
    $\Xi^-      $&$ \Lambda \pi $         & 99.89 \% & $5.49\times10^{-3}$    \\
    $\Xi^0      $&$ \Lambda \pi $         & 99.53 \% & $5.58\times10^{-3}$    \\
    $\Omega     $&$ \Lambda K $           & 67.8 \%  & $9.83\times10^{-4}$   \\                   
    \hline
  \end{tabular}
  \label{tab:thermal_model_secondaries}
\end{table}

A possible explanation for the difference between the \ppi\ ratio and the predictions from the thermal models is  antibaryon-baryon annihilation in the hadronic phase~\cite{Becattini:2012xb,Steinheimer:2012rd,Karpenko:2012yf}. The \ppi\ ratio depicted in Fig.~\ref{fig:ratios_vs_centr} suggests a decreasing trend with centrality, consistent with the antibaryon-baryon annhilation hypothesis: the  effect is expected to be less important for the more dilute system created in peripheral collisions. 
It should  be noted that all the available models incorporating a hadronic phase use the UrQMD~\cite{Bleicher:1999xi,Bass:1998ca} hadronic cascade model, and the effects induced by the hadronic phase are  model dependent. While this microscopic model includes annihilation processes, it does not implement the reverse process like $n\pi\to\mathrm{p\bar{p}}$. The effect of the reverse reactions was investigated in a recent calculation~\cite{Pan:2012ne}. It was found to be non-negligible, while the net suppression of baryons is still very significant.

The origin of the low proton yield with respect to the thermal model expectations is not yet established and alternative explanations exist in the literature.
Implementations of the thermal model incorporating non-equilibrium effects predict a reduction of the \ppi\ ratio, although the details depend on the exact value of model parameters. With the preferred set of parameters in~\cite{Rafelski:2010cw}, the authors could predict the correct value for p/pi but not for K/pi. This is explained in~\cite{Petran:2013lja} as a lower strangeness-over-entropy ratio as compared to the expectations.
An alternative explanation may come from the existence of flavor and mass dependent pre-hadronic bound states in the QGP phase, as suggested by recent lattice QCD calculation and QCD-inspired models~\cite{Ratti:2011au,Bellwied:2012kh}.
Additional constraints to discriminate between these different scenarios will be provided by a thermal analysis, including in particular strange and multistrange baryons.


\section{Conclusions}
\label{sec:conclusions}
In this paper we presented a comprehensive measurement of $\pi$, K, and p production in Pb-Pb collisions at \snn~=~2.76~TeV at the LHC. Antiparticle over particle ratios are compatible with unity at all \pt\ and at all centralities, as expected for LHC.  A clear evolution of all spectra with centrality is seen, with an almost exponential behavior at high \pt, and a flattening of the spectra at low \pt. These features are compatible with the development of a strong collective flow with centrality, which dominates the spectral shapes up to relatively high \pt\ in central collisions. The \avbT\ parameter extracted from fits to the blast-wave parametrization indicates a radial flow about 10\% higher than at RHIC at \snn~=~200~GeV in central collisions.
The integrated abundances of particles are almost independent of centrality. They are compared with expectations from thermal models.   While the \kpi\ ratio was found to agree with these expectations, the \ppi\ is a factor 1.5 lower.
The central collision data are successfully described by hydrodynamic models, but the low proton yield requires a refined description of the late fireball stages. These models~\cite{Werner:2012xh,Karpenko:2012yf} indicate a non-negligible baryon annihilation in the hadronic phase, which alters the thermal yields and leads to a lower proton yield. The origin of this effect, however, is not yet established and alternative explanations, such as non-equilibrium effects or flavor-dependent freeze-out, exist in the literature.
In more peripheral collisions, purely hydrodynamic models give a poor description of the data, indicating the limit of applicability of hydrodynamics.




\ifpreprint
\iffull
\newenvironment{acknowledgement}{\relax}{\relax}
\begin{acknowledgement}
\section*{Acknowledgements}
We are grateful to P. Bozek, R. Fries, U. Heinz, Y. Karpenko, H. Petersen, C. Shen, Y. Sinyukov, H. Song and  K. Werner for providing the theoretical calculations and for the useful discussion and to colleagues from the BRAHMS, PHENIX and STAR collaborations for the helpful discussions and clarifications on their measurements.

The ALICE collaboration would like to thank all its engineers and technicians for their invaluable contributions to the construction of the experiment and the CERN accelerator teams for the outstanding performance of the LHC complex.
\\
The ALICE collaboration acknowledges the following funding agencies for their support in building and
running the ALICE detector:
 \\
State Committee of Science,  World Federation of Scientists (WFS)
and Swiss Fonds Kidagan, Armenia,
 \\
Conselho Nacional de Desenvolvimento Cient\'{\i}fico e Tecnol\'{o}gico (CNPq), Financiadora de Estudos e Projetos (FINEP),
Funda\c{c}\~{a}o de Amparo \`{a} Pesquisa do Estado de S\~{a}o Paulo (FAPESP);
 \\
National Natural Science Foundation of China (NSFC), the Chinese Ministry of Education (CMOE)
and the Ministry of Science and Technology of China (MSTC);
 \\
Ministry of Education and Youth of the Czech Republic;
 \\
Danish Natural Science Research Council, the Carlsberg Foundation and the Danish National Research Foundation;
 \\
The European Research Council under the European Community's Seventh Framework Programme;
 \\
Helsinki Institute of Physics and the Academy of Finland;
 \\
French CNRS-IN2P3, the `Region Pays de Loire', `Region Alsace', `Region Auvergne' and CEA, France;
 \\
German BMBF and the Helmholtz Association;
\\
General Secretariat for Research and Technology, Ministry of
Development, Greece;
\\
Hungarian OTKA and National Office for Research and Technology (NKTH);
 \\
Department of Atomic Energy and Department of Science and Technology of the Government of India;
 \\
Istituto Nazionale di Fisica Nucleare (INFN) and Centro Fermi -
Museo Storico della Fisica e Centro Studi e Ricerche "Enrico
Fermi", Italy;
 \\
MEXT Grant-in-Aid for Specially Promoted Research, Ja\-pan;
 \\
Joint Institute for Nuclear Research, Dubna;
 \\
National Research Foundation of Korea (NRF);
 \\
CONACYT, DGAPA, M\'{e}xico, ALFA-EC and the HELEN Program (High-Energy physics Latin-American--European Network);
 \\
Stichting voor Fundamenteel Onderzoek der Materie (FOM) and the Nederlandse Organisatie voor Wetenschappelijk Onderzoek (NWO), Netherlands;
 \\
Research Council of Norway (NFR);
 \\
Polish Ministry of Science and Higher Education;
 \\
National Authority for Scientific Research - NASR (Autoritatea Na\c{t}ional\u{a} pentru Cercetare \c{S}tiin\c{t}ific\u{a} - ANCS);
 \\
Ministry of Education and Science of Russian Federation,
International Science and Technology Center, Russian Academy of
Sciences, Russian Federal Agency of Atomic Energy, Russian Federal
Agency for Science and Innovations and CERN-INTAS;
 \\
Ministry of Education of Slovakia;
 \\
Department of Science and Technology, South Africa;
 \\
CIEMAT, EELA, Ministerio de Educaci\'{o}n y Ciencia of Spain, Xunta de Galicia (Conseller\'{\i}a de Educaci\'{o}n),
CEA\-DEN, Cubaenerg\'{\i}a, Cuba, and IAEA (International Atomic Energy Agency);
 \\
Swedish Research Council (VR) and Knut $\&$ Alice Wallenberg
Foundation (KAW);
 \\
Ukraine Ministry of Education and Science;
 \\
United Kingdom Science and Technology Facilities Council (STFC);
 \\
The United States Department of Energy, the United States National
Science Foundation, the State of Texas, and the State of Ohio.

\end{acknowledgement}
\ifbibtex
\bibliographystyle{\bibstname}
\bibliography{biblio}{}
\else

\fi
\newpage
\appendix
\section{The ALICE Collaboration}
\label{app:collab}
\else
\ifbibtex
\bibliographystyle{\bibstname}
\bibliography{biblio}{}
\else
 
\fi
\fi
\else
\iffull
\vspace{0.5cm}
\ifbibtex
\bibliographystyle{\bibstname}
\bibliography{biblio}{}
\else
%

\fi
\else
\ifbibtex
\bibliographystyle{\bibstname}
\bibliography{biblio}{}

\begin{thebibliography}{70}%
\makeatletter
\providecommand \@ifxundefined [1]{%
 \@ifx{#1\undefined}
}%
\providecommand \@ifnum [1]{%
 \ifnum #1\expandafter \@firstoftwo
 \else \expandafter \@secondoftwo
 \fi
}%
\providecommand \@ifx [1]{%
 \ifx #1\expandafter \@firstoftwo
 \else \expandafter \@secondoftwo
 \fi
}%
\providecommand \natexlab [1]{#1}%
\providecommand \enquote  [1]{``#1''}%
\providecommand \bibnamefont  [1]{#1}%
\providecommand \bibfnamefont [1]{#1}%
\providecommand \citenamefont [1]{#1}%
\providecommand \href@noop [0]{\@secondoftwo}%
\providecommand \href [0]{\begingroup \@sanitize@url \@href}%
\providecommand \@href[1]{\@@startlink{#1}\@@href}%
\providecommand \@@href[1]{\endgroup#1\@@endlink}%
\providecommand \@sanitize@url [0]{\catcode `\\12\catcode `\$12\catcode
  `\&12\catcode `\#12\catcode `\^12\catcode `\_12\catcode `\%12\relax}%
\providecommand \@@startlink[1]{}%
\providecommand \@@endlink[0]{}%
\providecommand \url  [0]{\begingroup\@sanitize@url \@url }%
\providecommand \@url [1]{\endgroup\@href {#1}{\urlprefix }}%
\providecommand \urlprefix  [0]{URL }%
\providecommand \Eprint [0]{\href }%
\providecommand \doibase [0]{http://dx.doi.org/}%
\providecommand \selectlanguage [0]{\@gobble}%
\providecommand \bibinfo  [0]{\@secondoftwo}%
\providecommand \bibfield  [0]{\@secondoftwo}%
\providecommand \translation [1]{[#1]}%
\providecommand \BibitemOpen [0]{}%
\providecommand \bibitemStop [0]{}%
\providecommand \bibitemNoStop [0]{.\EOS\space}%
\providecommand \EOS [0]{\spacefactor3000\relax}%
\providecommand \BibitemShut  [1]{\csname bibitem#1\endcsname}%
\let\auto@bib@innerbib\@empty
\bibitem [{\citenamefont {Heinz}\ and\ \citenamefont
  {Jacob}(2000)}]{Heinz:2000bk}%
  \BibitemOpen
  \bibfield  {author} {\bibinfo {author} {\bibfnamefont {U.~W.}\ \bibnamefont
  {Heinz}}\ and\ \bibinfo {author} {\bibfnamefont {M.}~\bibnamefont {Jacob}},\
  }\href@noop {} {\  (\bibinfo {year} {2000})},\ \Eprint
  {http://arxiv.org/abs/0002042} {arXiv:0002042 [nucl-th]} \BibitemShut
  {NoStop}%
\bibitem [{\citenamefont {Arsene}\ \emph {et~al.}(2005)\citenamefont {Arsene}
  \emph {et~al.}}]{Arsene:2004fa}%
  \BibitemOpen
  \bibfield  {author} {\bibinfo {author} {\bibfnamefont {I.}~\bibnamefont
  {Arsene}} \emph {et~al.} (\bibinfo {collaboration} {BRAHMS Collaboration}),\
  }\href {\doibase 10.1016/j.nuclphysa.2005.02.130} {\bibfield  {journal}
  {\bibinfo  {journal} {Nucl.Phys.}\ }\textbf {\bibinfo {volume} {A757}},\
  \bibinfo {pages} {1} (\bibinfo {year} {2005})}\BibitemShut {NoStop}%
\bibitem [{\citenamefont {Adcox}\ \emph {et~al.}(2005)\citenamefont {Adcox}
  \emph {et~al.}}]{Adcox:2004mh}%
  \BibitemOpen
  \bibfield  {author} {\bibinfo {author} {\bibfnamefont {K.}~\bibnamefont
  {Adcox}} \emph {et~al.} (\bibinfo {collaboration} {PHENIX Collaboration}),\
  }\href {\doibase 10.1016/j.nuclphysa.2005.03.086} {\bibfield  {journal}
  {\bibinfo  {journal} {Nucl.Phys.}\ }\textbf {\bibinfo {volume} {A757}},\
  \bibinfo {pages} {184} (\bibinfo {year} {2005})}\BibitemShut {NoStop}%
\bibitem [{\citenamefont {Back}\ \emph {et~al.}(2005)\citenamefont {Back},
  \citenamefont {Baker}, \citenamefont {Ballintijn}, \citenamefont {Barton},
  \citenamefont {Becker} \emph {et~al.}}]{Back:2004je}%
  \BibitemOpen
  \bibfield  {author} {\bibinfo {author} {\bibfnamefont {B.}~\bibnamefont
  {Back}}, \bibinfo {author} {\bibfnamefont {M.}~\bibnamefont {Baker}},
  \bibinfo {author} {\bibfnamefont {M.}~\bibnamefont {Ballintijn}}, \bibinfo
  {author} {\bibfnamefont {D.}~\bibnamefont {Barton}}, \bibinfo {author}
  {\bibfnamefont {B.}~\bibnamefont {Becker}},  \emph {et~al.},\ }\href
  {\doibase 10.1016/j.nuclphysa.2005.03.084} {\bibfield  {journal} {\bibinfo
  {journal} {Nucl.Phys.}\ }\textbf {\bibinfo {volume} {A757}},\ \bibinfo
  {pages} {28} (\bibinfo {year} {2005})}\BibitemShut {NoStop}%
\bibitem [{\citenamefont {Adams}\ \emph {et~al.}(2005)\citenamefont {Adams}
  \emph {et~al.}}]{Adams:2005dq}%
  \BibitemOpen
  \bibfield  {author} {\bibinfo {author} {\bibfnamefont {J.}~\bibnamefont
  {Adams}} \emph {et~al.} (\bibinfo {collaboration} {STAR Collaboration}),\
  }\href {\doibase 10.1016/j.nuclphysa.2005.03.085} {\bibfield  {journal}
  {\bibinfo  {journal} {Nucl.Phys.}\ }\textbf {\bibinfo {volume} {A757}},\
  \bibinfo {pages} {102} (\bibinfo {year} {2005})}\BibitemShut {NoStop}%
\bibitem [{\citenamefont {Shuryak}(2005)}]{Shuryak:2004cy}%
  \BibitemOpen
  \bibfield  {author} {\bibinfo {author} {\bibfnamefont {E.~V.}\ \bibnamefont
  {Shuryak}},\ }\href {\doibase 10.1016/j.nuclphysa.2004.10.022} {\bibfield
  {journal} {\bibinfo  {journal} {Nucl.Phys.}\ }\textbf {\bibinfo {volume}
  {A750}},\ \bibinfo {pages} {64} (\bibinfo {year} {2005})}\BibitemShut
  {NoStop}%
\bibitem [{\citenamefont {Huovinen}\ and\ \citenamefont
  {Ruuskanen}(2006)}]{Huovinen:2006jp}%
  \BibitemOpen
  \bibfield  {author} {\bibinfo {author} {\bibfnamefont {P.}~\bibnamefont
  {Huovinen}}\ and\ \bibinfo {author} {\bibfnamefont {P.}~\bibnamefont
  {Ruuskanen}},\ }\href {\doibase 10.1146/annurev.nucl.54.070103.181236}
  {\bibfield  {journal} {\bibinfo  {journal} {Ann.Rev.Nucl.Part.Sci.}\ }\textbf
  {\bibinfo {volume} {56}},\ \bibinfo {pages} {163} (\bibinfo {year}
  {2006})}\BibitemShut {NoStop}%
\bibitem [{\citenamefont {M{\"u}ller}\ and\ \citenamefont
  {Nagle}(2006)}]{Muller:2006ee}%
  \BibitemOpen
  \bibfield  {author} {\bibinfo {author} {\bibfnamefont {B.}~\bibnamefont
  {M{\"u}ller}}\ and\ \bibinfo {author} {\bibfnamefont {J.~L.}\ \bibnamefont
  {Nagle}},\ }\href {\doibase 10.1146/annurev.nucl.56.080805.140556} {\bibfield
   {journal} {\bibinfo  {journal} {Ann.Rev.Nucl.Part.Sci.}\ }\textbf {\bibinfo
  {volume} {56}},\ \bibinfo {pages} {93} (\bibinfo {year} {2006})}\BibitemShut
  {NoStop}%
\bibitem [{\citenamefont {Schnedermann}\ \emph {et~al.}(1993)\citenamefont
  {Schnedermann}, \citenamefont {Sollfrank},\ and\ \citenamefont
  {Heinz}}]{Schnedermann:1993ws}%
  \BibitemOpen
  \bibfield  {author} {\bibinfo {author} {\bibfnamefont {E.}~\bibnamefont
  {Schnedermann}}, \bibinfo {author} {\bibfnamefont {J.}~\bibnamefont
  {Sollfrank}}, \ and\ \bibinfo {author} {\bibfnamefont {U.~W.}\ \bibnamefont
  {Heinz}},\ }\href {\doibase 10.1103/PhysRevC.48.2462} {\bibfield  {journal}
  {\bibinfo  {journal} {Phys.Rev.}\ }\textbf {\bibinfo {volume} {C48}},\
  \bibinfo {pages} {2462} (\bibinfo {year} {1993})}\BibitemShut {NoStop}%
\bibitem [{\citenamefont {Heinz}(2004)}]{Heinz:2004qz}%
  \BibitemOpen
  \bibfield  {author} {\bibinfo {author} {\bibfnamefont {U.~W.}\ \bibnamefont
  {Heinz}},\ }\href@noop {} {\ ,\ \bibinfo {pages} {165} (\bibinfo {year}
  {2004})},\ \Eprint {http://arxiv.org/abs/0407360} {arXiv:0407360 [hep-ph]}
  \BibitemShut {NoStop}%
\bibitem [{\citenamefont {Andronic}\ \emph {et~al.}(2011)\citenamefont
  {Andronic}, \citenamefont {Braun-Munzinger}, \citenamefont {Redlich},\ and\
  \citenamefont {Stachel}}]{Andronic:2011yq}%
  \BibitemOpen
  \bibfield  {author} {\bibinfo {author} {\bibfnamefont {A.}~\bibnamefont
  {Andronic}}, \bibinfo {author} {\bibfnamefont {P.}~\bibnamefont
  {Braun-Munzinger}}, \bibinfo {author} {\bibfnamefont {K.}~\bibnamefont
  {Redlich}}, \ and\ \bibinfo {author} {\bibfnamefont {J.}~\bibnamefont
  {Stachel}},\ }\href {\doibase 10.1088/0954-3899/38/12/124081} {\bibfield
  {journal} {\bibinfo  {journal} {J.Phys.G}\ }\textbf {\bibinfo {volume}
  {G38}},\ \bibinfo {pages} {124081} (\bibinfo {year} {2011})}\BibitemShut
  {NoStop}%
\bibitem [{\citenamefont {Andronic}\ \emph {et~al.}(2009)\citenamefont
  {Andronic}, \citenamefont {Braun-Munzinger},\ and\ \citenamefont
  {Stachel}}]{Andronic:2008gu}%
  \BibitemOpen
  \bibfield  {author} {\bibinfo {author} {\bibfnamefont {A.}~\bibnamefont
  {Andronic}}, \bibinfo {author} {\bibfnamefont {P.}~\bibnamefont
  {Braun-Munzinger}}, \ and\ \bibinfo {author} {\bibfnamefont {J.}~\bibnamefont
  {Stachel}},\ }\href {\doibase 10.1016/j.physletb.2009.02.014,
  10.1016/j.physletb.2009.06.021} {\bibfield  {journal} {\bibinfo  {journal}
  {Phys.Lett.}\ }\textbf {\bibinfo {volume} {B673}},\ \bibinfo {pages} {142}
  (\bibinfo {year} {2009})}\BibitemShut {NoStop}%
\bibitem [{\citenamefont {Becattini}\ and\ \citenamefont
  {Fries}(2010)}]{Becattini:2010lb}%
  \BibitemOpen
  \bibfield  {author} {\bibinfo {author} {\bibfnamefont {F.}~\bibnamefont
  {Becattini}}\ and\ \bibinfo {author} {\bibfnamefont {R.}~\bibnamefont
  {Fries}},\ }\enquote {\bibinfo {title} {{Landolt-Boernstein, Relativistic
  Heavy Ion Physics}},}\ \ (\bibinfo  {publisher} {Springer},\ \bibinfo {year}
  {2010})\ Chap.\ \bibinfo {chapter} {The QCD confinement transition: Hadron
  formation}\BibitemShut {NoStop}%
\bibitem [{\citenamefont {Cleymans}\ and\ \citenamefont
  {Redlich}(1998)}]{Cleymans:1998fq}%
  \BibitemOpen
  \bibfield  {author} {\bibinfo {author} {\bibfnamefont {J.}~\bibnamefont
  {Cleymans}}\ and\ \bibinfo {author} {\bibfnamefont {K.}~\bibnamefont
  {Redlich}},\ }\href {\doibase 10.1103/PhysRevLett.81.5284} {\bibfield
  {journal} {\bibinfo  {journal} {Phys.Rev.Lett.}\ }\textbf {\bibinfo {volume}
  {81}},\ \bibinfo {pages} {5284} (\bibinfo {year} {1998})}\BibitemShut
  {NoStop}%
\bibitem [{\citenamefont {Rapp}\ and\ \citenamefont
  {Shuryak}(2001)}]{Rapp:2000gy}%
  \BibitemOpen
  \bibfield  {author} {\bibinfo {author} {\bibfnamefont {R.}~\bibnamefont
  {Rapp}}\ and\ \bibinfo {author} {\bibfnamefont {E.~V.}\ \bibnamefont
  {Shuryak}},\ }\href {\doibase 10.1103/PhysRevLett.86.2980} {\bibfield
  {journal} {\bibinfo  {journal} {Phys.Rev.Lett.}\ }\textbf {\bibinfo {volume}
  {86}},\ \bibinfo {pages} {2980} (\bibinfo {year} {2001})}\BibitemShut
  {NoStop}%
\bibitem [{\citenamefont {Braun-Munzinger}\ \emph {et~al.}(2004)\citenamefont
  {Braun-Munzinger}, \citenamefont {Stachel},\ and\ \citenamefont
  {Wetterich}}]{BraunMunzinger:2003zz}%
  \BibitemOpen
  \bibfield  {author} {\bibinfo {author} {\bibfnamefont {P.}~\bibnamefont
  {Braun-Munzinger}}, \bibinfo {author} {\bibfnamefont {J.}~\bibnamefont
  {Stachel}}, \ and\ \bibinfo {author} {\bibfnamefont {C.}~\bibnamefont
  {Wetterich}},\ }\href {\doibase 10.1016/j.physletb.2004.05.081} {\bibfield
  {journal} {\bibinfo  {journal} {Phys.Lett.}\ }\textbf {\bibinfo {volume}
  {B596}},\ \bibinfo {pages} {61} (\bibinfo {year} {2004})}\BibitemShut
  {NoStop}%
\bibitem [{\citenamefont {Heinz}\ and\ \citenamefont
  {Kestin}(2006)}]{Heinz:2006ur}%
  \BibitemOpen
  \bibfield  {author} {\bibinfo {author} {\bibfnamefont {U.}~\bibnamefont
  {Heinz}}\ and\ \bibinfo {author} {\bibfnamefont {G.}~\bibnamefont {Kestin}},\
  }\href@noop {} {\bibfield  {journal} {\bibinfo  {journal} {PoS}\ }\textbf
  {\bibinfo {volume} {CPOD2006}},\ \bibinfo {pages} {038} (\bibinfo {year}
  {2006})}\BibitemShut {NoStop}%
\bibitem [{\citenamefont {Abelev}\ \emph
  {et~al.}(2012{\natexlab{a}})\citenamefont {Abelev} \emph
  {et~al.}}]{prl-spectra}%
  \BibitemOpen
  \bibfield  {author} {\bibinfo {author} {\bibfnamefont {B.}~\bibnamefont
  {Abelev}} \emph {et~al.} (\bibinfo {collaboration} {ALICE Collaboration}),\
  }\href@noop {} {\bibfield  {journal} {\bibinfo  {journal} {Phys.Rev.Lett.}\
  }\textbf {\bibinfo {volume} {109}},\ \bibinfo {pages} {252301} (\bibinfo
  {year} {2012}{\natexlab{a}})}\BibitemShut {NoStop}%
\bibitem [{\citenamefont {Abelev}\ \emph {et~al.}(2006)\citenamefont {Abelev}
  \emph {et~al.}}]{Abelev:2006jr}%
  \BibitemOpen
  \bibfield  {author} {\bibinfo {author} {\bibfnamefont {B.}~\bibnamefont
  {Abelev}} \emph {et~al.} (\bibinfo {collaboration} {STAR Collaboration}),\
  }\href {\doibase 10.1103/PhysRevLett.97.152301} {\bibfield  {journal}
  {\bibinfo  {journal} {Phys.Rev.Lett.}\ }\textbf {\bibinfo {volume} {97}},\
  \bibinfo {pages} {152301} (\bibinfo {year} {2006})}\BibitemShut {NoStop}%
\bibitem [{\citenamefont {Fries}\ \emph
  {et~al.}(2003{\natexlab{a}})\citenamefont {Fries}, \citenamefont {Muller},
  \citenamefont {Nonaka},\ and\ \citenamefont {Bass}}]{Fries:2003vb}%
  \BibitemOpen
  \bibfield  {author} {\bibinfo {author} {\bibfnamefont {R.}~\bibnamefont
  {Fries}}, \bibinfo {author} {\bibfnamefont {B.}~\bibnamefont {Muller}},
  \bibinfo {author} {\bibfnamefont {C.}~\bibnamefont {Nonaka}}, \ and\ \bibinfo
  {author} {\bibfnamefont {S.}~\bibnamefont {Bass}},\ }\href {\doibase
  10.1103/PhysRevLett.90.202303} {\bibfield  {journal} {\bibinfo  {journal}
  {Phys.Rev.Lett.}\ }\textbf {\bibinfo {volume} {90}},\ \bibinfo {pages}
  {202303} (\bibinfo {year} {2003}{\natexlab{a}})}\BibitemShut {NoStop}%
\bibitem [{\citenamefont {Greco}\ \emph {et~al.}(2003)\citenamefont {Greco},
  \citenamefont {Ko},\ and\ \citenamefont {Levai}}]{Greco:2003xt}%
  \BibitemOpen
  \bibfield  {author} {\bibinfo {author} {\bibfnamefont {V.}~\bibnamefont
  {Greco}}, \bibinfo {author} {\bibfnamefont {C.}~\bibnamefont {Ko}}, \ and\
  \bibinfo {author} {\bibfnamefont {P.}~\bibnamefont {Levai}},\ }\href
  {\doibase 10.1103/PhysRevLett.90.202302} {\bibfield  {journal} {\bibinfo
  {journal} {Phys.Rev.Lett.}\ }\textbf {\bibinfo {volume} {90}},\ \bibinfo
  {pages} {202302} (\bibinfo {year} {2003})}\BibitemShut {NoStop}%
\bibitem [{\citenamefont {Aamodt}\ \emph
  {et~al.}(2011{\natexlab{a}})\citenamefont {Aamodt} \emph
  {et~al.}}]{Aamodt:2011zj}%
  \BibitemOpen
  \bibfield  {author} {\bibinfo {author} {\bibfnamefont {K.}~\bibnamefont
  {Aamodt}} \emph {et~al.} (\bibinfo {collaboration} {ALICE Collaboration}),\
  }\href@noop {} {\bibfield  {journal} {\bibinfo  {journal} {Eur.Phys.J}\
  }\textbf {\bibinfo {volume} {C71}},\ \bibinfo {pages} {1} (\bibinfo {year}
  {2011}{\natexlab{a}})}\BibitemShut {NoStop}%
\bibitem [{\citenamefont {Alessandro}\ \emph {et~al.}(2006)\citenamefont
  {Alessandro} \emph {et~al.}}]{Alessandro:2006yt}%
  \BibitemOpen
  \bibfield  {author} {\bibinfo {author} {\bibfnamefont {B.}~\bibnamefont
  {Alessandro}} \emph {et~al.} (\bibinfo {collaboration} {ALICE
  Collaboration}),\ }\href {\doibase 10.1088/0954-3899/32/10/001} {\bibfield
  {journal} {\bibinfo  {journal} {J.Phys.G}\ }\textbf {\bibinfo {volume}
  {G32}},\ \bibinfo {pages} {1295} (\bibinfo {year} {2006})}\BibitemShut
  {NoStop}%
\bibitem [{\citenamefont {Aamodt}\ \emph {et~al.}(2008)\citenamefont {Aamodt}
  \emph {et~al.}}]{Aamodt:2008zz}%
  \BibitemOpen
  \bibfield  {author} {\bibinfo {author} {\bibfnamefont {K.}~\bibnamefont
  {Aamodt}} \emph {et~al.} (\bibinfo {collaboration} {ALICE Collaboration}),\
  }\href {\doibase 10.1088/1748-0221/3/08/S08002} {\bibfield  {journal}
  {\bibinfo  {journal} {JINST}\ }\textbf {\bibinfo {volume} {3}},\ \bibinfo
  {pages} {S08002} (\bibinfo {year} {2008})}\BibitemShut {NoStop}%
\bibitem [{\citenamefont {Abelev}\ \emph {et~al.}(2013)\citenamefont {Abelev}
  \emph {et~al.}}]{centrality-paper}%
  \BibitemOpen
  \bibfield  {author} {\bibinfo {author} {\bibfnamefont {B.}~\bibnamefont
  {Abelev}} \emph {et~al.} (\bibinfo {collaboration} {ALICE Collaboration}),\
  }\href@noop {} {\  (\bibinfo {year} {2013})},\ \Eprint
  {http://arxiv.org/abs/1301.4361} {arXiv:1301.4361 [nucl-ex]} \BibitemShut
  {NoStop}%
\bibitem [{\citenamefont {Aamodt}\ \emph
  {et~al.}(2011{\natexlab{b}})\citenamefont {Aamodt} \emph
  {et~al.}}]{Aamodt:2010cz}%
  \BibitemOpen
  \bibfield  {author} {\bibinfo {author} {\bibfnamefont {K.}~\bibnamefont
  {Aamodt}} \emph {et~al.} (\bibinfo {collaboration} {ALICE Collaboration}),\
  }\href {\doibase 10.1103/PhysRevLett.106.032301} {\bibfield  {journal}
  {\bibinfo  {journal} {Phys.Rev.Lett.}\ }\textbf {\bibinfo {volume} {106}},\
  \bibinfo {pages} {032301} (\bibinfo {year} {2011}{\natexlab{b}})}\BibitemShut
  {NoStop}%
\bibitem [{\citenamefont {Aamodt}\ \emph
  {et~al.}(2010{\natexlab{a}})\citenamefont {Aamodt} \emph
  {et~al.}}]{Aamodt:2010pb}%
  \BibitemOpen
  \bibfield  {author} {\bibinfo {author} {\bibfnamefont {K.}~\bibnamefont
  {Aamodt}} \emph {et~al.} (\bibinfo {collaboration} {ALICE Collaboration}),\
  }\href {\doibase 10.1103/PhysRevLett.105.252301} {\bibfield  {journal}
  {\bibinfo  {journal} {Phys.Rev.Lett.}\ }\textbf {\bibinfo {volume} {105}},\
  \bibinfo {pages} {252301} (\bibinfo {year} {2010}{\natexlab{a}})}\BibitemShut
  {NoStop}%
\bibitem [{\citenamefont {Abelev}\ \emph
  {et~al.}(2012{\natexlab{b}})\citenamefont {Abelev} \emph
  {et~al.}}]{ALICE:2012aa}%
  \BibitemOpen
  \bibfield  {author} {\bibinfo {author} {\bibfnamefont {B.}~\bibnamefont
  {Abelev}} \emph {et~al.} (\bibinfo {collaboration} {ALICE Collaboration}),\
  }\href {\doibase 10.1103/PhysRevLett.109.252302} {\bibfield  {journal}
  {\bibinfo  {journal} {Phys.Rev.Lett.}\ }\textbf {\bibinfo {volume} {109}},\
  \bibinfo {pages} {252302} (\bibinfo {year} {2012}{\natexlab{b}})}\BibitemShut
  {NoStop}%
\bibitem [{\citenamefont {Aamodt}\ \emph
  {et~al.}(2010{\natexlab{b}})\citenamefont {Aamodt} \emph
  {et~al.}}]{Aamodt:2010ft}%
  \BibitemOpen
  \bibfield  {author} {\bibinfo {author} {\bibfnamefont {K.}~\bibnamefont
  {Aamodt}} \emph {et~al.} (\bibinfo {collaboration} {ALICE Collaboration}),\
  }\href {\doibase 10.1140/epjc/s10052-010-1339-x} {\bibfield  {journal}
  {\bibinfo  {journal} {Eur.Phys.J.}\ }\textbf {\bibinfo {volume} {C68}},\
  \bibinfo {pages} {89} (\bibinfo {year} {2010}{\natexlab{b}})}\BibitemShut
  {NoStop}%
\bibitem [{\citenamefont {Beringer}\ \emph {et~al.}(2012)\citenamefont
  {Beringer} \emph {et~al.}}]{Beringer:1900zz}%
  \BibitemOpen
  \bibfield  {author} {\bibinfo {author} {\bibfnamefont {J.}~\bibnamefont
  {Beringer}} \emph {et~al.} (\bibinfo {collaboration} {Particle Data Group}),\
  }\href {\doibase 10.1103/PhysRevD.86.010001} {\bibfield  {journal} {\bibinfo
  {journal} {Phys.Rev.}\ }\textbf {\bibinfo {volume} {D86}},\ \bibinfo {pages}
  {010001} (\bibinfo {year} {2012})},\ \bibinfo {note} {chapter 30}\BibitemShut
  {NoStop}%
\bibitem [{\citenamefont {{X.-N. Wang and M. Gyulassy}}(1991)}]{hijing}%
  \BibitemOpen
  \bibfield  {author} {\bibinfo {author} {\bibnamefont {{X.-N. Wang and M.
  Gyulassy}}},\ }\href@noop {} {\bibfield  {journal} {\bibinfo  {journal}
  {Phys. Rev. D}\ }\textbf {\bibinfo {volume} {44}},\ \bibinfo {pages} {3501}
  (\bibinfo {year} {1991})}\BibitemShut {NoStop}%
\bibitem [{\citenamefont {Brun}\ \emph {et~al.}(1994)\citenamefont {Brun},
  \citenamefont {Carminati},\ and\ \citenamefont {Giani}}]{Geant:1994zzo}%
  \BibitemOpen
  \bibfield  {author} {\bibinfo {author} {\bibfnamefont {R.}~\bibnamefont
  {Brun}}, \bibinfo {author} {\bibfnamefont {F.}~\bibnamefont {Carminati}}, \
  and\ \bibinfo {author} {\bibfnamefont {S.}~\bibnamefont {Giani}},\
  }\href@noop {} {\  (\bibinfo {year} {1994})},\ \bibinfo {note}
  {{CERN-W5013}}\BibitemShut {NoStop}%
\bibitem [{\citenamefont {Agostinelli}\ \emph {et~al.}(2003)\citenamefont
  {Agostinelli} \emph {et~al.}}]{Agostinelli:2002hh}%
  \BibitemOpen
  \bibfield  {author} {\bibinfo {author} {\bibfnamefont {S.}~\bibnamefont
  {Agostinelli}} \emph {et~al.} (\bibinfo {collaboration} {GEANT4}),\ }\href
  {\doibase 10.1016/S0168-9002(03)01368-8} {\bibfield  {journal} {\bibinfo
  {journal} {Nucl.Instrum.Meth.}\ }\textbf {\bibinfo {volume} {A506}},\
  \bibinfo {pages} {250} (\bibinfo {year} {2003})}\BibitemShut {NoStop}%
\bibitem [{\citenamefont {Battistoni}\ \emph {et~al.}(2007)\citenamefont
  {Battistoni}, \citenamefont {Muraro}, \citenamefont {Sala}, \citenamefont
  {Cerutti}, \citenamefont {Ferrari} \emph {et~al.}}]{Battistoni:2007zzb}%
  \BibitemOpen
  \bibfield  {author} {\bibinfo {author} {\bibfnamefont {G.}~\bibnamefont
  {Battistoni}}, \bibinfo {author} {\bibfnamefont {S.}~\bibnamefont {Muraro}},
  \bibinfo {author} {\bibfnamefont {P.~R.}\ \bibnamefont {Sala}}, \bibinfo
  {author} {\bibfnamefont {F.}~\bibnamefont {Cerutti}}, \bibinfo {author}
  {\bibfnamefont {A.}~\bibnamefont {Ferrari}},  \emph {et~al.},\ }\href
  {\doibase 10.1063/1.2720455} {\bibfield  {journal} {\bibinfo  {journal} {AIP
  Conf.Proc.}\ }\textbf {\bibinfo {volume} {896}},\ \bibinfo {pages} {31}
  (\bibinfo {year} {2007})}\BibitemShut {NoStop}%
\bibitem [{\citenamefont {Aamodt}\ \emph
  {et~al.}(2010{\natexlab{c}})\citenamefont {Aamodt} \emph
  {et~al.}}]{Aamodt:2010dx}%
  \BibitemOpen
  \bibfield  {author} {\bibinfo {author} {\bibfnamefont {K.}~\bibnamefont
  {Aamodt}} \emph {et~al.} (\bibinfo {collaboration} {ALICE Collaboration}),\
  }\href {\doibase 10.1103/PhysRevLett.105.072002} {\bibfield  {journal}
  {\bibinfo  {journal} {Phys.Rev.Lett.}\ }\textbf {\bibinfo {volume} {105}},\
  \bibinfo {pages} {072002} (\bibinfo {year} {2010}{\natexlab{c}})}\BibitemShut
  {NoStop}%
\bibitem [{\citenamefont {Lee}\ and\ \citenamefont
  {Redwine}(2002)}]{Lee:2002eq}%
  \BibitemOpen
  \bibfield  {author} {\bibinfo {author} {\bibfnamefont {T.}~\bibnamefont
  {Lee}}\ and\ \bibinfo {author} {\bibfnamefont {R.}~\bibnamefont {Redwine}},\
  }\href {\doibase 10.1146/annurev.nucl.52.050102.090713} {\bibfield  {journal}
  {\bibinfo  {journal} {Ann.Rev.Nucl.Part.Sci.}\ }\textbf {\bibinfo {volume}
  {52}},\ \bibinfo {pages} {23} (\bibinfo {year} {2002})}\BibitemShut {NoStop}%
\bibitem [{\citenamefont {Friedman}\ \emph {et~al.}(1997)\citenamefont
  {Friedman}, \citenamefont {Gal}, \citenamefont {Weiss}, \citenamefont
  {Aclander}, \citenamefont {Alster} \emph {et~al.}}]{Friedman:1997eq}%
  \BibitemOpen
  \bibfield  {author} {\bibinfo {author} {\bibfnamefont {E.}~\bibnamefont
  {Friedman}}, \bibinfo {author} {\bibfnamefont {A.}~\bibnamefont {Gal}},
  \bibinfo {author} {\bibfnamefont {R.}~\bibnamefont {Weiss}}, \bibinfo
  {author} {\bibfnamefont {J.}~\bibnamefont {Aclander}}, \bibinfo {author}
  {\bibfnamefont {J.}~\bibnamefont {Alster}},  \emph {et~al.},\ }\href
  {\doibase 10.1103/PhysRevC.55.1304} {\bibfield  {journal} {\bibinfo
  {journal} {Phys.Rev.}\ }\textbf {\bibinfo {volume} {C55}},\ \bibinfo {pages}
  {1304} (\bibinfo {year} {1997})}\BibitemShut {NoStop}%
\bibitem [{\citenamefont {Ashery}\ \emph {et~al.}(1981)\citenamefont {Ashery},
  \citenamefont {Navon}, \citenamefont {Azuelos}, \citenamefont {Walter},
  \citenamefont {Pfeiffer} \emph {et~al.}}]{Ashery:1981tq}%
  \BibitemOpen
  \bibfield  {author} {\bibinfo {author} {\bibfnamefont {D.}~\bibnamefont
  {Ashery}}, \bibinfo {author} {\bibfnamefont {I.}~\bibnamefont {Navon}},
  \bibinfo {author} {\bibfnamefont {G.}~\bibnamefont {Azuelos}}, \bibinfo
  {author} {\bibfnamefont {H.}~\bibnamefont {Walter}}, \bibinfo {author}
  {\bibfnamefont {H.}~\bibnamefont {Pfeiffer}},  \emph {et~al.},\ }\href
  {\doibase 10.1103/PhysRevC.23.2173} {\bibfield  {journal} {\bibinfo
  {journal} {Phys.Rev.}\ }\textbf {\bibinfo {volume} {C23}},\ \bibinfo {pages}
  {2173} (\bibinfo {year} {1981})}\BibitemShut {NoStop}%
\bibitem [{\citenamefont {Carlson}(1996)}]{Carlson199693}%
  \BibitemOpen
  \bibfield  {author} {\bibinfo {author} {\bibfnamefont {R.}~\bibnamefont
  {Carlson}},\ }\href {\doibase 10.1006/adnd.1996.0010} {\bibfield  {journal}
  {\bibinfo  {journal} {Atomic Data and Nuclear Data Tables}\ }\textbf
  {\bibinfo {volume} {63}},\ \bibinfo {pages} {93 } (\bibinfo {year}
  {1996})}\BibitemShut {NoStop}%
\bibitem [{\citenamefont {Bendiscioli}\ and\ \citenamefont
  {Kharzeev}(1994)}]{Bendiscioli:1994uv}%
  \BibitemOpen
  \bibfield  {author} {\bibinfo {author} {\bibfnamefont {G.}~\bibnamefont
  {Bendiscioli}}\ and\ \bibinfo {author} {\bibfnamefont {D.}~\bibnamefont
  {Kharzeev}},\ }\href@noop {} {\bibfield  {journal} {\bibinfo  {journal}
  {Riv.Nuovo Cim.}\ }\textbf {\bibinfo {volume} {17N6}},\ \bibinfo {pages} {1}
  (\bibinfo {year} {1994})}\BibitemShut {NoStop}%
\bibitem [{\citenamefont {Sollfrank}\ \emph {et~al.}(1990)\citenamefont
  {Sollfrank}, \citenamefont {Koch},\ and\ \citenamefont
  {Heinz}}]{Sollfrank:1990qz}%
  \BibitemOpen
  \bibfield  {author} {\bibinfo {author} {\bibfnamefont {J.}~\bibnamefont
  {Sollfrank}}, \bibinfo {author} {\bibfnamefont {P.}~\bibnamefont {Koch}}, \
  and\ \bibinfo {author} {\bibfnamefont {U.~W.}\ \bibnamefont {Heinz}},\ }\href
  {\doibase 10.1016/0370-2693(90)90870-C} {\bibfield  {journal} {\bibinfo
  {journal} {Phys.Lett.}\ }\textbf {\bibinfo {volume} {B252}},\ \bibinfo
  {pages} {256} (\bibinfo {year} {1990})}\BibitemShut {NoStop}%
\bibitem [{\citenamefont {Abelev}\ \emph {et~al.}(2009)\citenamefont {Abelev}
  \emph {et~al.}}]{Abelev:2008ez}%
  \BibitemOpen
  \bibfield  {author} {\bibinfo {author} {\bibfnamefont {B.}~\bibnamefont
  {Abelev}} \emph {et~al.} (\bibinfo {collaboration} {STAR Collaboration}),\
  }\href {\doibase 10.1103/PhysRevC.79.034909} {\bibfield  {journal} {\bibinfo
  {journal} {Phys.Rev.}\ }\textbf {\bibinfo {volume} {C79}},\ \bibinfo {pages}
  {034909} (\bibinfo {year} {2009})}\BibitemShut {NoStop}%
\bibitem [{\citenamefont {Adler}\ \emph {et~al.}(2005)\citenamefont {Adler}
  \emph {et~al.}}]{Adler:2004zn}%
  \BibitemOpen
  \bibfield  {author} {\bibinfo {author} {\bibfnamefont {S.}~\bibnamefont
  {Adler}} \emph {et~al.} (\bibinfo {collaboration} {PHENIX Collaboration}),\
  }\href {\doibase 10.1103/PhysRevC.71.049901, 10.1103/PhysRevC.71.034908}
  {\bibfield  {journal} {\bibinfo  {journal} {Phys.Rev.}\ }\textbf {\bibinfo
  {volume} {C71}},\ \bibinfo {pages} {034908} (\bibinfo {year}
  {2005})}\BibitemShut {NoStop}%
\bibitem [{\citenamefont {Bearden}\ \emph {et~al.}(2002)\citenamefont {Bearden}
  \emph {et~al.}}]{Bearden:2001qq}%
  \BibitemOpen
  \bibfield  {author} {\bibinfo {author} {\bibfnamefont {I.}~\bibnamefont
  {Bearden}} \emph {et~al.} (\bibinfo {collaboration} {BRAHMS Collaboration}),\
  }\href {\doibase 10.1103/PhysRevLett.88.202301} {\bibfield  {journal}
  {\bibinfo  {journal} {Phys.Rev.Lett.}\ }\textbf {\bibinfo {volume} {88}},\
  \bibinfo {pages} {202301} (\bibinfo {year} {2002})}\BibitemShut {NoStop}%
\bibitem [{\citenamefont {Adler}\ \emph {et~al.}(2004)\citenamefont {Adler}
  \emph {et~al.}}]{Adler:2003cb}%
  \BibitemOpen
  \bibfield  {author} {\bibinfo {author} {\bibfnamefont {S.~S.}\ \bibnamefont
  {Adler}} \emph {et~al.} (\bibinfo {collaboration} {PHENIX Collaboration}),\
  }\href {\doibase 10.1103/PhysRevC.69.034909} {\bibfield  {journal} {\bibinfo
  {journal} {Phys.Rev.}\ }\textbf {\bibinfo {volume} {C69}},\ \bibinfo {pages}
  {034909} (\bibinfo {year} {2004})}\BibitemShut {NoStop}%
\bibitem [{\citenamefont {Bozek}\ and\ \citenamefont
  {Wyskiel-Piekarska}(2012)}]{Bozek:2012qs}%
  \BibitemOpen
  \bibfield  {author} {\bibinfo {author} {\bibfnamefont {P.}~\bibnamefont
  {Bozek}}\ and\ \bibinfo {author} {\bibfnamefont {I.}~\bibnamefont
  {Wyskiel-Piekarska}},\ }\href@noop {} {\  (\bibinfo {year} {2012})},\ \Eprint
  {http://arxiv.org/abs/1203.6513} {arXiv:1203.6513 [nucl-th]} \BibitemShut
  {NoStop}%
\bibitem [{\citenamefont {Karpenko}\ \emph {et~al.}(2013)\citenamefont
  {Karpenko}, \citenamefont {Sinyukov},\ and\ \citenamefont
  {Werner}}]{Karpenko:2012yf}%
  \BibitemOpen
  \bibfield  {author} {\bibinfo {author} {\bibfnamefont {I.}~\bibnamefont
  {Karpenko}}, \bibinfo {author} {\bibfnamefont {Y.}~\bibnamefont {Sinyukov}},
  \ and\ \bibinfo {author} {\bibfnamefont {K.}~\bibnamefont {Werner}},\ }\href
  {\doibase 10.1103/PhysRevC.87.024914} {\bibfield  {journal} {\bibinfo
  {journal} {Phys.Rev.}\ }\textbf {\bibinfo {volume} {C87}},\ \bibinfo {pages}
  {024914} (\bibinfo {year} {2013})}\BibitemShut {NoStop}%
\bibitem [{\citenamefont {Werner}\ \emph {et~al.}(2012)\citenamefont {Werner},
  \citenamefont {Karpenko}, \citenamefont {Bleicher}, \citenamefont {Pierog},\
  and\ \citenamefont {Porteboeuf-Houssais}}]{Werner:2012xh}%
  \BibitemOpen
  \bibfield  {author} {\bibinfo {author} {\bibfnamefont {K.}~\bibnamefont
  {Werner}}, \bibinfo {author} {\bibfnamefont {I.}~\bibnamefont {Karpenko}},
  \bibinfo {author} {\bibfnamefont {M.}~\bibnamefont {Bleicher}}, \bibinfo
  {author} {\bibfnamefont {T.}~\bibnamefont {Pierog}}, \ and\ \bibinfo {author}
  {\bibfnamefont {S.}~\bibnamefont {Porteboeuf-Houssais}},\ }\href@noop {}
  {\bibfield  {journal} {\bibinfo  {journal} {Phys. Rev. C}\ }\textbf {\bibinfo
  {volume} {85}},\ \bibinfo {pages} {064907} (\bibinfo {year}
  {2012})}\BibitemShut {NoStop}%
\bibitem [{\citenamefont {Shen}\ \emph {et~al.}(2011)\citenamefont {Shen},
  \citenamefont {Heinz}, \citenamefont {Huovinen},\ and\ \citenamefont
  {Song}}]{Shen:2011eg}%
  \BibitemOpen
  \bibfield  {author} {\bibinfo {author} {\bibfnamefont {C.}~\bibnamefont
  {Shen}}, \bibinfo {author} {\bibfnamefont {U.}~\bibnamefont {Heinz}},
  \bibinfo {author} {\bibfnamefont {P.}~\bibnamefont {Huovinen}}, \ and\
  \bibinfo {author} {\bibfnamefont {H.}~\bibnamefont {Song}},\ }\href {\doibase
  10.1103/PhysRevC.84.044903} {\bibfield  {journal} {\bibinfo  {journal}
  {Phys.Rev.}\ }\textbf {\bibinfo {volume} {C84}},\ \bibinfo {pages} {044903}
  (\bibinfo {year} {2011})}\BibitemShut {NoStop}%
\bibitem [{\citenamefont {Karpenko}\ and\ \citenamefont
  {Sinyukov}(2011)}]{Karpenko:2011qn}%
  \BibitemOpen
  \bibfield  {author} {\bibinfo {author} {\bibfnamefont {Y.}~\bibnamefont
  {Karpenko}}\ and\ \bibinfo {author} {\bibfnamefont {Y.}~\bibnamefont
  {Sinyukov}},\ }\href {\doibase 10.1088/0954-3899/38/12/124059} {\bibfield
  {journal} {\bibinfo  {journal} {J.Phys.G}\ }\textbf {\bibinfo {volume}
  {G38}},\ \bibinfo {pages} {124059} (\bibinfo {year} {2011})}\BibitemShut
  {NoStop}%
\bibitem [{\citenamefont {Bozek}(2012)}]{Bozek:2011ua}%
  \BibitemOpen
  \bibfield  {author} {\bibinfo {author} {\bibfnamefont {P.}~\bibnamefont
  {Bozek}},\ }\href {\doibase 10.1103/PhysRevC.85.034901} {\bibfield  {journal}
  {\bibinfo  {journal} {Phys.Rev.}\ }\textbf {\bibinfo {volume} {C85}},\
  \bibinfo {pages} {034901} (\bibinfo {year} {2012})}\BibitemShut {NoStop}%
\bibitem [{\citenamefont {Bozek}(2011)}]{Bozek:2011gq}%
  \BibitemOpen
  \bibfield  {author} {\bibinfo {author} {\bibfnamefont {P.}~\bibnamefont
  {Bozek}},\ }\href@noop {} {\  (\bibinfo {year} {2011})},\ \Eprint
  {http://arxiv.org/abs/1111.4398} {arXiv:1111.4398 [nucl-th]} \BibitemShut
  {NoStop}%
\bibitem [{\citenamefont {Ryu}\ \emph {et~al.}(2012)\citenamefont {Ryu},
  \citenamefont {Jeon}, \citenamefont {Gale}, \citenamefont {Schenke},\ and\
  \citenamefont {Young}}]{Ryu:2012at}%
  \BibitemOpen
  \bibfield  {author} {\bibinfo {author} {\bibfnamefont {S.}~\bibnamefont
  {Ryu}}, \bibinfo {author} {\bibfnamefont {S.}~\bibnamefont {Jeon}}, \bibinfo
  {author} {\bibfnamefont {C.}~\bibnamefont {Gale}}, \bibinfo {author}
  {\bibfnamefont {B.}~\bibnamefont {Schenke}}, \ and\ \bibinfo {author}
  {\bibfnamefont {C.}~\bibnamefont {Young}},\ }\href@noop {} {\  (\bibinfo
  {year} {2012})},\ \Eprint {http://arxiv.org/abs/1210.4588} {arXiv:1210.4588
  [hep-ph]} \BibitemShut {NoStop}%
\bibitem [{\citenamefont {Petersen}(2011)}]{Petersen:2011sb}%
  \BibitemOpen
  \bibfield  {author} {\bibinfo {author} {\bibfnamefont {H.}~\bibnamefont
  {Petersen}},\ }\href {\doibase 10.1103/PhysRevC.84.034912} {\bibfield
  {journal} {\bibinfo  {journal} {Phys.Rev.}\ }\textbf {\bibinfo {volume}
  {C84}},\ \bibinfo {pages} {034912} (\bibinfo {year} {2011})}\BibitemShut
  {NoStop}%
\bibitem [{\citenamefont {Fries}\ \emph
  {et~al.}(2003{\natexlab{b}})\citenamefont {Fries}, \citenamefont {Muller},
  \citenamefont {Nonaka},\ and\ \citenamefont {Bass}}]{Fries:2003kq}%
  \BibitemOpen
  \bibfield  {author} {\bibinfo {author} {\bibfnamefont {R.}~\bibnamefont
  {Fries}}, \bibinfo {author} {\bibfnamefont {B.}~\bibnamefont {Muller}},
  \bibinfo {author} {\bibfnamefont {C.}~\bibnamefont {Nonaka}}, \ and\ \bibinfo
  {author} {\bibfnamefont {S.}~\bibnamefont {Bass}},\ }\href {\doibase
  10.1103/PhysRevC.68.044902} {\bibfield  {journal} {\bibinfo  {journal}
  {Phys.Rev.}\ }\textbf {\bibinfo {volume} {C68}},\ \bibinfo {pages} {044902}
  (\bibinfo {year} {2003}{\natexlab{b}})}\BibitemShut {NoStop}%
\bibitem [{\citenamefont {Fries}(2012)}]{FriesPrivateComm}%
  \BibitemOpen
  \bibfield  {author} {\bibinfo {author} {\bibfnamefont {R.}~\bibnamefont
  {Fries}},\ }\href@noop {} {\bibfield  {journal} {\bibinfo  {journal} {Private
  Communication}\ } (\bibinfo {year} {2012})}\BibitemShut {NoStop}%
\bibitem [{\citenamefont {Lamont}(2007)}]{Lamont:2007ce}%
  \BibitemOpen
  \bibfield  {author} {\bibinfo {author} {\bibfnamefont {M.}~\bibnamefont
  {Lamont}},\ }\href {\doibase 10.1140/epjc/s10052-006-0090-9} {\bibfield
  {journal} {\bibinfo  {journal} {Eur.Phys.J.}\ }\textbf {\bibinfo {volume}
  {C49}},\ \bibinfo {pages} {35} (\bibinfo {year} {2007})}\BibitemShut
  {NoStop}%
\bibitem [{HMP(1998)}]{HMPIDTDR}%
  \BibitemOpen
  \href@noop {} {\emph {\bibinfo {title} {ALICE TDR 1}}},\ \bibinfo {type}
  {Tech. Rep.}\ \bibinfo {number} {LHCC 9819}\ (\bibinfo  {institution}
  {CERN},\ \bibinfo {year} {1998})\BibitemShut {NoStop}%
\bibitem [{\citenamefont {Cleymans}\ \emph {et~al.}(2006)\citenamefont
  {Cleymans}, \citenamefont {Kraus}, \citenamefont {Oeschler}, \citenamefont
  {Redlich},\ and\ \citenamefont {Wheaton}}]{Cleymans:2006xj}%
  \BibitemOpen
  \bibfield  {author} {\bibinfo {author} {\bibfnamefont {J.}~\bibnamefont
  {Cleymans}}, \bibinfo {author} {\bibfnamefont {I.}~\bibnamefont {Kraus}},
  \bibinfo {author} {\bibfnamefont {H.}~\bibnamefont {Oeschler}}, \bibinfo
  {author} {\bibfnamefont {K.}~\bibnamefont {Redlich}}, \ and\ \bibinfo
  {author} {\bibfnamefont {S.}~\bibnamefont {Wheaton}},\ }\href {\doibase
  10.1103/PhysRevC.74.034903} {\bibfield  {journal} {\bibinfo  {journal}
  {Phys.Rev.}\ }\textbf {\bibinfo {volume} {C74}},\ \bibinfo {pages} {034903}
  (\bibinfo {year} {2006})}\BibitemShut {NoStop}%
\bibitem [{\citenamefont {Andronic}\ \emph {et~al.}(2006)\citenamefont
  {Andronic}, \citenamefont {Braun-Munzinger},\ and\ \citenamefont
  {Stachel}}]{Andronic:2005yp}%
  \BibitemOpen
  \bibfield  {author} {\bibinfo {author} {\bibfnamefont {A.}~\bibnamefont
  {Andronic}}, \bibinfo {author} {\bibfnamefont {P.}~\bibnamefont
  {Braun-Munzinger}}, \ and\ \bibinfo {author} {\bibfnamefont {J.}~\bibnamefont
  {Stachel}},\ }\href {\doibase 10.1016/j.nuclphysa.2006.03.012} {\bibfield
  {journal} {\bibinfo  {journal} {Nucl.Phys.}\ }\textbf {\bibinfo {volume}
  {A772}},\ \bibinfo {pages} {167} (\bibinfo {year} {2006})}\BibitemShut
  {NoStop}%
\bibitem [{\citenamefont {Becattini}\ \emph {et~al.}(2010)\citenamefont
  {Becattini}, \citenamefont {Castorina}, \citenamefont {Milov},\ and\
  \citenamefont {Satz}}]{Becattini:2010sk}%
  \BibitemOpen
  \bibfield  {author} {\bibinfo {author} {\bibfnamefont {F.}~\bibnamefont
  {Becattini}}, \bibinfo {author} {\bibfnamefont {P.}~\bibnamefont
  {Castorina}}, \bibinfo {author} {\bibfnamefont {A.}~\bibnamefont {Milov}}, \
  and\ \bibinfo {author} {\bibfnamefont {H.}~\bibnamefont {Satz}},\ }\href
  {\doibase 10.1140/epjc/s10052-010-1265-y} {\bibfield  {journal} {\bibinfo
  {journal} {Eur.Phys.J.}\ }\textbf {\bibinfo {volume} {C66}},\ \bibinfo
  {pages} {377} (\bibinfo {year} {2010})}\BibitemShut {NoStop}%
\bibitem [{\citenamefont {Becattini}\ \emph {et~al.}(2012)\citenamefont
  {Becattini}, \citenamefont {Bleicher}, \citenamefont {Kollegger},
  \citenamefont {Schuster}, \citenamefont {Steinheimer} \emph
  {et~al.}}]{Becattini:2012xb}%
  \BibitemOpen
  \bibfield  {author} {\bibinfo {author} {\bibfnamefont {F.}~\bibnamefont
  {Becattini}}, \bibinfo {author} {\bibfnamefont {M.}~\bibnamefont {Bleicher}},
  \bibinfo {author} {\bibfnamefont {T.}~\bibnamefont {Kollegger}}, \bibinfo
  {author} {\bibfnamefont {T.}~\bibnamefont {Schuster}}, \bibinfo {author}
  {\bibfnamefont {J.}~\bibnamefont {Steinheimer}},  \emph {et~al.},\
  }\href@noop {} {\  (\bibinfo {year} {2012})},\ \Eprint
  {http://arxiv.org/abs/1212.2431} {arXiv:1212.2431 [nucl-th]} \BibitemShut
  {NoStop}%
\bibitem [{\citenamefont {Steinheimer}\ \emph {et~al.}(2012)\citenamefont
  {Steinheimer}, \citenamefont {Aichelin},\ and\ \citenamefont
  {Bleicher}}]{Steinheimer:2012rd}%
  \BibitemOpen
  \bibfield  {author} {\bibinfo {author} {\bibfnamefont {J.}~\bibnamefont
  {Steinheimer}}, \bibinfo {author} {\bibfnamefont {J.}~\bibnamefont
  {Aichelin}}, \ and\ \bibinfo {author} {\bibfnamefont {M.}~\bibnamefont
  {Bleicher}},\ }\href@noop {} {\  (\bibinfo {year} {2012})},\ \Eprint
  {http://arxiv.org/abs/1203.5302} {arXiv:1203.5302 [nucl-th]} \BibitemShut
  {NoStop}%
\bibitem [{\citenamefont {Bleicher}\ \emph {et~al.}(1999)\citenamefont
  {Bleicher}, \citenamefont {Zabrodin}, \citenamefont {Spieles}, \citenamefont
  {Bass}, \citenamefont {Ernst} \emph {et~al.}}]{Bleicher:1999xi}%
  \BibitemOpen
  \bibfield  {author} {\bibinfo {author} {\bibfnamefont {M.}~\bibnamefont
  {Bleicher}}, \bibinfo {author} {\bibfnamefont {E.}~\bibnamefont {Zabrodin}},
  \bibinfo {author} {\bibfnamefont {C.}~\bibnamefont {Spieles}}, \bibinfo
  {author} {\bibfnamefont {S.}~\bibnamefont {Bass}}, \bibinfo {author}
  {\bibfnamefont {C.}~\bibnamefont {Ernst}},  \emph {et~al.},\ }\href {\doibase
  10.1088/0954-3899/25/9/308} {\bibfield  {journal} {\bibinfo  {journal}
  {J.Phys.G}\ }\textbf {\bibinfo {volume} {G25}},\ \bibinfo {pages} {1859}
  (\bibinfo {year} {1999})}\BibitemShut {NoStop}%
\bibitem [{\citenamefont {Bass}\ \emph {et~al.}(1998)\citenamefont {Bass},
  \citenamefont {Belkacem}, \citenamefont {Bleicher}, \citenamefont
  {Brandstetter}, \citenamefont {Bravina} \emph {et~al.}}]{Bass:1998ca}%
  \BibitemOpen
  \bibfield  {author} {\bibinfo {author} {\bibfnamefont {S.}~\bibnamefont
  {Bass}}, \bibinfo {author} {\bibfnamefont {M.}~\bibnamefont {Belkacem}},
  \bibinfo {author} {\bibfnamefont {M.}~\bibnamefont {Bleicher}}, \bibinfo
  {author} {\bibfnamefont {M.}~\bibnamefont {Brandstetter}}, \bibinfo {author}
  {\bibfnamefont {L.}~\bibnamefont {Bravina}},  \emph {et~al.},\ }\href
  {\doibase 10.1016/S0146-6410(98)00058-1, 10.1016/S0146-6410(98)00058-1}
  {\bibfield  {journal} {\bibinfo  {journal} {Prog.Part.Nucl.Phys.}\ }\textbf
  {\bibinfo {volume} {41}},\ \bibinfo {pages} {255} (\bibinfo {year}
  {1998})}\BibitemShut {NoStop}%
\bibitem [{\citenamefont {Pan}\ and\ \citenamefont {Pratt}(2012)}]{Pan:2012ne}%
  \BibitemOpen
  \bibfield  {author} {\bibinfo {author} {\bibfnamefont {Y.}~\bibnamefont
  {Pan}}\ and\ \bibinfo {author} {\bibfnamefont {S.}~\bibnamefont {Pratt}},\
  }\href@noop {} {\  (\bibinfo {year} {2012})},\ \Eprint
  {http://arxiv.org/abs/1210.1577} {arXiv:1210.1577 [nucl-th]} \BibitemShut
  {NoStop}%
\bibitem [{\citenamefont {Rafelski}\ and\ \citenamefont
  {Letessier}(2011)}]{Rafelski:2010cw}%
  \BibitemOpen
  \bibfield  {author} {\bibinfo {author} {\bibfnamefont {J.}~\bibnamefont
  {Rafelski}}\ and\ \bibinfo {author} {\bibfnamefont {J.}~\bibnamefont
  {Letessier}},\ }\href {\doibase 10.1103/PhysRevC.83.054909} {\bibfield
  {journal} {\bibinfo  {journal} {Phys.Rev.}\ }\textbf {\bibinfo {volume}
  {C83}},\ \bibinfo {pages} {054909} (\bibinfo {year} {2011})}\BibitemShut
  {NoStop}%
\bibitem [{\citenamefont {Petran}\ \emph {et~al.}(2013)\citenamefont {Petran},
  \citenamefont {Letessier}, \citenamefont {Petracek},\ and\ \citenamefont
  {Rafelski}}]{Petran:2013lja}%
  \BibitemOpen
  \bibfield  {author} {\bibinfo {author} {\bibfnamefont {M.}~\bibnamefont
  {Petran}}, \bibinfo {author} {\bibfnamefont {J.}~\bibnamefont {Letessier}},
  \bibinfo {author} {\bibfnamefont {V.}~\bibnamefont {Petracek}}, \ and\
  \bibinfo {author} {\bibfnamefont {J.}~\bibnamefont {Rafelski}},\ }\href@noop
  {} {\  (\bibinfo {year} {2013})},\ \Eprint {http://arxiv.org/abs/1303.2098}
  {arXiv:1303.2098 [hep-ph]} \BibitemShut {NoStop}%
\bibitem [{\citenamefont {Ratti}\ \emph {et~al.}(2012)\citenamefont {Ratti},
  \citenamefont {Bellwied}, \citenamefont {Cristoforetti},\ and\ \citenamefont
  {Barbaro}}]{Ratti:2011au}%
  \BibitemOpen
  \bibfield  {author} {\bibinfo {author} {\bibfnamefont {C.}~\bibnamefont
  {Ratti}}, \bibinfo {author} {\bibfnamefont {R.}~\bibnamefont {Bellwied}},
  \bibinfo {author} {\bibfnamefont {M.}~\bibnamefont {Cristoforetti}}, \ and\
  \bibinfo {author} {\bibfnamefont {M.}~\bibnamefont {Barbaro}},\ }\href
  {\doibase 10.1103/PhysRevD.85.014004} {\bibfield  {journal} {\bibinfo
  {journal} {Phys.Rev.}\ }\textbf {\bibinfo {volume} {D85}},\ \bibinfo {pages}
  {014004} (\bibinfo {year} {2012})}\BibitemShut {NoStop}%
\bibitem [{\citenamefont {Bellwied}(2012)}]{Bellwied:2012kh}%
  \BibitemOpen
  \bibfield  {author} {\bibinfo {author} {\bibfnamefont {R.}~\bibnamefont
  {Bellwied}},\ }\href@noop {} {\  (\bibinfo {year} {2012})},\ \Eprint
  {http://arxiv.org/abs/1205.3625} {arXiv:1205.3625 [hep-ph]} \BibitemShut
  {NoStop}%
\end{thebibliography}
\else

\fi
\fi
\fi
\end{document}